\documentclass[useAMS,usenatbib]{mn2e}
\usepackage{graphicx}

\newcommand{\logg}{$\log g$}

\newcommand{\simlt}{\lower.5ex\hbox{$\; \buildrel < \over \sim \;$}}
\newcommand{\simgt}{\lower.5ex\hbox{$\; \buildrel > \over \sim \;$}}

\newcommand{\mgfep}{$\rm [MgFe]'$}
\newcommand{\mgfe}{$\rm [MgFe]$}
\newcommand{\fe}{$\langle\mbox{Fe}\rangle$}

\newcommand{\hbo}{$\rm H\beta_o$}
\newcommand{\hb}{$\rm H\beta$}
\newcommand{\hdf}{$\rm H\delta_F$}
\newcommand{\mgb}{\rm Mg$b$}
\newcommand{\hgf}{$\rm H\gamma_F$}

\newcommand{\kms}{\,km\,s$^{-1}$}

\def\lesssim{\mathrel{\hbox{\rlap{\hbox{\lower4pt\hbox{$\sim$}}}\hbox{$<$}}}}
\def\gtrsim{\mathrel{\hbox{\rlap{\hbox{\lower4pt\hbox{$\sim$}}}\hbox{$>$}}}}
\newcommand{\Msol}{${\rm M}_{\odot}$}
\newcommand{\Rsol}{${\rm R}_{\odot}$}
\newcommand{\Teff}{${\rm T}_{\mbox{\scriptsize eff}}$}
\newcommand{\Feh}{\mbox{$\mbox{[Fe/H]}$}}
\newcommand{\Mh}{\mbox{$\mbox{[M/H]}$}}
\newcommand{\Zh}{\mbox{$\mbox{[Z/H]}$}}
\newcommand{\Mgh}{\mbox{$\mbox{[Mg/H]}$}}
\newcommand{\MgFe}{\mbox{$\mbox{[Mg/Fe]}$}}
\newcommand{\CFe}{\mbox{$\mbox{[C/Fe]}$}}
\newcommand{\OFe}{\mbox{$\mbox{[O/Fe]}$}}
\newcommand{\NFe}{\mbox{$\mbox{[N/Fe]}$}}
\newcommand{\TiFe}{\mbox{$\mbox{[Ti/Fe]}$}}
\newcommand{\CaFe}{\mbox{$\mbox{[Ca/Fe]}$}}
\newcommand{\NaFe}{\mbox{$\mbox{[Na/Fe]}$}}
\newcommand{\aFe}{\mbox{$\mbox{[$\alpha$/Fe]}$}}
\newcommand{\proxy}{\mbox{$\mbox{[$Z_{Mg}/Z_{Fe}$]}$}}
\newcommand{\alfa}{$\alpha$}
\newcommand{\afepSS}{$\rm [Z_{Mg}/Z_{Fe}]_{SS(BaSTI)}$}
\newcommand{\afepB}{$\rm [Z_{Mg}/Z_{Fe}]_{Base(BaSTI)}$}
\newcommand{\afepBP}{$\rm [Z_{Mg}/Z_{Fe}]_{Base(Padova00)}$}
\def \aj {AJ}
\def \mnras {MNRAS}
\def \apj {ApJ}
\def \apjs {ApJS}

\def \aap {A\&A}
\def \aaps {A\&AS}

\def \araa {ARAA}
\def \pasp {PASP}
\def \pasj {PASJ}
\def \aps {AAPS}

\title[Stellar population synthesis with MILES - scaled-solar and \alfa-enhanced models]
{Evolutionary stellar population synthesis with MILES - II. Scaled-solar and
\alfa-enhanced models}

\author[Vazdekis et~al.]{A. Vazdekis$^{1,2}$\thanks{E-mail:vazdekis@iac.es}, 
P. Coelho$^{3}$, S. Cassisi$^{4}$, E. Ricciardelli$^{5}$, J. Falc\'on-Barroso$^{1,2}$,
\newauthor
P. S\'anchez-Bl\'azquez$^{6}$, F. La Barbera$^{7}$, M. A. Beasley$^{1,2}$, A. Pietrinferni$^{4}$\\
$^{1}$Instituto de Astrof\'\i sica de Canarias (IAC), E-38200 La Laguna, Tenerife, Spain\\
$^{2}$Departamento de Astrof\'\i sica, Universidad de La Laguna, E-38205, Tenerife, Spain\\
$^{3}$Instituto de Astronomia, Geof\'isica e Ci\^encias Atmosf\'ericas, Univ. de S\~ao Paulo, Rua do Mat\~ao 1226, 05508-090, S\~ao Paulo, Brazil\\
$^{4}$INAF-Osservatorio Astronomico di Collurania, via Mentore Maggini, 64100 Teramo, Italy\\
$^{5}$Departament d'Astronomia i Astrof\'{\i}sica, Universitat de Valencia, C/Dr. Moliner 50, E-46100, Burjassot, Valencia, Spain\\
$^{6}$Universidad Aut\'onoma de Madrid, Departamento de F\'isica Te\'orica, E28049 Cantoblanco, Madrid, Spain\\
$^{7}$INAF -- Osservatorio Astronomico di Capodimonte, Napoli, Italy\\ 
}
\begin{document}

\date{}

\pagerange{\pageref{firstpage}--\pageref{lastpage}} \pubyear{2002}

\maketitle

\label{firstpage}

\begin{abstract} 

We present models that predict spectra of old- and intermediate-aged stellar
populations at $2.51$\,\AA\,(FWHM) with varying \aFe\ abundance. The models are
based on the MILES library and on corrections from theoretical stellar spectra.
The models employ recent \MgFe\ determinations for the MILES stars and BaSTI
scaled-solar and \alfa-enhanced isochrones. We compute models for a suite of IMF
shapes and slopes, covering a wide age/metallicity range. Using BASTI, we also 
compute "base models" matching The Galactic abundance pattern. We confirm that the
\alfa-enhanced models show a flux excess with respect to the scaled-solar models
blue-ward $\sim$4500\,\AA, which increases with age and metallicity. We also
confirm that both \mgfe\ and \mgfep\ indices are \aFe-insensitive. We show
that the sensitivity of the higher order Balmer lines to \aFe\ resides in their
pseudo-continua, with narrower index definitions yielding lower sensitivity. We
confirm that the \alfa-enhanced models yield bluer (redder) colours in the blue
(red)  spectral range. To match optical colours of massive galaxies we require
both \alfa-enhancement and a bottom-heavy IMF. The comparison of Globular Cluster
line-strengths with our predictions match the \MgFe\ determinations from their
individual stars. We obtain good fits to both full spectra and indices of galaxies
with varying \aFe. Using thousands of SDSS galaxy spectra we obtain a linear
relation between a proxy for the abundance, \afepSS, using solely scaled-solar
models and the \MgFe\ derived with models with varying abundance
(\MgFe$=0.59$\afepSS). Finally we provide a user-friendly, web-based facility, which
allows composite populations with varying IMF and \aFe.

\end{abstract}



\section{Introduction}
\label{intro}

The integrated light emitted by stellar systems such as star clusters and galaxies
provide us with the necessary information to understand their fundamental properties,
and how these properties arise and evolve. From observables like colours,
line-strength indices or full spectral energy distributions (SEDs) we can derive
the Star Formation History (SFH), mean age, metal content, abundance pattern,
stellar Initial Mass Function (IMF) and dust properties of these systems. 

For a quantitative
study of their properties we require stellar population synthesis
models. The method consists in comparing the observables to the predictions of
these models (e.g., \citealt{Tinsley80}). Stellar population synthesis 
models are also employed in other type of studies, e.g., to provide the 
necessary templates for kinematic or redshift measurement, or to determine galaxy 
stellar masses via mass-to-light ratios (M/L), just to mention two applications.   
Evolutionary population synthesis models, which are regarded nowadays as standard, 
take into account the contributions of all possible stars, in proportions prescribed by
stellar evolution theory (e.g., \citealt{Tinsley80}). These models combine at
least three main ingredients that determine the quality of the predictions: a
prescription for the IMF, a set of stellar evolutionary isochrones and stellar
spectral libraries to predict a variety  of observables in several bands such as
fluxes, colours, M/L, or low resolution spectra (e.g.,
\citealt*{Bruzual83, ArimotoYoshii86, GuiderdoniVolmerange87, Bressan94, Fritze94, Worthey94, Vazdekis96, Pegase, KodamaArimoto97, Maraston98, Leitherer99, Maraston05, ConroyGunn10})
and also surface brightness fluctuations (SBFs) (e.g.,
\citealt*{Liu00, Blakeslee01, Liu02, Cantiello03}).

In the last two decades important advances have been made in the field with the
aid of models that predict absorption line-strength indices for single-age,
single-metallicity stellar populations (SSPs) (e.g., \citealt{Worthey94}) and
more complex stellar populations, built as compositions of SSPs (e.g.,
\citealt{Vazdekis96}). These models employ fitting functions that relate
measured line-strength indices to the atmospheric parameters (\Teff, \logg,
\Feh) of the library stars. The most widely used fitting functions are those
obtained with the Lick/IDS stellar library
\citep{Burstein84,Faber85,Gorgas93,Wortheyetal94,WortheyOttaviani97}. The Lick/IDS
system comprises definitions of $25$ absorption line-strengths in the optical
spectral range with varying, low, resolutions (FWHM$>$8-11\,\AA). There are
alternative fitting functions in the optical range (e.g.,
\citealt{Buzzoni95,Gorgas99};\citealt*{Tantalo07};\citealt{Schiavon07}) and in other spectral ranges
(e.g., \citealt{CATIII,MarmolQueralto08,Cenarro09}). The majority of these model predictions
are implicitly based on a scaled-solar\footnote{Some authors use the term ``solar-scaled''
as opposed to scaled-solar, and we regard these terms as interchangeable.} 
heavy element mixture around solar metallicity. This is due to the fact
that these libraries are composed of stars that are imprinted with the abundance
pattern of the solar neighborhood, as a result of their experienced SFH. Such types
of predictions have also been developed in other spectral ranges (e.g.,
\citealt{Saglia02,CATIV,Cenarro09}). 

There are a variety of approaches that have been employed in the literature to
extract the information contained in spectral line indices. The most popular method
is to build key diagnostic model grids by plotting an age-sensitive (e.g., \hb)
versus a metallicity-sensitive indicator (e.g., \mgb, \fe), to estimate these
parameters (e.g.,
\citealt{Rose94,Worthey94,Vazdekis96,Kuntschner06,Schiavon07}),
including estimations of the abundance ratios of individual elements (e.g.,
\citealt*{Trager00,Thomas03,Schiavon07}).
There are other methods that employ as many Lick indices as possible and
simultaneously fit them in a $\chi^2$ sense (e.g.,
\citealt*{Vazdekis97,Proctor04}). An alternative method is the use of Principal
Component Analysis (e.g., \citealt*{Covino95,Wild09}). 

Unlike colours, these indices are insensitive to the effects of dust extinction
(e.g., \citealt{MacArthur05}) and allow us to study the age (as a proxy for the
SFH) and metallicity, \Mh\ (i.e., the total metal content). In addition they
also allow us to study the abundances of various elements from the strengths of
these absorption lines (e.g., \citealt{Rose85,Rose94,SB03}; \citealt*{Carretero04};
\citealt{Trager98,Trager00,ProctorSansom02,Thomas05,Schiavon07,CvD12}).
In fact, one of the most important results has been the finding that massive
early-type galaxies (ETGs) show an overabundance of \MgFe\ (e.g.,
\citealt*{Peletier89,Worthey92}). Such departure from the solar abundance pattern
is generally attributed to short formation time-scales ($<$1\,Gyr) for the bulk
of their stellar populations (see the reviews of e.g.,
\citealt{Trager98,Trager00,Renzini06,Conroy13rev}, and references therein). The
Magnesium, an \alfa-element, is ejected into the ISM by type-II supernovae
resulting from massive stars on time-scales of $\sim$10\,Myr, whereas most 
of the iron is released on time-scales of $\sim$1\,Gyr, when SNIa explosions 
of intermediate-mass stars in binary systems occur. 
Although other \alfa-elements are expected to follow
magnesium there are exceptions to the rule such as calcium, which nearly
tracks Fe (e.g., \citealt{Vazdekis97,Worthey98,Saglia02,Cenarro03,FalconBarroso03a},
\citealt*{Thomas03b,Yamada06,Worthey11,CvD12,Johansson12},\citealt{LaBarbera13}).
These abundances and the abundance ratios of other elements can therefore be
used to investigate the starformation timescales of ETGs (e.g., \citealt*{Conroy14}). 

Aiming at obtaining more quantitative determinations, models with varying
abundance ratios were developed with the aid of theoretical stellar spectra
\citep*{TripiccoBell95,Korn05}. The responses obtained for the Lick indices to
variations of the individual element abundances, i.e. the so called response
functions, were used to correct the model predictions based on the empirical
stellar libraries (e.g., \citealt{Vazdekis97}; \citealt*{Tantalo98}; 
\citealt{Trager00,TantaloChiosi04,Thomas03,Schiavon07,Chung13}). These
models allowed us to achieve a better understanding of what these line-strengths
are made of, as none of these rather strong atomic and molecular lines are
contributed to by a single element. Although with the use of these indices it has
been possible to partially lift the age/metallicity degeneracy affecting the
colours, these indices are not completely free from
this degeneracy \citep{Worthey94,Arimoto96}. 

This degeneracy is due to the fact that an age increase causes
an effect on the theoretical isochrone and on line-strength indices that can be
mimicked - at constant age - by a metallicity increase. The effects of this
degeneracy are stronger when low resolution indices are used, as the
metallicity lines appear blended. For such a reason  alternative indices -- that
were thought to work at higher spectral resolutions, such as those of e.g.,
\citet*{Rose85,Rose94,Serven05} --, have been designed to alleviate the problem.

New means for improving the stellar population analysis have been achieved by
predicting full spectral energy distributions (SEDs) at moderately high
resolution, rather than just a number of indices. Such modelling has been
possible with the development of extensive empirical stellar spectral libraries
with flux-calibrated spectral response and good atmospheric parameters coverage.
Among the most widely employed stellar libraries in the optical range are those
of \citet{Jones99}, ELODIE \citep{ELODIE01}, STELIB \citep{LeBorgne03}, Indo-US
\citep{Valdes04} and MILES \citep{MILESI} and more recently XSL \citep{Chen14}.
Models that employ such libraries are, e.g.,
\citet{Vazdekis99,Schiavon02,BC03,LeBorgne04,MILESIII,Maraston11,CvD12models}. There are also
empirical libraries in other spectral ranges such as, e.g.
\citet*{CATI,CATII,Cushing05,Raynier09}, that are also implemented in these
models \citep{CATIV,CvD12models}. Alternatively, theoretical stellar libraries
at high spectral resolutions have also been developed for this purpose (e.g.,
\citealt{Barbuy03,MurphyMeiksin04};\citealt*{Zwitter04};
\citealt{Coelho05,Martins05,Munari05,RodriguezMerino05,Fremaux06,MartinsCoelho07,Leitherer10,Palacios10,Kirby11,deLaverny12,Coelho14}).
Models that make use of these libraries are, e.g., \citet{Leitherer99},
\citet*{Schiavon00},\citet{GonzalezDelgado05,Coelho07,Buzzoni09,Lee09,Percival09,Maraston11}. 

These models provide ways to alleviate the main degeneracies hampering
stellar population studies. For example, with these SEDs it has been possible to
define new indices with enhanced abilities to disentangle the age from the
metallicity (e.g., \citealt{VazdekisArimoto99,BC03,CervantesVazdekis09}), or
even the IMF (e.g., \citealt{Schiavon00,CATI,CATIV,CvD12models,MIUSCATI,LaBarbera13,Spiniello14}). It is becoming common practice to use these model SEDs with newly developed 
algorithms that aim at constraining and recovering the SFH by means of full
spectrum-fitting approach (e.g., \citealt*{Panter03},
\citealt{Cid05,Ocvirk06a,Ocvirk06b,Chilingarian07,Koleva09,Tojeiro11}). This method
represents an alternative to the more classical approach of fitting a selected
number of line-strength indices (e.g., \citealt{Trager98}). Furthermore these
model SEDs have been used in a variety of applications. For example, these SSP
SEDs have been shown to improve the analysis of galaxy kinematics for both
absorption and emission lines (e.g., \citealt{FalconBarroso03b,Sarzi06}). Models
based on empirical libraries are free from the uncertainties in the underlying
model atmospheric calculations and tend to provide good fits to both, absorption
line-strengths and spectra (e.g.
\citealt{MILESI,Yamada06,Maraston11,MIUSCATI,Conroy13rev}) and also photometric data
(e.g., \citealt{Maraston09,Peacock11,MIUSCATII}). However, unlike the models
based on theoretical libraries, these models are generally restricted in their
SSP parameters coverage, as most of the stars come from the solar neighbourhood,
with a characteristic, and limited, elemental abundance ratio coverage
(\citealt*{Milone11}; see also figure~1 in \citealt{Walcher09} which illustrates the 
effect of the solar-neighbourhood pattern on models based on empirical libraries). 
The quality of the resulting model SEDs relies to a
great extent on the atmospheric parameters coverage of the library. 

As most of the stars of these empirical libraries belong to the solar
neighborhood, the resulting model SEDs have a characteristic abundance pattern:
at low metallicity the predicted spectra are enhanced in \alfa-elements, whereas
around the solar value the models are nearly scaled-solar (e.g.,
\citealt{Edvardsson93,McWilliam97}). In addition, most of these models employ scaled-solar isochrones
and therefore they are not self-consistent at low metallicities. These models, which can
also be regarded as "base models", present important limitations for studying
massive ETGs (or any other system that deviates strongly from the chemical
pattern of the solar neighbourhood), which reside in a high metallicity regime
and show enhanced \MgFe\
abundance. It has been shown that by using \alfa-enhanced stellar
evolution models (e.g., \citealt*{Salaris00}) and the stellar theoretical spectra (e.g.,
\citealt{TripiccoBell95,Korn05,Coelho07}), line indices such as \mgb\ and \fe\
are affected. However, it is worth noting that the base models have been
successfully applied to these galaxies. In fact it is possible to obtain a good
proxy for the \MgFe\ abundance, if appropriate indices are employed for the
analysis  
\citep{Vazdekis01Virgo,SB06b,Yamada06,delaRosa07,Michielsen08,MILESIII,LaBarbera13}.
This proxy leads to a linear relation with the abundance ratio estimated with
the aid of models that specifically take into account such non-solar element
mixtures (e.g., \citealt{Tantalo98,Thomas03,LeeWorthey05,GravesSchiavon08}).

There are some examples of models predicting SSP spectra at moderately high
resolution with varying abundance ratios. Spectral SSP models that include varying 
abundance ratios both at the isochrones and stellar spectra have been presented
in \citet{Coelho07,Lee09,Percival09}. 
Also \citet{Cervantes07,Prugniel07,Walcher09,CvD12models} present models with varying abundances
but considering only the effect of the stellar spectra. 
There are also examples where only \alfa-enhanced isochrones 
are employed, ignoring the effects of the stellar spectra in the line-strengths (e.g.,
\citealt*{Weiss95,CATIV}). Self-consistent models can only be achieved
when both the spectral and the evolutionary effects are correctly taken into account, 
and thus require the use of both theoretical spectra (e.g., 
\citealt{Coelho05,Munari05}), and isochrones with varying \aFe\ abundance (e.g.,
\citealt{SalarisWeiss98,Salasnich00,VandenBerg00,VandenBerg14,Kim02,Pietrinferni06,Coelho07,Dotter08}). 
To our knowledge, the three self-consistent models for a fair range of ages and metallicities
available in the literature are \citet{Coelho07,Lee09} and \citet{Percival09}. 
However, both these models carry a caveat: the stellar spectral libraries employed are theoretical, 
and as such they carry wavelength-dependent 
systematic deviations when compared to observed spectra \citep{MartinsCoelho07,Bertone08,Coelho14}.
Inspired by the work of \citet{Cervantes07} and \citet{Prugniel07}, and in order to bring together 
the predictive power of fully-theoretical models and the accuracy of the base models, 
\citet{Walcher09} proposed a method to use the predictions of fully-theoretical SSP models 
only differentially.  

Here we further develop our models to predict SSP spectra with
varying \aFe\ ratios based on the MILES library \citep{MILESI,MILESII}. For this
purpose we apply a differential correction to our empirical SSP models which is 
obtained from the theoretical stellar spectral library of
\citet{Coelho05,Coelho07}. We also employ the scaled-solar and \alfa-enhanced
isochrones of BaSTI \citep{Pietrinferni04,Pietrinferni06}. This work is the
second of the series that we started with \citet{MILESIII}, hereafter
Paper~I, where we presented our base model predictions that were fully based on the MILES library. As
the parameter coverage of MILES constitutes a significant improvement over
previous empirical stellar libraries, these models are safely extended to
intermediate-aged stellar populations and to lower and higher metallicities. 

The layout of the paper aims to permit the reader to follow the model
construction from the very first steps, through the characterisation of the model 
outputs and finally to provide some illustrative applications. In
Section~\ref{sec:ingredients} we describe the main model ingredients, which
include the IMF, the isochrones and the empirical and theoretical
libraries employed here. In Section~\ref{sec:SSPs} we describe our approach for
constructing these SSP SEDs. The quality and 
behaviour of these SSP spectra are assessed in
Section~\ref{sec:SSP_spectra}. In Section~\ref{sec:lines} and
Section~\ref{sec:SSPcolours} we show the behaviour of the line-strength indices
and broad-band colours as measured on the resulting SSP spectra, respectively.
In Section~\ref{sec:othermodels} we compare our models to predictions from other
authors. In Section~\ref{sec:data} we apply these models to data of Galactic
Globular Clusters and ETGs to illustrate the predictive power of these new models. We
compare the colours, the full spectra and the line-strengths of ETGs with our
models. We also calibrate a proxy to estimate the \MgFe\
abundance ratio, relying solely on scaled-solar predictions. 
In Section~\ref{web} we present the new features that we have
introduced in our web-page to make it publicly available these predictions, including more
complex models with arbitrary star formation histories. 
Finally in Section~\ref{summary} we provide a summary. 


\section{Model ingredients}
\label{sec:ingredients}

In this Section we describe the main ingredients in our new stellar
population synthesis models, namely the IMF, the isochrones and the stellar
spectral libraries. We summarise the chief aspects of these ingredients that are
particularly relevant to this work. We also describe the details and the
treatment that have been required for implementing these ingredients in the
models, which include new isochrones that were specifically computed for this
work.

\subsection{IMF}
\label{sec:IMF}

Several IMFs are considered: the two power-law IMFs described in
\citet{Vazdekis96}, i.e unimodal and bimodal, both characterised by the
logarithmic slope, $\Gamma$ and $\Gamma_b$\footnote{Note that we have 
changed our notation for the IMF slope from that we used
in our previous papers, i.e. $\mu$ (e.g.,
\citealt{Vazdekis96,CATIV},Paper~I,\citet{MIUSCATI}). We now adopt $\Gamma$ and
$\Gamma_b$ for the slopes of the unimodal and bimodal IMF shapes, respectively,
to be consistent with recent works on the variation of the IMF. Note also that in the new 
notation we  distinguish the slopes of these two IMF shapes as they apply to
different functional forms.}, 
respectively, as a free parameter, and the multi-part power-law IMFs of
\citet{Kroupa01}, i.e. universal and revised. The \citet{Salpeter55} IMF is
obtained by adopting the unimodal case with slope
$\Gamma=1.3$\footnote{In reality it would be more accurate to use $\Gamma=1.35$ to match
the Salpeter slope}. Although not identical, the Kroupa
Universal IMF is very similar to a bimodal IMF with slope $\Gamma_b=1.3$. 
An extensive
description of these definitions is given in \citet{CATIV} (Appendix~A). We set
the lower and upper mass-cutoff of the IMF to $0.1$ and $100$\,\Msol,
respectively. It is worth noting that the very low-mass ($<0.6$\,\Msol)
tapered "bimodal" IMF has been found to be consistent with the line-strengths and
mass-to-light (M/L) ratios of massive ETGs (e.g.,
\citealt{Ferreras13,LaBarbera13,Spiniello14}). On the contrary, the single
power-law "unimodal" IMF leads to extremely high M/L values that are not
supported by independent gravitational lensing and dynamical estimates (e.g.,
\citealt{Treu10,Spiniello12}).

\subsection{Isochrones}
\label{sec:isochrones}

We employ two sets of theoretical isochrones. The scaled-solar isochrones of
\citet{Girardi00} (hereafter Padova00) and \citet{Pietrinferni04} and the
\alfa-enhanced models of \citet{Pietrinferni06}. The latter two will be
regarded hereafter as BaSTI. 

The Padova isochrones cover a wide range of ages, from $0.063$ to
$17.8$\,Gyr, and six metallicity bins ($Z=0.0004$, $0.001$, $0.004$, $0.008$,
$0.019$ and
$0.03$), where $0.019$ represents the solar value. The isochrones also include the later
stages of stellar evolution, using a simple synthetic prescription for
incorporating the thermally pulsing AGB regime to the point of complete envelope
ejection. The range of initial stellar masses extends from $0.15$ to $7$\,\Msol. The
input physics of these models was updated with respect to \citet{Bertelli94}
with an improved version of the equation of state, the opacities of
\citet{AlexanderFerguson94} and a milder convective overshoot scheme. A helium
fraction was adopted according to the relation: $Y\approx0.23+2.25Z$. 

We also use here the BaSTI theoretical isochrones\footnote{The BaSTI database
is publicly available at http://www.oa-teramo.inaf.it/BASTI.}. The original sets
of models have been supplemented by additional computations specifically
performed for this project (see below for more details). A description
of the complete BaSTI database as well as of the adopted physical inputs and
numerical assumptions can be found in
\citet{Pietrinferni04,Pietrinferni06,Pietrinferni09,Pietrinferni13} and
\citet{Cordier07}. We refer the interested reader to these references. Here
it is sufficient to note that the BaSTI archive is based on an updated
physical framework and that it has been extensively tested with many
observational constraints based on both resolved (mainly eclipsing binaries and
cluster color-magnitude diagrams) and unresolved  stellar populations (see
\citealt{Pietrinferni04} and \citealt{Percival09}). In this work we have adopted
the non-canonical BaSTI models with the mass loss efficiency set to
$\eta=0.4$\footnote{We note that $\eta$ represents the free parameter present in
the Reimers mass loss law \citep{Reimers} adopted in the BaSTI stellar model computations.}
for both the scaled-solar and \alfa-enhanced heavy elements mixture. The initial He
mass fraction ranges from $0.245$, for the more metal-poor composition, up to
$0.303$ for the more metal-rich composition, with $\Delta Y / \Delta Z \approx 1.4$. All
the metallicity grid points available in the archive have been used. However in
order to better sample the whole metallicity range from very metal-poor to the
most metal-rich stellar populations we have added an additional grid point
corresponding to $Z=0.024$, $Y=0.279$ (i.e. $\Mh=+0.15$) for both scaled-solar and
\alfa-enhanced mixture. This metallicity is particularly relevant for studying
massive early-type galaxies (e.g. \citealt{LaBarbera13}).  Therefore the 
BaSTI metallicity grid adopted in present analysis was based on 12 grid points:
$Z=0.0001$, $0.0003$, $0.0006$, $0.001$, $0.002$, $0.004$, $0.008$, $0.0100$, $0.0198$, $0.0240$, $0.0300$ and $0.0400$. For each metallicity, the isochrone age range covers
the interval from $0.03$ to $14$\,Gyr, with a fine age grid.

We note that the atomic diffusion of both helium and metals was properly accounted
for when computing the solar model in order to be able to match accurately the
helioseismological constraints. The best match to the depth of the convective
envelope ($0.716$\,\Rsol) of the present solar envelope He abundance (Y$=0.244$),
and of the actual (Z/X) ratio [(Z/X$=0.0244$], lead to an initial He abundance and
metallicity (Y$_{\odot}=0.2734$, Z$_{\odot}=0.0198$), which was adopted for the
solar metallicity model set.  

We have used here the \lq{AGB-extended}\rq\ version of the BaSTI isochrones that
account for the Asymptotic Giant Branch (AGB) stage as provided by
\citet{Cordier07}. This phase includes the entire Thermally Pulsing AGB (TP-AGB)
regime through a synthetic-AGB treatment following the technique of
\citet{IbenTruran78}. In this approach the mass-loss processes are accounted for
by using the formulae of \citet{VassiliadisWood93}. The evolution with time of
the CO core mass and surface luminosity is accounted for by adopting the
analytical relationships provided by \cite{Wagenhuber98}, while the trend with
time of the effective temperature is provided according to the relationship of
\cite{Wagenhuber96}. We wish to emphasise that our synthetic-AGB treatment does
account\footnote{We would like to emphasise that the claim provided by
\cite{Marigo08} that the AGB-extended version of the BaSTI isochrones does not
explicitly account for AGB nucleosynthesis, i.e. the effects of the third
dredge-up and hot-bottom burning, in the underlying AGB models, is not correct.
The analytical relationships for the trend of the surface luminosity and
effective temperature as a function of the He core mass provided by
\citet{Wagenhuber98} and \citet{Wagenhuber96} are adopted in the BaSTI
synthetic-AGB treatment, do account for the occurrence of these physical
processes during the AGB stage. In order to provide a correct description
of the situation, we wish to note that the computations by
\cite{Wagenhuber96} were performed without accounting for low-temperature
radiative opacities with various C/O ratios - because these opacity tables were
not available at that time. Notwithstanding, the effect of the change in the
envelope C/O ratio induced by the third dredge-up was mimicked
by adopting a radiative opacity at a constant heavy elements distribution, but
allowing the global metallicity to change. \citet{Cristallo07} and
\citet{WeissFerguson09} have shown that, although the use of opacity tables that
properly include the enhancements of C is more appropriate, the
predictions of the behaviour with time of \Teff\ as provided  by fully
consistent AGB models are not significantly different from those predicted by models
that account for the opacity effects of the third dredge-up by adopting a
variable global metallicity but constant C/O ratio.} for
the occurrence - and, hence for the related evolutionary effects - of the third
dredge-up and hot bottom burning when appropriate. 
We notice that this simplified model for the
TP-AGB  phase has been shown to be adequate for matching several integrated
properties in the near-IR as discussed in \citet{Cordier07}. In addition, in
order to assess the reliability of the theoretical framework adopted in the
present work, we would like to call the attention of the reader to the recent
analysis performed by \citet{Salaris14}. They presented stellar population
synthesis models based on fully evolutionary AGB computations
\citep{WeissFerguson09} (i.e. without the need of adopting analytical relations
as in the synthetic-AGB treatment) and compared their model predictions with
integrated colours from other models. They found significant discrepancies with
results from \citet{Marigo08} and \citet{Maraston05}\footnote{A comparison among
these population synthesis models and observations has been already performed by
\citet{Noel13}, and it was found that these models disagree -- being in general
too red -- with the integrated colours of superclusters for ages between
$\sim10^8$ and $\sim10^9$ years.}, whereas an excellent agreement was found with
the BaSTI AGB-extended models.

The lower mass limit in the standard BaSTI isochrones is $0.5$\,\Msol. However,
for the aim of the present study, we have extended the isochrones to the very
low-mass (VLM) regime by including additional stellar models down to a lower
mass limit of $0.1$\,\Msol. The adopted VLM models are the same presented in
\citet{Cassisi00}, which are based on an updated physical framework
for the evolutionary computations of VLM stellar models. We have made a 
significant effort to obtain a smooth transition when passing from
the low mass regime (BaSTI models) to the VLM ones. This has been obtained by
computing some additional stellar models in the transition mass regime (around
$\approx0.5-0.6$\,\Msol) by using the same physical inputs adopted for the BaSTI
model computations but the same outer boundary conditions used in the VLM
stellar model computations (see \citealt{Cassisi00} for more details on this
issue). Our temperatures for these stars are cooler than those of Padova00, and
warmer than those of \citet{Pols95}, which were implemented in
\citet{Vazdekis96}, and those of \citet{Dotter08} (see also \citealt{An09}). As
their temperatures are higher the Padova00 VLM stars are brighter than the 
ones we employ here for the BASTI models (see figure~5 in \citealt{MIUSCATI}).

\begin{figure*}
\includegraphics[width=\textwidth]{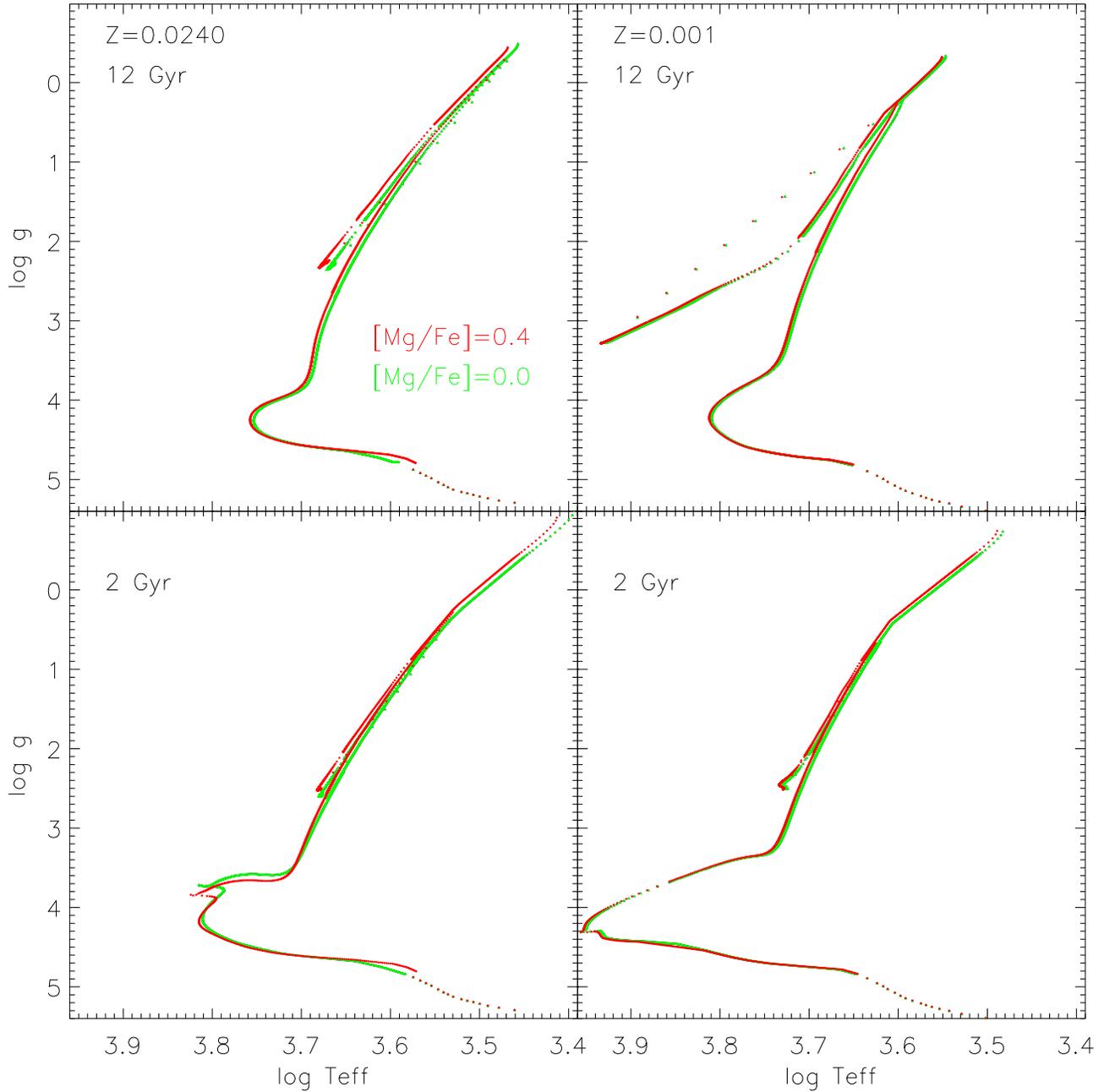}
\caption{Comparison of \alfa-enhanced ($\aFe=+0.4$)and scaled-solar BaSTI isochrones. Two
representative metallicities Z$=0.024$ (left) and Z$=0.001$ (right) and two ages
$12$\,Gyr (top) and $2$\,Gyr (bottom) are shown.}
\label{ISOC_SS_AA}
\end{figure*}

\begin{figure*}
\includegraphics[width=\textwidth]{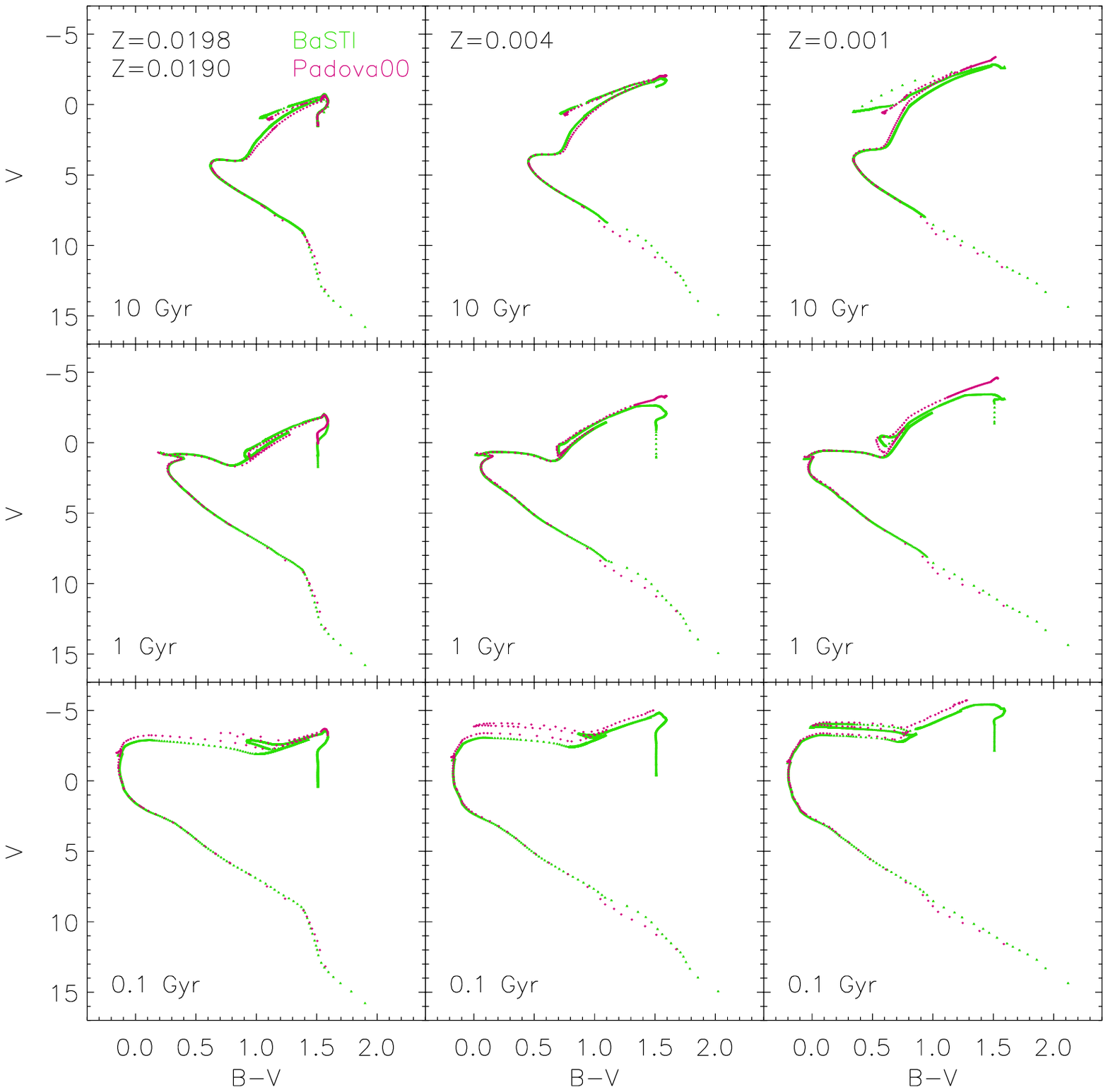}
\caption{Comparison of the Padova00 and BaSTI isochrones for three representative
metallicities (Z$=0.001$, Z$=0.004$ and solar), as quoted within the upper panels,
and three different ages, $10$, $1$ and $0.1$\,Gyr from top to bottom panels. Note
that the reference solar metallicity value is slightly different for these two
models. There are differences in the RGB, HB and AGB phases as well as
in the very low-mass dwarf stars.
}
\label{PAD_TER_isochrones}
\end{figure*}

In Fig.~\ref{ISOC_SS_AA} we plot a set of BaSTI isochrones for two
selected metallicities and ages. We see that the higher the metallicity, the greater the
difference between the scaled-solar and the \alfa-enhanced models, with the
scaled-solar isochrones being slightly cooler than the \alfa-enhanced isochrones (see
\citealt{Pietrinferni06}). These differences arise since we are comparing the
two sets of isochrones at constant global metallicity; when comparing
the two sets at constant iron content, the \alfa-enhanced isochrones would
appear cooler that the corresponding scaled-solar isochrones as a consequence of their
larger global metallicity and, hence, higher opacity at low temperatures. 
(see, e.g., figures~1 and 2 in \citealt{Coelho07} 
and figure~1 in \citealt*{Lee09b}). A more elaborated discussion on these
effects, including those when varying the abundance ratio of individual \alfa-elements 
can be found in \citet*{Lee10}.
While there are clearly differences between the two sets of isochrones, it is worth noting that 
these differences are always small.

We use the theoretical parameters of these two sets of isochrones (\Teff, $\log
g$, \Mh) to obtain stellar fluxes on the basis of empirical relations between
colours and stellar parameters (temperature, gravity and metallicity), instead
of using theoretical stellar fluxes. We mainly use the
metallicity-dependent empirical relations of \citet*{Alonso96,Alonso99};
respectively, for dwarfs and giants. Each of these libraries are composed of
$\sim500$ stars and the temperature scales are based on the IR-Flux
method, i.e. only marginally dependent on model atmospheres. We use the
empirical compilation of \citet*{Lejeune97, Lejeune98} (and references therein)
for the coolest dwarfs (\Teff$\lesssim$4000\,K) and giants (\Teff$\lesssim$3500\,K) for
solar metallicity, and also for stars with temperatures above $\sim$8000\,K. We
use a semi-empirical approach for these low temperature stars for other
metallicities. For this purpose we combine these relations and the model
atmosphere predictions of \citet{Bessell89,Bessell91} and the library of \citet{Fluks94}.
We also employ the metal-dependent bolometric corrections given by
\citet*{Alonso95} and \citet{Alonso99} for dwarfs and giants, respectively. Finally, we
adopt $BC_{\odot}=-0.12$.  By assuming $V_\odot=26.75$ \citep{Hayes85}, the
solar absolute magnitude is $M_{V_\odot}=4.82$,  therefore, for the adopted
normalization, $M_{{bol}_{\odot}}$ is given by $M_{V_{\odot}}+BC_{V_{\odot}}=
4.70$.

In the present work the integrated colors and spectral indices based on
the BaSTI stellar models are compared to those obtained on the basis of the Padova
models. Therefore we consider it worthwhile to briefly discuss some of the 
principal differences between the two theoretical evolutionary frameworks. 
Fig.~\ref{PAD_TER_isochrones} shows a
comparison for three representative metallicities (Z=0.001, Z=0.004 and solar)
and three ages ($10$, $1$ and $0.1$\,Gyr). A detailed comparison between the
scaled-solar BaSTI models and the Padova00 isochrones has been performed in
\citet{Pietrinferni04}, and we refer the reader to section~6 of this paper
for details. Here it is sufficient to note that there is a systematic offset between
the BaSTI and Padova models in the RGB ${\rm T_{eff}}$ scale,
with BaSTI being hotter models at solar metallicity, whereas at lower
metallicities the trend is reversed and BaSTI is cooler than Padova00. 
There are also significant differences in
the brightness of the isochrones (at a given age) for ages lower than about
$2$\,Gyr. These differences can be fully understood when accounting for the
differences in the adopted physical inputs and in the different treatment of
core convective overshooting during the core H-burning stage (see the discussion in
\citealt{Pietrinferni04}). We also see important differences in the AGB stellar
evolutionary phase, as illustrated by the middle panels of
Fig.~\ref{PAD_TER_isochrones}. These late evolutionary stages are brighter in
the Padova models at $1$\,Gyr, whereas the BaSTI AGB extends towards cooler
temperatures, with such trends depending on the metallicity. This difference is
quite well known as these stars reach the temperature regimes where the $B-V$
colour saturates (see, e.g., \citealt{WortheyLee11}). Note that this difference
is further emphasised at younger ages.  Similarly, the two sets of models
show important differences during the core He-burning stage: for old ages (i.e.
larger than $\sim3-4$\,Gyr) this is due to both the differences in the adopted
physical framework and (mainly) in the adopted efficiency of mass loss along the
RGB stage; while for younger ages the origin of the differences in the \lq{blue
loops}\rq\ -- characteristic of the core He-burning stage of intermediate-mass
stars -- could be related to many different contributions (physical inputs,
treatment of convection, the details of the heavy elements mixtures, etc.) as
discussed in detail by \citet{Cassisi04}.

Another interesting feature that is relevant to IMF studies is the
difference in effective temperature between these two sets of models for the VLM
stars, with the Padova models being hotter in all the panels of Fig.~\ref{PAD_TER_isochrones}.
This is related to the different physical inputs to the models - mainly in
the equation of state and outer boundary conditions - adopted for computing the
two sets of VLM stellar models  (see \citealt{Cassisi00} and references therein
for more details on this topic).

\subsection{Stellar spectral libraries}
\label{sec:libraries}

\subsubsection{MILES}
\label{sec:miles}

The new models presented in this work, for both the scaled-solar and
\alfa-enhanced mixture, are computed -- in the optical range -- on the basis of
the MILES stellar spectral library\footnote{The MILES library is publicly
available at http://miles.iac.es} \citep{MILESI}. This library was prepared
carefully to be implemented in our models (Paper~I). For a detailed description
we refer the reader to these two papers and also to \citet{MILESII}, for the
determination of the stellar parameters. Here we briefly summarise the most relevant
aspects of the library. 

The MILES library is composed of 985 stars observed at the INT 2.5m telescope on El
Roque de Los Muchachos observatory at La Palma, Spain. Two gratings were
employed to cover the full optical range at $2.51$\,\AA\ (FWHM) resolution, which is 
constant with wavelength (see \citealt{FalconBarroso11} for a full description).
The spectra have very high signal-to-noise (S/N) which is usually well above
$100$, with the exception for the stars that belong to globular clusters such as M\,71, which
were included to improve the parameter space coverage. The spectra were
carefully flux-calibrated; each individual star was also observed through a
wide slit in order to avoid selective flux losses due to differential refraction
effects, in addition to the higher resolution setups used  to achieve the blue
and red parts of the stellar spectra. Careful attention was also paid to the cleaning
of telluric absorptions that were present in the redder part of the spectra.

Our models make use of a subsample of $925$ stars from the original library since we
removed those stars whose spectra were not properly representative of a given
set of atmospheric parameters. These included spectroscopic binaries (identified
with SIMBAD), or stars with unexpectedly high signal of variability (according to
the CDS electronically-readable version of the \citealt{Kholopov98}
catalogue) for which there were alternative stars in the MILES database with
similar parameters. Other reasons for removing stars were the presence of
emission lines, very low S/N in the blue, problems in the continuum and 
the lack of atmospheric parameters determinations. For this process, each
stellar spectrum was compared to a synthetic spectrum of similar atmospheric
parameters computed with the interpolation algorithm described in Paper~I after
exclusion of the targeted star. Finally, $75$ stellar spectra with some, milder,
problems were retained in the final sample as they were found to be useful for
improving the coverage of certain regions of the parameter space. For these
stars we decreased their relative weight in comparison to the other stars.

The parameters for MILES stars were obtained from an extensive compilation from
the literature, homogenized by taking as a reference the stars in common with
the high resolution stars of \citet*{Soubiran98} (see \citealt{MILESII}). One of
the most relevant features of this library is the significant improvement of the
parameter coverage in comparison to other widely employed stellar spectral
libraries (see figure~2 in Paper~I). MILES spans a fairly wide range in
metallicity, including stars with \Mh$<-2$ as well as super-solar, for the
temperature range that is relevant for intermediate and old stellar populations,
both for dwarf and  giant stars. Note that, unlike in other stellar libraries
(see Paper~I), MILES includes a number of metal-poor dwarf stars with  \Teff$ >
6000$\,K, which are relevant for predicting stellar populations with ages
smaller than $\sim 10$\,Gyr at low metallicities. This library also includes
a number of very low mass dwarfs,  which are relevant when varying the IMF
slope. Although not optimised for very young stellar populations  (see Paper~I)
the library has MS stars as hot as $\sim30000$\,K, mostly around  solar
metallicity. It also includes giants with  temperatures above $6000$\,K, such as
metal-poor Horizontal Branch (HB) stars, Post AGB stars and variable stars of
the type  $\delta$ Scuti, RR Lyrae or Cepheids along the instability strip.

An important improvement here is the consideration of the \MgFe\ estimates for most
of the MILES stars, which have been recently determined by \citet{Milone11}.
These abundance ratios were obtained through a compilation of values from the
literature using abundances from high-resolution spectroscopic studies and from
a robust spectroscopic analysis using the MILES spectra.
These \MgFe\ values were carefully calibrated to a single uniform scale with a 
typical uncertainty on the estimates of $\sim0.1$\,dex. As
expected, the MILES library follows the abundance pattern of the Galaxy (see
figure~10 in \citealt{Milone11}), although there is some dispersion at any given
metallicity. There were $75$ stars in the selected MILES subsample for which
\citet{Milone11} could not determine their \MgFe\ values (the majority of them
have effective temperatures either above $\sim8000$\,K or below $\sim3500$\,K,
and also stars belonging to stellar clusters). To retain these stars for our model
computations we assigned the mean abundance ratio expected for the \Feh\ value
of the star according to the pattern found by \citet{Milone11} for the field
stars. For the cluster stars we adopted the mean value determined by
\citet{Milone11} for the other stars of the same cluster.

\begin{figure}
\includegraphics[width=\columnwidth]{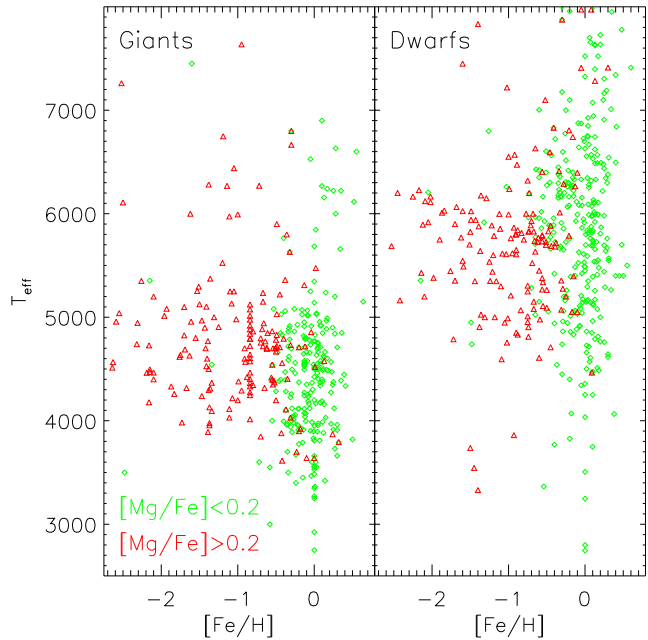}
\caption{Stellar parameters coverage of the MILES library. Red triangles and
green diamonds represent stars with \MgFe\ values larger and smaller than 0.2,
respectively. For this plot we split the database in giants (left panel) 
and dwarfs (right panel) at \logg$=3.0$.}
\label{milone}
\end{figure}

Fig.~\ref{milone} shows the stellar parameter coverage of the subsample composed
of $925$ stars selected from the MILES database. Stars with \MgFe\ abundance ratio
smaller (larger) than $+0.2$ are plotted in green (red). 
Note that we only plot parameters for stars with
temperatures below $8000$\,K, as the abundance estimates for higher temperatures
is more uncertain (see for details \citealt{Milone11}) and the net effect of the
abundance on the
spectra is small (see figure~9 of \citealt{Sansom13}). As expected, the stars
enhanced in \MgFe\ also have \Feh$\lesssim-0.5$, whereas the higher metallicity regime
is populated with scaled-solar stars. We also see that MILES stars show a good
coverage of temperature within the metallicity regime 
$-0.5\lesssim\Feh\lesssim+0.2$. Despite a significant decrease in the number of
stars for the more metal-poor and metal-rich regimes, there are sufficient stars 
to compute models with \Feh$\simeq-2.3$ and \Feh$\simeq+0.4$, although within a
limited age range. A quantitative analysis of the quality of the models 
based on the atmospheric parameters coverage of the library stars is 
shown in Section~\ref{sec:quality}.  

\subsubsection{Theoretical stellar spectral library}
\label{sec:coelho}

Theoretical spectra are employed to determine a
differential correction to be applied to the reference models computed on the
basis of MILES (see Section~\ref{sec:SSPs}). 
In order to compute these ``corrected'' SSPs we adopt the theoretical stellar library of
\citet{Coelho05}, with the extension to very cool giants and the spectrophotometric
calibration presented in \citet{Coelho07}. 
These stars were computed with {\sc
Pfant} spectral synthesis code, and are based on refined atomic and molecular line lists
\citep[e.g.][]{Cayrel91,Barbuy03}. Model atmospheres from
\citet{CastelliKurucz03} are adopted for effective temperatures \Teff$\ge3500$\,K, and from
\citet{Plez92} for \Teff$<3500$\,K.  We refer the reader to
\cite{Coelho05,Coelho07} for more details.

The \citet{Coelho05} theoretical library covers effective temperatures, \Teff,
between $3500$ and $7000$\,K, surface
gravities, \logg, between $+0.0$ and $+5.0$, iron metallicities, \Feh, between
$-2.5$ and $+0.5$ and \alfa-element over iron abundances, \aFe$=+0.0$ and \aFe$=+0.4$.
The extension to cool giants from \citet{Coelho07} covers \Teff\ down
to $2800$\,K, \logg\ between $-0.5$ and $1.5$, \Feh$=-0.5$, $0.0$ and $+0.2$,
and \aFe$=+0.0$ and \aFe$=+0.4$. The scaled-solar spectra (\aFe$=0.0$) have the solar
abundances from \citet{GrevesseSauval98}. The \alfa-enhanced spectra
(\aFe$=+0.4$) adopt a uniform enhancement [X/Fe]$=+0.4$ for the elements O, Ne, Mg,
Si, S, Ca and Ti, and the other elements follow the solar abundances.

The accuracy of this library has been assessed in \citet{MartinsCoelho07}, who
showed that this library had the highest success in reproducing
spectral features of observed stars among the theoretical libraries analysed in
that work. The spectra have higher spectral resolution and wavelength coverage
than MILES, therefore they were convolved with a gaussian kernel and were
resampled to match MILES spectral resolution, dispersion and wavelength coverage.

\section{SSP spectral synthesis modelling}
\label{sec:SSPs}

To compute the spectra of single-burst stellar populations, i.e. SSPs, we follow the method
described in detail in Paper~I. This approach consists of integrating the spectra of the
stars along the isochrone taking into account their number per mass bin
according to the adopted IMF. In the present study we distinguish and provide three types of
models for a given total metallicity, \Mh~\footnote{We use \Mh, rather than \Zh\
(as used in Paper~I), because Z denotes a mass fraction, and therefore \Mh\ is less
confusing.}. The first type of model SSPs are what we call {\it base models}, 
which combine scaled-solar isochrones
of a given total metallicity with the MILES stars of similar \Feh\ (rather than
\Mh), irrespective of their \MgFe\ abundance ratio. As MILES follows the
abundance pattern of The Galaxy, these models are effectively scaled-solar at solar
metallicity. However, for lower metallicities these models are not self-consistent 
in the sense that observed stars are enhanced in \MgFe, whereas the isochrones are
scaled-solar. Therefore the SSP SEDs of Paper~I, which combine MILES with the Padova00
isochrones, should be regarded as base models. The second type of models
presented here are termed {\it scaled-solar} as both the stellar spectra and the isochrones
are scaled-solar throughout the whole metallicity range. 
Finally, the third type of models computed here are the
{\it \alfa-enhanced} models, which combine \alfa-enhanced stellar spectra with \alfa-enhanced
isochrones. For the scaled-solar and \alfa-enhanced models, which are computed on the basis
of the total metallicity, \Mh, we take into account both the recent \MgFe\
determinations of \citet{Milone11} and the \Feh\ estimates of the individual library stars.

\subsection{Base models}
\label{sec:baseSSPs}

The term "base model" refers to the commonly used approach that is employed to
compute SSP models, as we describe in this Section. It should not be confused with
the terms BaSTI- or Padova00-based models, which refer to the use of
a particular set of isochrones used to generate the SSP models. For the base models we
always employ scaled-solar isochrones. We compute here two sets
of models that, to be precise, should be termed BaSTI-based base models and
Padova00-based base models. For brevity henceforth  we will refer to these
models as BasTI base models and Padova00 base models.

The base SSP model spectra are computed according to the following expression

\begin{eqnarray}
S_{\lambda}(t,\Mh,\Phi,I_{0.0})&=&\int_{m_{\rm l}}^{m_{\rm t}}
S_{\lambda_V}(m,t,\Feh)\times \nonumber \\
&&{F_{V}}(m,t,\Feh)N_{\Phi}(m,t)dm,
\label{eq:SSP}
\end{eqnarray}

\noindent where $S_{\lambda_V}(m,t,\Feh)$ is the empirical stellar spectrum, 
irrespective of its \MgFe\ abundance, normalised in the $V$ band, corresponding
to a star of mass $m$ and metallicity \Feh, which is alive at the age $t$ of the
stellar population. Its spectrum is also characterised by the \Teff\ and \logg\ parameters as
determined by the adopted scaled-solar isochrones, $I_{0.0}$. $F_{V}(m,t,\Feh)$ is the flux of the star
in the $V$-band. For these base models we assume that $\Mh=\Feh$. This is why we
write in the left term of this equation \Mh, whereas in the right one we use the
\Feh\ abundance measurement. Therefore, the base
models in the low metallicity regime are not self-consistent since the MILES stars are
enhanced in \MgFe, and therefore the true \Mh\ value is in reality larger than
the adopted \Feh\ value (see Section~\ref{sec:aassSSPs}). $N_{\Phi}(m,t)$ is the
number of such stars in the mass bin ($m,m+dm$), which depends on the adopted
IMF, $\Phi(m)$. The lowest and highest mass stars that are alive at
this age are $m_{\rm l}$ and $m_{\rm t}$. Note that $m_{\rm t}$ is determined by
the isochrone.

To normalise the stellar spectra in the $V$ band we convolve with the filter
response of \citet{BuserKurucz78}. These spectra are then scaled according to
the predicted flux in this band, following the empirical photometric libraries
described in Section~\ref{sec:isochrones}. In order to obtain the absolute flux in the $V$-band we
follow the method described in \citet{FalconBarroso11}, which is based on the
calibration of \citet*{Fukugita95}:

\begin{equation}
F_{V} = 10^{-0.4 (V + Zp_{V} - V_{\alpha Lyr})}
\end{equation}

\noindent where $Zp_{V}$ and $V_{\alpha Lyr}$ are the adopted zero-point and $V$
magnitude for the reference Vega spectrum, respectively. To compute the
zero-point, we used the \citet{Hayes85} spectrum of Vega with a flux of
$3.44\times10^{-9}$\,erg cm$^{-2}$\,s$^{-1}$\,\AA$^{-1}$ at 5556\,\AA. The $V$
magnitude of Vega is set to $0.03$\,mag, which is consistent with the value
adopted by \citet{Alonso95}, i.e. in agreement with our prescriptions to
transform the theoretical parameters of the isochrones to the observational
plane. This scheme allowed us to achieve absolute magnitudes derived directly
from the synthesised SSP spectra, which were fully consistent with those computed
from the photometric libraries for the same SSPs. These uncertainties are well
below the observational uncertainties ($\lesssim0.01$\,mag) for all ages, metallicities, and
IMFs. We refer the reader to \citet{FalconBarroso11} for more details on the
achieved accuracy.

\subsection{Truly scaled-solar and \alfa-enhanced models}
\label{sec:aassSSPs}

To compute both the truly scaled-solar and the \alfa-enhanced SSP spectra, with
$\aFe=0.0$ and $+0.4$, respectively, we feed the code with stars according to
their estimated \Feh\ and \MgFe\ values. We therefore re-write Eq.~\ref{eq:SSP}
as follows

\begin{eqnarray}
S_{\lambda}(t,\Mh,\aFe,\Phi,I_{\alpha})&=&\int_{m_{\rm l}}^{m_{\rm t}}
S_{\lambda}(m,t,\Feh,\MgFe)\nonumber \\
&& \times {F_{V}}(m,t,\Feh,\MgFe)\nonumber \\
&& \times N_{\Phi}(m,t)dm,
\label{eq:SSPE}
\end{eqnarray}

\noindent In this case the isochrone, $I_{\alpha}$, is characterised by the \aFe\
abundance ratio, which in the present study can take the value of $0.0$ and $+0.4$.
This notation can be generalised to any abundance ratio, including any particular
choice of element ratios, by substituting \alfa\ and \MgFe\ by the desired
elemental abundance(s) ratio(s) in this equation. Conversely, this notation is
not appropriate for the base models, which do not take into account any
consideration for the abundance ratios. 

Note that the only abundance ratio provided in \citet{Milone11} is that of
\MgFe, but not those corresponding to the other \alfa-elements. Hereafter we
write \MgFe\ when referring to the abundance ratio of the stars, which is a
measured quantity, whilst for the models we use \aFe . This is because we use 
the \MgFe\ as a proxy for the \aFe\ abundance ratio of the star (see below and
see also Section~\ref{sec:interpolation}) and because we employ theoretical
isochrones with varying abundance for all the \alfa-elements, i.e., \aFe. 

We take into account the following relation between the iron abundance and the
total metallicity

\begin{equation}
\Feh = \Mh - A\times\MgFe
\label{eq:MhFehrelation}
\end{equation}

\noindent The conversion between \Feh\ and \MgFe\ is dependent on the exact 
solar chemical mixture adopted. For the present study we adopt $A=0.75$, in order
to be consistent with the theoretical stellar spectra of Section~\ref{sec:coelho}
which follow the solar mixture of \citet{GrevesseSauval98}. 

For the high metallicity regime, i.e. with $\Mh\geq0$, where most of the MILES
stars have $\MgFe\sim0.0$, we use Eq.~\ref{eq:SSPE} to combine MILES with
scaled-solar isochrones to compute scaled-solar SSP spectra of total metallicity
$\Mh=\Feh$. Therefore the resulting model, $S_{\lambda}(t,\Mh,0.0,\Phi,I_{0.0})$,
does not differ significantly from the corresponding base model,
$S_{\lambda}(t,\Mh,\Phi),I_{0.0})$, around solar metallicity. Note that in this
case only MILES library star spectra are employed to compute these models.

To synthesise an \alfa-enhanced SSP spectrum with $\MgFe=+0.4$ and total
metallicity, \Mh, we first compute a reference SSP spectrum,
$S_{\lambda_{MILES}}(t,\Mh,0.0,\Phi,I_{0.4})$, which combines an \alfa-enhanced
isochrone, $I_{0.4}$, with MILES scaled-solar stars. We use the BaSTI database
for this purpose. Note that this reference model is not self-consistent as the
abundance ratios of the isochrones and the stellar spectra do differ. In
addition, using the same \alfa-enhanced isochrone we employ the theoretical
stellar spectral library to compute two fully-theoretical SSP spectra: one 
scaled-solar and the other \alfa-enhanced,
$S_{\lambda_{theo}}(t,\Mh,0.0,\Phi,I_{0.4})$  and
$S_{\lambda_{theo}}(t,\Mh,0.4,\Phi,I_{0.4})$, respectively. With these models we
obtain a differential correction as follows

\begin{equation}
\gamma_{\lambda}(t,\Mh,\phi)=\frac
{S_{\lambda_{theo}}(t,\Mh,0.4,\Phi,I_{0.4})}
{S_{\lambda_{theo}}(t,\Mh,0.0,\Phi,I_{0.4})}
\label{eq:difcorr}
\end{equation}

\noindent This correction is then applied to the reference model,
 $S_{\lambda_{MILES}}(t,\Mh,0.0,\Phi,I_{0.4})$, thereby giving a semi-empirical,
 MILES-based, self-consistent\footnote{It is worth noticing that, being similar, the theoretical stellar
spectra adopt a uniform enhancement for all the \alfa-elements, whereas the
isochrones employ a different mixture for these elements, which on average yields
the same enhancement. The variation of the abundance of individual \alfa-elements
has been shown to affect the temperature of the stars (see \citealt{Lee10} for an
extended discussion). In addition, as described in Section~\ref{sec:miles}, we only
take into consideration the \Feh\ and \MgFe\ (which is used as a proxy for the
overall \alfa-enhancement) estimates that are available for the MILES stars.
Therefore our models cannot be regarded as fully self-consistent as this 
is only possible for fully theoretical models that combine theoretical stellar
spectra and isochrones computed with exactly the same element partition.  With
these important caveats in mind, our models are hereafter regarded as
self-consistent (although not fully self-consistent) in the sense that the
average \alfa-enhancement value of the main ingredients of these models is similar.
}
, \alfa-enhanced SSP spectrum

\begin{eqnarray}
S_{\lambda}(t,\Mh,0.4,\Phi,I_{0.4})&=&
{S_{\lambda_{MILES}}(t,\Mh,0.0,\Phi,I_{0.4})}\nonumber \\
&&\times\gamma_{\lambda}(t,\Mh,\Phi).
\label{eq:difcorrMILES}
\end{eqnarray}

\begin{figure}
\includegraphics[angle=0,width=\columnwidth]{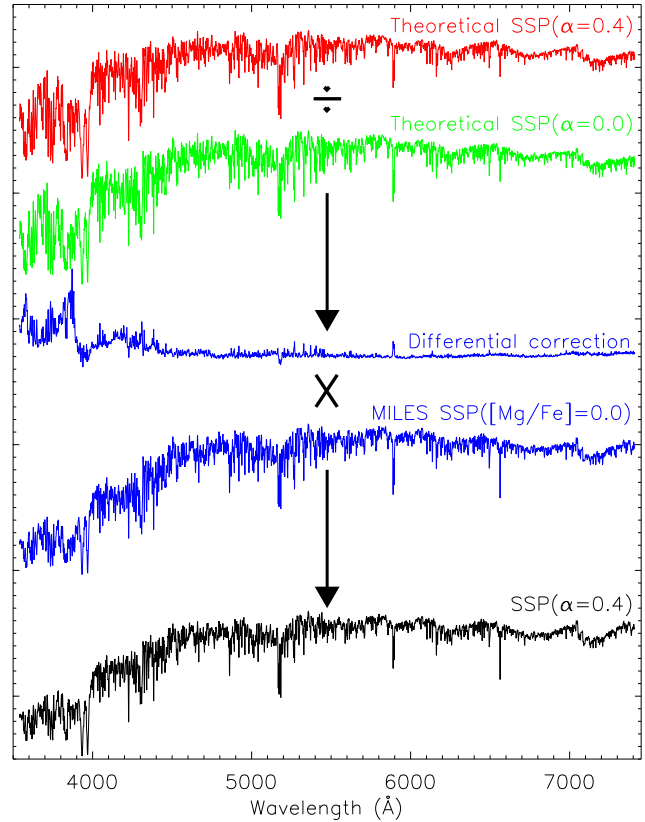}
\caption{The method followed for computing an \alfa-enhanced SSP spectrum with
$\Mh=+0.06$. Fully-theoretical \alfa-enhanced ($\MgFe=+0.4$; shown
in red) and scaled-solar (shown in green) SSP spectra, both computed with an
\alfa-enhanced isochrone, are employed to obtain a differential correction (in
blue). This correction is applied to a reference SSP spectrum (also shown in
blue), computed with MILES scaled-solar stars and the same \alfa-enhanced
isochrone. Therefore we obtain a self-consistent SSP spectrum (shown in black)
that is based on MILES, yet incorporates the \aFe\ abundance ratio effects through
this theoretical differential correction. All the models have $\Mh=+0.06$,
$10$\,Gyr and Kroupa Universal IMF. See the text for a detailed description.
} 
\label{method}
\end{figure}

\noindent The key steps of the method are illustrated in Fig.~\ref{method}. 

For low metallicities, i.e. with $\Mh<0.4$, where the MILES stars have
$\MgFe\sim+0.4$, we use Eq.~\ref{eq:SSPE} to compute fully-consistent models
employing the MILES stellar spectra and \alfa-enhanced isochrones, with no aid
from theoretical stellar spectra. The MILES stars are selected to have
$\MgFe=+0.4$ and $\Feh=\Mh-0.3$, according to Eq.~\ref{eq:MhFehrelation}. To
compute the scaled-solar models we follow a similar approach to that employed
above for obtaining an \alfa-enhanced SSP spectrum at high metallicity. However,
in this case we apply a differential correction, $\gamma_{\lambda}(t,\Mh,\Phi)$,
that allows us to transform the reference MILES-based model with $\aFe=+0.4$,
$S_{\lambda_{MILES}}(t,\Mh,0.4,\Phi,I_{0.0})$ to a self-consistent SSP spectrum
with $\aFe=0.0$, $S_{\lambda}(t,\Mh,0.0,\Phi,I_{0.0})$.

For intermediate metallicities, namely $\Mh=-0.35$ and $\Mh=-0.25$,
where most of the MILES stars have $\MgFe\sim0.2$, we obtain for each \Mh\
metallicity and each \aFe\ abundance two SSP spectra using the two approaches
described above. On one hand, we compute self-consistent scaled-solar SSP models
that are fully based on MILES making use of the stars that have $\MgFe\sim0.0$
and $\Feh=-0.35$ and $\Feh=-0.25$, respectively. Similarly, for the corresponding
\alfa-enhanced models we select stars with $\Feh=-0.65$ and \Feh$=-0.55$,
respectively. Notice that at such low \Feh\ abundances we take the advantage that
most of the stars of MILES have $\MgFe\sim+0.4$. On the other hand, for each of
these two \Mh\ and two \aFe\ values, we also compute a reference model based on
MILES as well as two models based on the theoretical stellar spectral library.
The latter are used to obtain the differential correction that allows us to bring
the reference model onto the desired \aFe\ abundance ratio.
The use of these two self-consistent SSP models for each \Mh, \aFe\ combination,
allows us to significantly increase the total number of the MILES stars that are
employed for the computation of the final SSP spectrum, which, therefore, has a
higher quality than either of the above two SSP spectra, separately. 
It is worth reiterating that we employ this approach as our isochrones only have
two possible \aFe\ values, i.e. $0.0$ and $+0.4$, whereas the MILES stars mostly
have $\MgFe\sim+0.2$. Finally, the last step consists of combining these two SSP
spectra using equal weights. This equal-weighting is motivated by the fact that 
the quality parameters of these two SSP spectra (described in Section~\ref{sec:quality}) 
are very similar for the vast majority of the cases, with differences being 
smaller than $\sim10\,\%$.

Our approach differs from other methods that apply a
differential correction on the individual stellar spectra before synthesising
the SSP spectra (e.g., \citealt{Prugniel07,CvD12models}). In our approach
$\gamma_{\lambda}(t,\Mh,\phi)$ is obtained from the SSP spectra calculated
specifically for this purpose, without modifying the stellar spectra in our
models, similar to the method proposed in \citet{Cervantes07} and \citet{Walcher09}.
This differs from the differential models built in \citet{Walcher09} in that all our
intermediate SSP spectra, either fully-theoretical or reference MILES-based, are
computed with the same population synthesis code. We also differ from
e.g.,\citet{CvD12models} and \citet{Cervantes07}, in that we employ \alfa-enhanced
isochrones to obtain the \alfa-enhanced SSP spectra. Our models also differ from
the fully-theoretical SSP SEDs of, e.g. \citet{Coelho07}, in that we always use MILES
as our input library either completely, or in the form of the reference models. 
To our knowledge, this is the first time that SSPs using the 
differential approach and empirical stellar spectra have been built self-consistently.

\begin{figure}
\includegraphics[angle=0,width=\columnwidth]{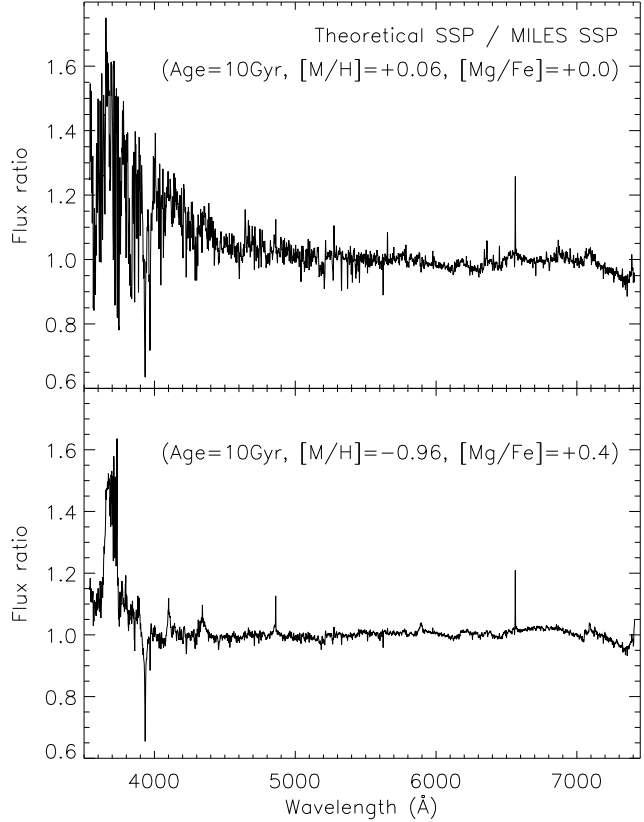}
\caption{Flux-ratio obtained by dividing an SSP SED computed with theoretical
stellar spectra by the corresponding SSP model computed with MILES stars. 
In the top panel the models are scaled-solar and have $\Mh=+0.06$, whereas in the
lower panel the models are \alfa-enhanced with $\Mh=-0.96$. All the models have
$10$\,Gyr and a Kroupa Universal IMF.
}
\label{AvsM}
\end{figure}

Fig.~\ref{AvsM} shows the flux-ratios obtained by dividing fully-theoretical
(i.e., no MILES stars have been used here) SSP spectra versus the
corresponding models computed with MILES (i.e., with no aid of theoretical
stellar spectra) for the same SSP parameters. These ratios correspond to two
representative metallicity and abundance ratio regimes, where it is possible to
compute self-consistent SSP models based solely on the MILES stars: low-metallicity,
\alfa-enhanced and high-metallicity, scaled-solar. The two flux-ratios show a
large-scale variation of the continuum in the blue spectral region. These
residuals include relevant features, which are related to carbon and some
also to nitrogen, such as CNO at $\sim3860$\,\AA, Ca{\sc
II}\,H-K lines around $\sim3950$\,\AA, CN at $\sim4150$\,\AA, Ca at
$\sim4227$\,\AA, CH at $\sim4300$\,\AA\ and C$_2$ at $\sim4670$\,\AA\ (see
\citealt{TripiccoBell95}). Note that all these residuals are emphasised in the flux-ratio 
obtained for the high metallicity regime. These features are attributed, at least in part, to
possible differences in the CN abundance between the theoretical and the MILES
stars. Note that the CNO group is an important contributor to the total metal
content and integrated opacity in a stellar photosphere, whose patterns do not
necessarily follow the \alfa-elements, iron-peak elements or global metallicity,
but have their own significant contributions \citep[e.g.][]{McWilliam08}.
\citet{MartinsCoelho07} find for the C-dominated features that the difference
between the theoretical and empirical stars is emphasised for stars cooler than
$\simeq4500$\,K. These stars have a large impact on the old SSPs represented in
Fig.~\ref{AvsM}.

We also see in Fig.~\ref{AvsM} significant residuals for all the hydrogen
lines, e.g. H$\delta$ at $\sim4100$\,\AA, H$\gamma$ at $\sim4340$\,\AA, H$\beta$
at $\sim4860$\,\AA, H$\alpha$ at $\sim6563$\,\AA. This is a well known problem
in the theoretical models; the modelling of these lines with the assumption 
of LTE is able to reproduce the wings, but cannot match the core of the Balmer lines
which form in the chromosphere and are not accounted for in the theoretical stellar 
spectra (see \citealt{MartinsCoelho07}). 

Residuals for many metallic lines are seen along the whole spectral range in the
high metallicity regime. There are also large-scale residuals
red-ward of $\sim6000$\,\AA, which are associated with molecular bands (mainly TiO).
The contribution from these bands is only relevant for cool stars
($\leq4500$\,K), but is a contribution that increases very rapidly with decreasing temperature. 
The modelling of these bands is challenging in part due to an incomplete knowledge of
the molecular (and line) lists. Difficulties also arise since there may be differences
between the temperature scales of the theoretical and empirical stars (see a
discussion in \citealt{MartinsCoelho07} and \citealt{Coelho14}).

We are aware that our method is not completely free from these residuals that
are related to the use of the theoretical stellar spectra. However, their effects
on our models are minimised as we only employ a differential correction in
passing from, e.g., a scaled-solar to an \alfa-enhanced abundance ratio. This
approach avoids most of the uncertainties coming from the prescriptions adopted
in the theoretical models such as the treatment of convection, turbulence, N-LTE
effects, incomplete line lists, etc. See \citet{Coelho07} for a detailed
discussion on these differences, including a comparison of the resulting
line-strength indices.

Finally, the very large residuals seen at wavelengths smaller than
$\sim3800$\,\AA\ are artificial in the sense that they are a result of the stellar 
spectra calculations performed by \citet{Coelho05} who only included the first 
ten Hydrogen Balmer transitions. This results in artificially high fluxes in the wavelength range
$\lambda\lambda$\ $3650$ -- $3790$\,\AA\ in stars of spectral types F and G, and
artificial peaks centred around $3700$\,\AA\ in the differential spectra
computed from integrated light models. Obviously this problem is far more severe
for the young and/or metal-poor stellar populations. For this reason we do not
recommend the use of the spectral range blue-ward of $\sim3800$\,\AA\ for our model
predictions that make use of the differential correction, i.e. the scaled-solar
SSPs at low metallicities, and the \alfa-enhanced SSPs at high metallicities.

\subsection{Stellar spectra interpolation scheme}
\label{sec:interpolation}

To compute each requested star spectrum, following Eq.~\ref{eq:SSP}, we use the
local interpolation algorithm described in \citet{CATIV} (Appendix B) and
Paper~I. For a given set of stellar atmospheric parameters, $\theta_{0}$ (i.e.
$5040/$\Teff), \logg$_{0}$, \Feh$_{0}$, we interpolate the spectra of adjacent
stars. The code identifies the MILES stars whose parameters are enclosed within
a given box around the point in parameter space. If no star is found this box
is enlarged in each of these directions until suitable
stars are found.  This is done by dividing the original box into eight cubes, all
with one corner at that point. This scheme is particularly useful for minimising
biases in presence of gaps and asymmetries in the distribution of stars around the
point. The larger the density of stars around the requested point, the smaller
the box. The sizes of the smallest boxes are determined by the typical
uncertainties in the determination of the parameters according to \citet{CATII}.
In each of the boxes, the stars are combined taking into account their
parameters and the S/N of their spectra. Stars with similar parameters and higher
S/N are given more weight according to a gaussian function that is applied on 
each parameter axis. 

Finally, the combined spectra in the different boxes, with stellar
parameters close to the requested ones, are used to obtain the final
MILES-based synthetic spectrum. In order to obtain the required atmospheric parameters we
employ a system of linear equations. Note that in some cases, usually in
scarcely populated regions of parameter space, when a mathematical but unphysical (i.e.
with at least one star weighting negatively) solution is found we simply combine
these stars according to the same weighting scheme applied previously for the
individual boxes. This generally provides a star with parameters very close to 
those initially requested. We refer the reader to \citet{CATIV} for a full description of
the algorithm. In the present work we have further minimised these unphysical cases by
repeating the whole process after varying the initial set of parameters
($\theta_{0}$ and \logg$_{0}$) in steps of $1$\,K and $0.1$\,dex for 
temperature and gravity, respectively, until a solution is found. We only allow a
very modest deviation of these two parameters, i.e. $10\%$ of the initial box
size, as estimated for that parametric point. We do not vary \Feh$_{0}$ as in
doing so we could obtain solutions that would be biased towards the solar metallicity
direction, particularly for the lowest and highest metallicity bins covered by
the models.

We have extended this $3$-dimensional interpolation algorithm to be able to 
synthesise the star spectra taking into account the \MgFe\ parameter, as a proxy
for \aFe. However we are constrained within the limits imposed by the library
coverage of this parameter. For this reason, following the abundance pattern of
MILES \citep{Milone11}, at low metallicities our \alfa-enhanced models are fully
based on MILES, whereas for obtaining truly scaled-solar models we require first
MILES-based reference models (that are then modified by theoretical differential
corrections). At high metallicities we make use of the reference models to
obtain our \alfa-enhanced predictions, whereas the scaled-solar models are fully
based on MILES. Therefore our approach, which makes use of the \Feh\ and \MgFe\
estimates available for the stars of MILES, is fully optimised to exploit the
potential of this library (and other empirical libraries). 

To synthesise a stellar spectrum with given \MgFe$_o$ and \Feh$_o$, we run 
our $3$-dimensional interpolator twice: in one case, $\alpha_l$, we restrict the
database to stars with $\MgFe\leq\MgFe_{o}+\Delta\MgFe$ and in the other,
$\alpha_h$, we use the stars with  $\MgFe\geq\MgFe_{o}-\Delta\MgFe$.
$\Delta\MgFe$ represents a fraction of the typical uncertainty in the determination
of this parameter according to \citet{Milone11}. By adopting larger
$\Delta\MgFe$ values we increase the number of stars involved in the process,
but at the expense of increasing the probability of obtaining two stars with lower or
higher \MgFe\ values than requested. After performing several tests, where we
varied  $\Delta\MgFe$ for a representative set of stellar
parametric regions, we adopted $\Delta\MgFe=+0.05$. Note that this corresponds
to approximately half the typical uncertainty in the determination of
this abundance (see \citealt{Milone11}). To estimate the \MgFe\ abundance of
these stars, i.e. $\alpha_l$ and $\alpha_h$, we take into account the \MgFe\
estimates for all the stars that were involved in their calculation through the
3-dimensional interpolation scheme. We then apply the same weights obtained
during this process, which depend on their proximity to $\theta_{0}$,
\logg$_{0}$, \Feh$_{0}$. Note that these three stellar parameters are already in
place for $\alpha_l$ and $\alpha_h$ through this interpolation. Finally we
combine them to obtain the desired stellar spectrum with \MgFe$_o$. In those
cases where $\alpha_l$ and $\alpha_h$ stars have either lower or higher
abundance ratio than \MgFe$_o$ we choose the star whose abundance value is the
closest to the requested one. We note that such cases only occur in poorly
populated parametric regions (e.g. at metallicity \Mh$=+0.4$ or \Mh$=-2.3$).

\section{SSP spectra}
\label{sec:SSP_spectra}

\begin{table*}
\centering
\caption{\label{tab:SEDproperties}Spectral properties of the model SEDs.}
\begin{tabular}{ll}
\hline                   
Spectral range     & $\lambda\lambda$ $3540.5^{1}-7409.6$\,\AA \\	
Spectral resolution& FWHM$=2.51$\,\AA\ ($\sigma=64$\,\kms\ at $5000$\,\AA)\\	
Linear dispersion  & $0.9$\,\AA/pix\\
Continuum shape    & Flux-scaled \\
Telluric absorption & Cleaned \\
Units &F$_{\lambda}$/L$_{\odot}$\AA$^{-1}$\Msol$^{-1}$, L$_{\odot}=3.826\times10^{33}{\rm erg.s}^{-1}$ \\
\hline
\\
\end{tabular}
\centering
\begin{tabular}{l}

$^{1}$ Unsafe blue-ward of $3800$\,\AA\
for the scaled-solar models with $\Mh<0.06$ and \alfa-enhanced models with
$\Mh>-0.4$ (see Section~\ref{sec:aassSSPs}).
\\
\end{tabular}
\end{table*}

\begin{table*}
\centering
\caption{\label{tab:SEDcoverage}SSP parameter coverage.}
\begin{tabular}{ll}
\multicolumn{2}{c}{IMF}\\
\hline
Mass range& $0.1$ -- $100$\,M$_{\odot}$ \\
IMF shape    & Unimodal, Bimodal, Kroupa Universal$^{1}$, Kroupa Revised$^{1}$\\
Unimodal IMF slope & $0.3\leq\Gamma\leq3.3$$^2$ (Salpeter IMF $^{3}$ if $\Gamma=1.3$)\\
Bimodal IMF slope & $0.3\leq\Gamma_{b}\leq3.3$\\
\hline
\\
\multicolumn{2}{c}{Age range (Gyr)}\\
\hline
BaSTI-based models   & $0.030$ -- $14.0$\\
Padova00-based models& $0.063$ -- $17.8$\\
\hline
\\
\multicolumn{2}{c}{Total metallicity (\Mh)}\\
\hline
BaSTI-based models & $-2.27^{4}$,$-1.79^{5,6,7}$,$-1.49^{8,9,10,11,12}$,$-1.26^{13,14,15,16,17}$,$-0.96^{18}$,$-0.66$,$-0.35$,$-0.25$,$+0.06$,$+0.15$,$+0.26$,$+0.40^{4}$\\
Padova00-based models & $-2.32^{4}$,$-1.71^{19,20}$,$-1.31$,$-0.71$,$-0.41$,$+0.0$,$+0.22$\\
\hline
\\
\multicolumn{2}{c}{Abundance ratio (\aFe)}\\
\hline
Base models    & MILES pattern\\
scaled-solar$^{21}$   & $+0.0$\\
\alfa-enhanced$^{21}$ & $+0.4$\\
\hline
\\
\end{tabular}
\centering
\begin{tabular}{l}

$^{\ 1}$ Same unsafe models as for Bimodal IMF with $\Gamma_{b}=1.3$\\
$^{\ 2}$ Unimodal IMF models unsafe for $\Gamma > 2.3$\\
$^{\ 3}$ Same unsafe models as for Unimodal IMF with $\Gamma=1.3$\\
$^{\ 4}$ Unsafe models\\

$^{\ 5}$ Unimodal BaSTI base models unsafe for $t<0.09$\\
$^{\ 6}$ Bimodal BaSTI base models unsafe for ($t<0.1$,$\Gamma_{b}<2.3$), ($t<0.09$,$2.3\leq\Gamma_{b}\leq2.5$), ($t<0.08$,$2.8\leq\Gamma_{b}\leq3.0$), ($t<0.06$,$\Gamma_{b}=3.3$)\\
$^{\ 7}$ scaled-solar and \alfa-enhanced models unsafe\\

$^{\ 8}$ Unimodal BaSTI base models unsafe for $t<0.07$\\
$^{\ 9}$ Bimodal BaSTI base models unsafe for ($t<0.07$,$\Gamma_{b}<2.8$), ($t<0.06$,$\Gamma_{b}=2.8$), ($t<0.05$,$\Gamma_{b}=3.3$)\\
$^{10}$ Bimodal scaled-solar models unsafe for ($t<0.08$,$\Gamma_{b}\leq2.5$), ($t<0.07$,$2.8\leq\Gamma_{b}\leq3.0$), ($t<0.06$,$\Gamma_{b}=3.3$)\\
$^{11}$ Bimodal \alfa-enhanced models unsafe for ($t<0.08$,$\Gamma_{b}\leq2.8$), ($t<0.07$,$3.0\leq\Gamma_{b}\leq3.3$)\\
$^{12}$ Unimodal scaled-solar and \alfa-enhanced models unsafe for $t<0.08$\\

$^{13}$ Unimodal BaSTI base models unsafe for $t<0.04$\\
$^{14}$ Bimodal BaSTI base models unsafe for ($t<0.04$,$\Gamma_{b}<2.8$)\\
$^{15}$ Bimodal scaled-solar models unsafe for ($t<0.07$,$\Gamma_{b}\leq1.8$), ($t<0.06$,$2.0\leq\Gamma_{b}\leq3.0$), ($t<0.05$,$\Gamma_{b}=3.3$)\\
$^{16}$ Bimodal \alfa-enhanced models unsafe for ($t<0.07$,$\Gamma_{b}\leq2.0$), ($t<0.06$,$2.3\leq\Gamma_{b}\leq3.0$), ($t<0.05$,$\Gamma_{b}=3.3$)\\
$^{17}$ Unimodal scaled-solar and \alfa-enhanced models unsafe for $t<0.07$\\

$^{18}$ Unimodal scaled-solar and \alfa-enhanced models unsafe for $t<0.05$\\

$^{19}$ Bimodal Padova00 base models unsafe for ($t<0.075$,$\Gamma_{b}<2.5$), ($t<0.07$,$2.5\leq\Gamma_{b}\leq3.0$)\\
$^{20}$ Unimodal Padova00 base models unsafe for $t<0.075$\\

$^{21}$ See the caveats in Table~\ref{tab:SEDproperties} for the spectral range blue-ward of $\sim3800$\,\AA.\\

\end{tabular}
\end{table*}

Table~\ref{tab:SEDproperties} summarises the spectral properties of the newly
synthesized SSP SEDs. The nominal resolution of the models is FWHM$=2.51$\,\AA\
\citep{FalconBarroso11}. This resolution is constant along the whole spectral
range $\lambda\lambda$ $3540.5-7409.6$\,\AA. The models have flux-calibrated
response and are cleaned from telluric absorption residuals \citep{MILESI}. For
computing the SEDs we have adopted a total initial mass of $1$\,\Msol.
See also the caveats
in Table~\ref{tab:SEDproperties} and Section~\ref{sec:aassSSPs} regarding the
spectral range blue-ward of $\sim3800$\,\AA.

\subsection{Quality of the SSP spectra}
\label{sec:quality}

Table~\ref{tab:SEDcoverage} lists the covered SSP parameters and also gives the
ranges where these predictions cannot be considered safe. 

\begin{figure}
\includegraphics[angle=0,width=\columnwidth]{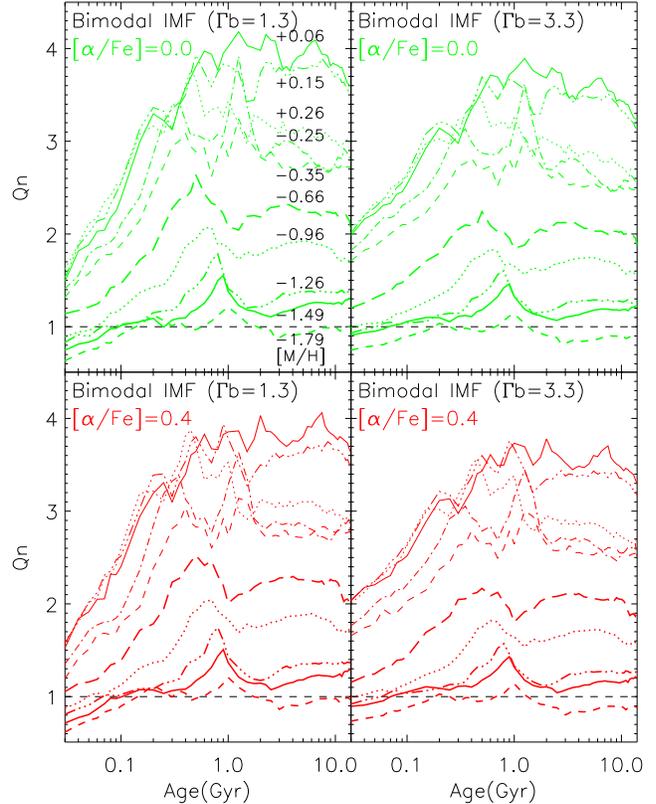}
\caption{Reliability of the SSP spectra synthesised with a bimodal IMF with slope
$\Gamma_{b}=1.3$ (i.e. Kroupa-like; left panels) and $\Gamma_{b}=3.3$ (right panels), for
scaled-solar (upper panels) and \alfa-enhanced (lower panels) models. The SSP
age increases from left to right, whereas the total metallicity, \Mh, is shown
as different line-types and thicknesses as given in the first panel. Only
Q$_n$ values above $1$ are acceptable, as indicated by the horizontal black
dashed-line.}
\label{fig:Q_BI}
\end{figure}

\begin{figure}
\includegraphics[angle=0,width=\columnwidth]{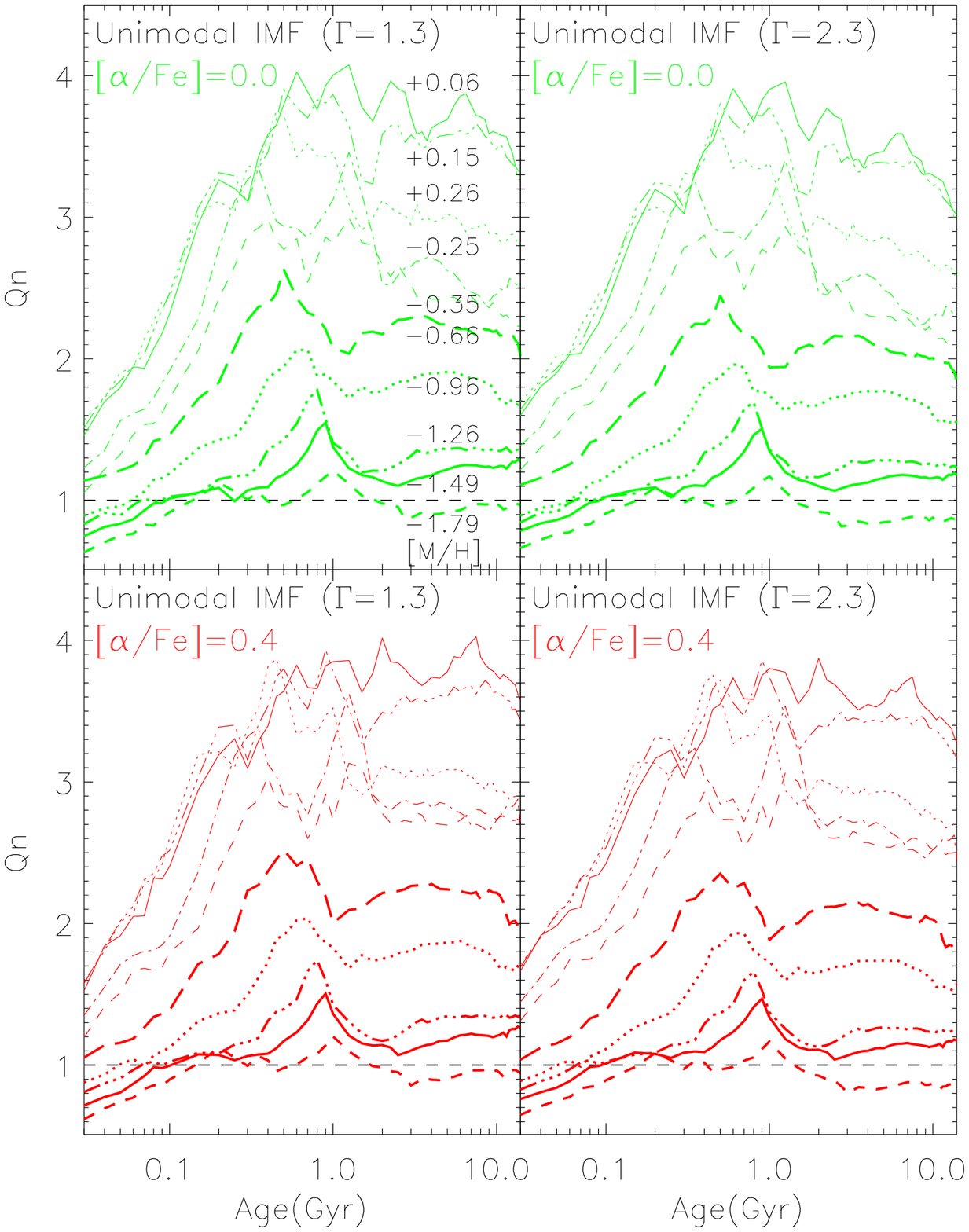}
\caption{Same as in Fig.~\ref{fig:Q_BI} for models with unimodal IMF with two
slopes ($\Gamma=1.3$, i.e. Salpeter, and $\Gamma=2.3$).}
\label{fig:Q_UN}
\end{figure}

To estimate the
reliability of the synthesised SSP spectra we follow the quantitative approach
described in Paper~I. For each SSP calculation we use the same interpolation
scheme described in Section~\ref{sec:interpolation}, which synthesises a
representative stellar spectrum for a given set of atmospheric parameters. We
compute a normalised quality parameter $Q_n$ following Equations 4--6 of
Paper~I. Essentially, the higher the density of stars around the requested
parametric point of each star along the isochrone, the higher the quality of the
SSP spectrum. $Q_n$ is determined by the MILES library, either when the model is computed
on the basis of this library alone or when we make use of the MILES reference
model. Our quality parameter is normalised with respect to a minimum acceptable
value, which comes from a poor, but still acceptable, parameter coverage. We
refer the interested reader to Paper~I for a full description of the method. The
only difference in the present work is that in the process of synthesising each requested
stellar spectrum we obtain two values for this parameter, which correspond to
the two stars required by our extended interpolation algorithm to match the
\MgFe\ abundance ratio. We simply combine these values in the same way we did for
these stars. 

The reliability of the scaled-solar and \alfa-enhanced SSPs with bimodal and
unimodal IMFs is assessed in Fig.\,{\ref{fig:Q_BI}} and
Fig.\,{\ref{fig:Q_UN}}, respectively. Each panel shows how $Q_n$ varies with
the SSP age and total metallicity. SSPs with $Q_n$ values above $1$ can be
considered of sufficient quality, and therefore they can be safely used. We use
the same scale for all the panels of these two figures. The figures show that
the $Q_n$ values corresponding to the scaled-solar and \alfa-enhanced SSPs
are very similar. The subtle differences seen between the upper and lower panels
are attributed to differences in the temperatures of the selected stars
along the isochrones, i.e. scaled-solar vs. \alfa-enhanced), and the \Feh\
metallicity according to Eq.~\ref{eq:MhFehrelation}. 

Although not shown, the $Q_n$ values of the SSP models computed with a Kroupa
Universal and Kroupa revised IMFs are similar to those obtained for a
bimodal IMF with slope $1.3$. The quality of the scaled-solar models is lower
than that obtained for the base models, i.e. above $5$ (see Figure~6 of
Paper~I). This is because in the computation of truly scaled-solar models we
select stars according to both \Feh and \MgFe, whereas for the base models we
only take into account \Feh. In summary, the base models, either based on
Padova00 or BaSTI have in general higher quality than the corresponding
scaled-solar and \alfa-enhanced models.

As expected, the higher $Q_n$ values are obtained at solar metallicity. $Q_n$
reaches a value of $\sim4$ for stellar populations in the range $1$ -- $10$\,Gyr
for the SSP models with standard IMF slope (i.e. the left panels of
Fig.\,{\ref{fig:Q_BI}} and Fig.\,{\ref{fig:Q_UN}}). 

We also see that the quality of the bimodal models is slightly higher than that
computed with a single power-law IMF. This is due to the small number of very
low-mass dwarfs with temperatures below $\sim3500$\,K. This also explains the
decay in quality for ages above $\sim10$\,Gyr, as well as the drop seen for the
models with a bottom-heavy IMF. Note that, as expected, this drop is more
pronounced for the unimodal SSP models. Note also that the models with bimodal
IMF and slope $3.3$ yield very similar values to those obtained with unimodal
IMF and slope $2.3$. This is due to the fact that the relative fraction of
dwarfs with masses below $0.5$\,\Msol\ is very similar for these two IMFs (see
\citealt{LaBarbera13}). 

Finally the models with a {\it unimodal} IMF with
$\Gamma>2.3$ are unsafe due to an emphasised contribution of very low-mass
stars with temperatures cooler than $3500$\,K, which are scarce in the MILES
library. We refer the interested reader to \citet{MIUSCATI} for the details
regarding this limitation. Despite the claim of a steepening of the IMF slope
with galaxy velocity dispersion, such extremely bottom-heavy IMF models do not
seem to be required for fitting even the most massive galaxies, always with
$\Gamma<2.3$ (e.g. \citealt{LaBarbera13,Spiniello14}).

The variation of the $Q_n$ values with age at a given metallicity
originates from the varying number density of stars with temperature, such as the Turn-Off
stars, which become cooler with increasing age. This is illustrated by
looking at the number density of stars around solar metallicity (i.e. along a
vertical line in the right panel of Fig.~\ref{milone}). For ages smaller than
$1$\,Gyr we also see a decrease in quality due to a decrease in the number of
stars with high temperatures. 

These figures also show that the derived quality depends on metallicity. In
general, the more we depart from solar the lower the $Q_n$ value. Unlike the
base models (see Figure~6 of Paper~I) the SSPs with supersolar metallicity show
higher $Q_n$ values than the models with \Mh$=-0.35$ and $-0.25$ shown here.
This is because at
these intermediate metallicities the required reference models, either with
\aFe$=+0.0$ ($S_{\lambda_{MILES}}(t,\Mh,0.0,\Phi,I_{0.4})$) or $\MgFe=+0.4$
($S_{\lambda_{MILES}}(t,\Mh,0.4,\Phi,I_{0.0})$), are composed of 
significantly fewer stars, since the majority have \MgFe$\simeq+0.2$. 
For the base model we neglect this parameter and therefore
all these stars are taken into account in the interpolation scheme. The drop in
quality is more pronounced for lower metallicities. Overall, the very youngest
models are not reliable for metallicities below \Mh$\sim-1.0$. The minimum
acceptable metallicity for the \alfa-enhanced and scaled-solar models is reached
for models with $\Mh=-1.49$. Note that this limit is lower for the base models,
which are safe for $\Mh=-1.79$ (BaSTI) with ages above $\sim0.1$\,Gyr and for
$\Mh=-1.71$ (Padova00) with ages above $\sim0.08$\,Gyr. Finally we also see that
the \alfa-enhanced and scaled-solar predictions are unsafe for ages younger than
$\sim0.07$\,Gyr. In Table~\ref{tab:SEDcoverage} we provide more detailed information
on which regions of parameter space (age, metallicity, IMF slope and IMF shape) cannot 
be considered fully safe. 

Models outside the safe ranges discussed here should not be considered suitable
for detailed spectroscopic studies that employ line-strengths or a full
spectrum-fitting approach. However these models might be still useful, with
caveats, for some applications such as, e.g., broad-band colours or for more
complex populations, which are dominated by SSPs within the safe ranges.
We also computed scaled-solar and \alfa-enhanced models with \Mh$=+0.4$ and
\Mh$=-2.3$, which are unsafe, but that could help us to visualise trends with
metallicity of e.g. line-strengths. 

\subsection{Effects of varying the abundance on the SSP spectra}
\label{sec:MILES_SSPs}

\begin{figure} 
\includegraphics[angle=0,width=\columnwidth]{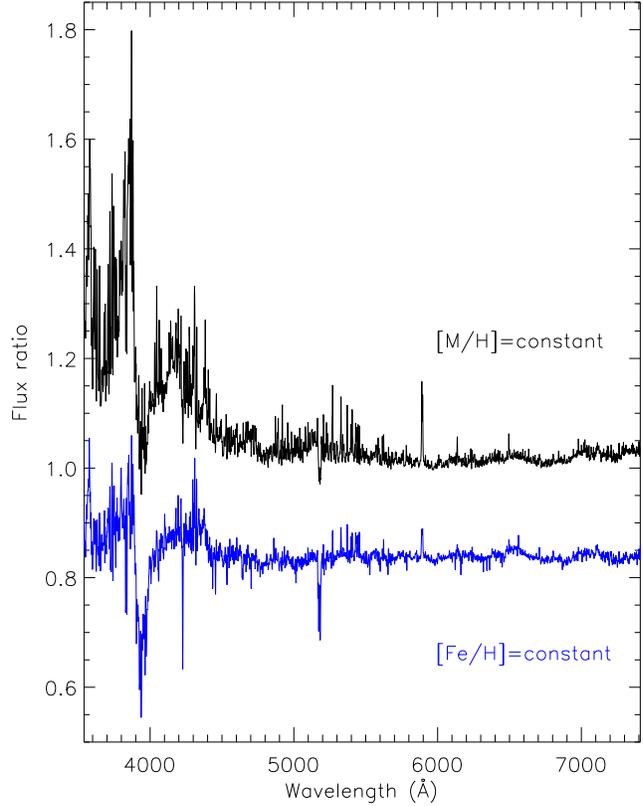}
\caption{Flux-ratios obtained by dividing an \alfa-enhanced versus a
scaled-solar SSP model with constant iron content ($\Feh=-0.25$, in
blue), and constant total metallicity ($\Mh=+0.06$, in black).  Both
flux-ratios correspond to models with $10$\,Gyr and Kroupa Universal IMF.}
\label{fig:mhfehconstant} 
\end{figure}

\begin{figure}
\includegraphics[angle=0,width=\columnwidth]{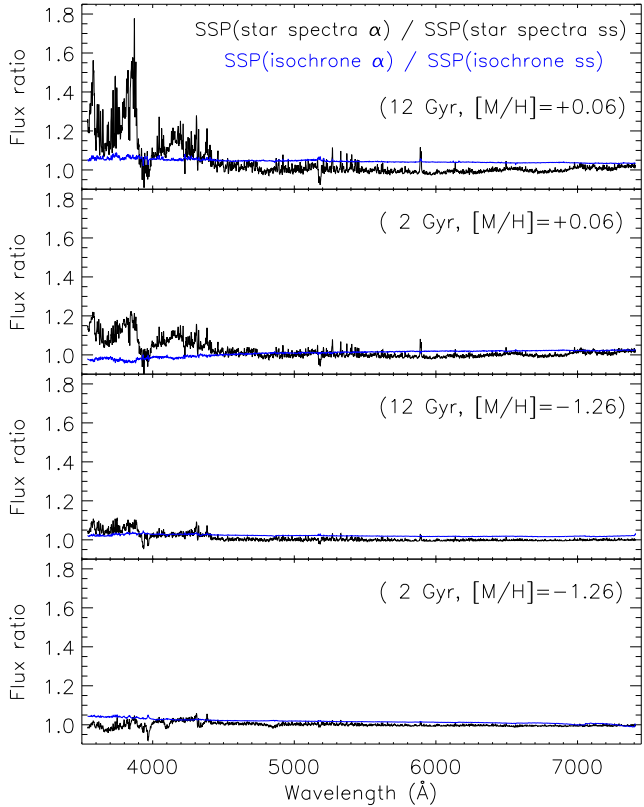}
\caption{Flux-ratios obtained by dividing two SSP spectra computed with the same
isochrones but in one case employing \alfa-enhanced spectra and in the other
scaled-solar stellar spectra, shown in black. The ratios resulting from dividing
two SSP spectra computed with the same stellar spectra but varying the abundance
of the isochrones are shown in blue. The results for two total metallicity
values ($+0.06$ and $-1.26$) and two ages ($2$ and $12$\,Gyr) are shown as quoted within
the panels.
} 
\label{fig:isocspectra}
\end{figure}

\begin{figure}
\includegraphics[angle=0,width=\columnwidth]{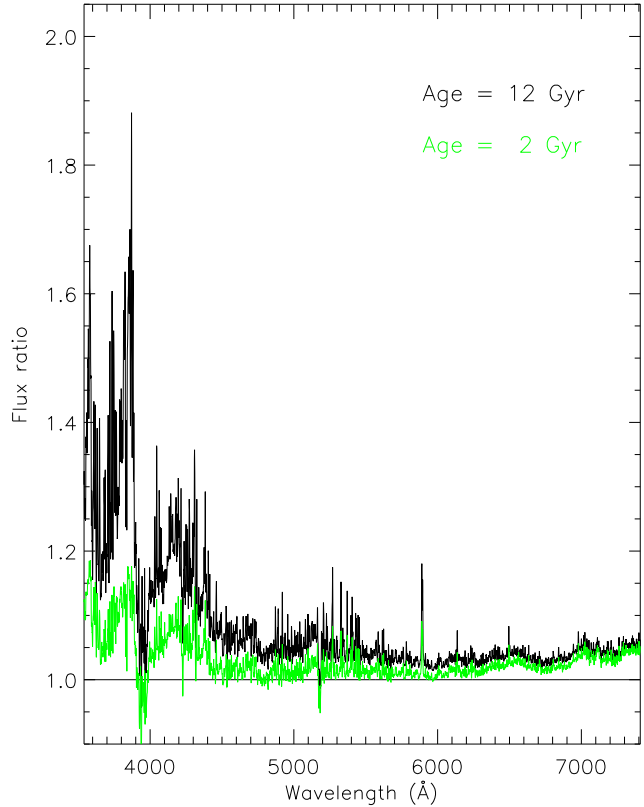}
\caption{Flux-ratios obtained by dividing \alfa-enhanced versus
scaled-solar SSP spectra of $2$\,Gyr (green) and $12$\,Gyr (black). All the flux-ratios 
correspond to models with solar (total) metallicity and Kroupa Universal IMF.}
\label{fig:ageratio}
\end{figure}

\begin{figure}
\includegraphics[angle=0,width=\columnwidth]{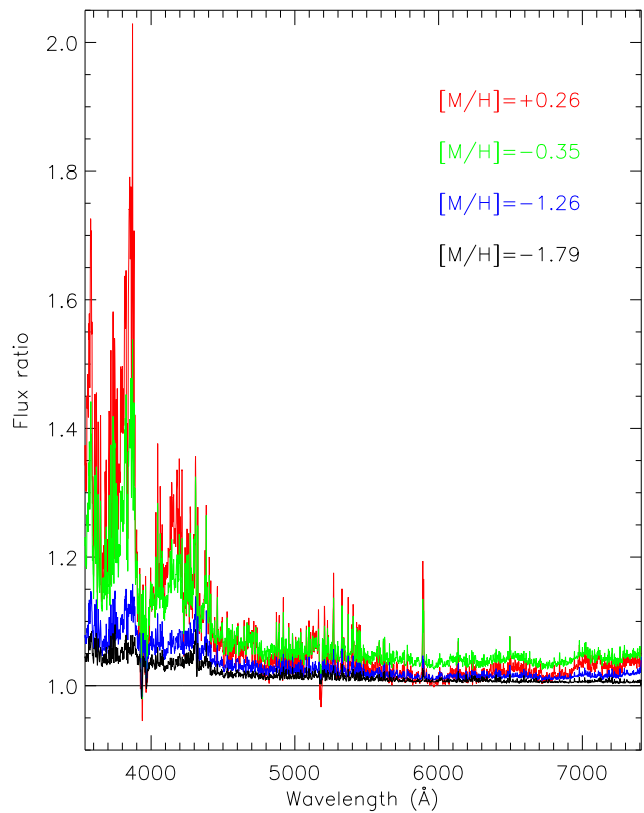}
\caption{Flux-ratios obtained by dividing \alfa-enhanced by
scaled-solar SSP spectra with constant total metallicity. The flux-ratios are
shown for four metallicity values: $\Mh=-1.79$ (black), $-1.26$
(blue),$-0.35$ (green) and $+0.26$ (red). All the flux-ratios 
correspond to models of $10$\,Gyr with a Kroupa Universal IMF.}
\label{fig:metratio}
\end{figure}

The effects of varying the \alfa-elements abundance ratio at constant total
metallicity, \Mh, and at constant \Feh\ abundance can be assessed from the
flux-ratios shown in Fig.~\ref{fig:mhfehconstant}. When the total metallicity
is held constant we see a significant increase in flux at wavelengths shorter than about
$4500$\,\AA. This behaviour is also seen when dividing stellar spectra with
varying \MgFe, both for empirical stars from the MILES library and for
theoretical stellar spectra. We refer the reader to the extensive study performed in
\citet{Sansom13} for more details (see also \citealt{Cassisietal04}). 
This excess of blue flux is attributed to the fact that the \alfa-enhanced element 
mixture has a decreased iron abundance, and therefore lower opacity, with respect to the scaled-solar
model of similar total metallicity (irrespective of the exact value of the term
multiplying the \MgFe\ abundance in Eq.~\ref{eq:MhFehrelation}). This is
confirmed by the modest excess in blue flux seen in the bottom spectral ratio of
Fig.~\ref{fig:mhfehconstant}, which corresponds to models with similar \Feh\
abundance. 

In addition, we see similar spectral features blue-ward of $\sim4500$\,\AA\ in both 
the spectral ratios, although generally these are less pronounced when \Feh\  is 
held constant. These residuals
are mainly related to the Ca{\sc II}\,H-K and Ca{\sc I} features around
$\sim3950$\,\AA\ and $\sim4227$\,\AA, respectively, the CN molecular absorptions
at $\sim4150$\,\AA\ and in the complex blue-ward of $\sim3900$\,\AA\, and the CH at
$\sim4300$\,\AA\ and also within this blue complex. These residuals likely 
originate in intrinsic differences between the C, N and O and Ca abundances
between the MILES stars and the theoretical stellar spectra of \citet{Coelho05}.
In our SSP model computation we only have considered the \MgFe\ and \Feh\
estimates for the MILES stars, as the
measurements for the other elements are incomplete for these stars.
However there are indications of such varying abundance patterns among the
MILES and the theoretical stars. For example, applying the relation between
\CFe\ and \Feh\ of \citet{TakedaHonda05}, i.e. \CFe$=-0.21$\Feh$+0.014$, derived for
a sample of $160$ FGK dwarfs/subgiants, with metallicities around the solar value,
we obtain \CFe$\sim+0.07$ for \Mh$=+0.06$ and \aFe$=+0.4$, whilst the theoretical
stellar spectra adopt a scaled-solar abundance for carbon.
Moreover, unlike in the theoretical stellar spectra, for which \OFe\ tracks \MgFe, observationally
the abundances of these two \alfa-elements are slightly decoupled around solar
metallicity (e.g. \citealt{SoubiranGirard05}). In addition, the CNO group
represents an important fraction of the total metallicity, contributing
significantly to the integrated opacity in a stellar photosphere (see the
extensive discussion in \citealt{Sansom13}).

Fig.~\ref{fig:mhfehconstant} also shows that the Fe lines at, e.g., $\sim5270$\,\AA\ and
$5335$\,\AA, are deeper in the scaled-solar models 
than the \alfa-enhanced models (positive residuals in the ratios plotted). 
This happens also at fixed \Feh, as noted in \citet{Barbuy03,Coelho05}: 
Mg is an important electron donor and as such can act to lower the true continuum. This effect
is also seen for the Na{\sc I} doublet at $\sim5895$\,\AA\ (see, e.g., 
\citealt{Vazdekis97,Coelho07}) and, to a lesser degree, in the variations of the
molecular features red-ward of $\sim6000$\,\AA. In general these features, although
weaker, are also present in the flux-ratio obtained when \Feh\ is kept
constant. The exception here is the Mg feature at $\sim5175$\,\AA, where the \Mgh\
abundance of the two SSP spectra differ by $-0.3$ dex. See Section~\ref{sec:lines}
for more details.

In Fig.~\ref{fig:isocspectra} we separate the effects of the
isochrones from those of the stellar spectra on our SEDs. We show the results for two ages
($2$ and $12$\,Gyr) and two total metallicities with \Mh$=+0.06$ and $-1.26$.
For each age and metallicity we compute SSP spectra with two different isochrones
(\alfa-enhanced, and scaled-solar), but without changing the abundance ratios of the
stellar spectra. In addition, we also show the flux-ratios obtained when using the same
isochrones, but varying the abundance ratios of the stellar spectra. The
flux-ratios obtained indicate that the effects of the stellar spectra are
significantly more relevant than the isochrones, particularly for solar metallicity and old stellar
populations, confirming the previous findings by \citet[][Section~6.4
]{Coelho07} and \citet{Lee09}. The impact of the stellar spectra is mostly seen
in the excess of flux in the blue, the strengths of many features and, to a
lesser degree, on some molecular bands. However the effects of the isochrones
are mostly reflected in the continuum over the whole spectral range covered by
the models. Although smaller than the effects of the stellar spectra, the impact
of the isochrones is significant for the SSPs of $2$\,Gyr. Note also the small
difference in flux between the two flux-ratios. 

Fig.~\ref{fig:ageratio} shows the flux-ratios obtained by dividing SSP
spectra with the same total metallicity (solar), but with two different \aFe\
abundances, for two ages ($2$ and $12$\,Gyr). We see that the impact 
on the spectra due to the  \aFe\ abundance is significantly greater at old
ages. Such behaviour is expected since the SSPs with
$2$\,Gyr include hotter stars, for which the effects of varying the \aFe\
abundance are considerably smaller (see Figure~9 of \citealt{Sansom13}).

Finally, Fig.~\ref{fig:metratio} illustrates the effects of varying the \aFe\
abundance as a function of the total metallicity, \Mh, while keeping constant
the age. We see both a larger excess in blue flux and a larger line feature
residuals, for higher metallicities. As was the case when varying the SSP
age, this behaviour is also expected when varying the metallicity since 
the stellar temperatures decrease with increasing metallicity. 

\section{Line-strength indices}
\label{sec:lines}

In this section we study the behaviour of the line-strength indices predicted
by our new models. We measure the indices on the synthesised SSP spectra at
resolution $8.4$\,\AA\ (FWHM) to match the resolution of the LIS-8.4\,\AA\
system that was introduced in Paper~I. Unlike the Lick/IDS system
\citep{WortheyOttaviani97}, all these indices are measured with the same
resolution on a flux-scaled response. 

To illustrate the main behaviours and differences of our models 
we first compare a number of representative
line-strengths from our base models computed with the Padova00 and BaSTI
isochrones. In the following subsections we compare the predictions from our
scaled-solar and \alfa-enhanced models for a larger number of commonly used
line indices, which include those of Lick. We have organised these indices
according to their sensitivities to relevant parameters of the stellar
populations.  

\subsection{Line-strengths of the base models}
\label{sec:indiceBastiPadova}

\begin{figure}
\resizebox{1.0\columnwidth}{!}{\includegraphics[angle=-90]{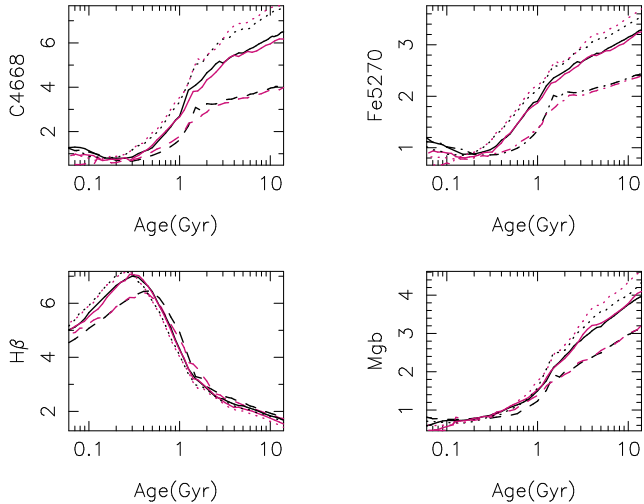}}
\caption{Base models with two different isochrones; black are
BaSTI and magenta lines are Padova00. Three different metallicities are plotted
for BaSTI (Padova00), $\Mh=-0.35 (-0.40), +0.06 (0.00)$, and $+0.26 (+0.22)$, with 
dashed, solid and dotted lines, respectively. All the models have a Kroupa
Universal IMF. Note that whereas the models based on the BaSTI isochrones
reach ages as young as $30$\,Myr the Padova00-based models stop at $63$\,Myr 
(see Table~\ref{tab:SEDcoverage}). All the index measurements were obtained in
the LIS-8.4\,\AA\ system (Paper~I), i.e. at FWHM$=8.4$\,\AA\ and flux-calibrated response.}
\label{isochrones:fig}
\end{figure}

Fig.~\ref{isochrones:fig} shows the predictions of some line-strength indices as
a function of age for the base models, computed with BaSTI and Padova00
isochrones. The predictions are shown for three different metallicities. As can
be seen, the differences found using these two sets of isochrones
are very small. For old stellar populations the BaSTI models give index
values that are slightly higher for the Balmer lines (around $0.1-0.2$\,\AA\ for old stellar
populations), and slightly lower for the
three metallicity indices shown in the figure. Note that the given \Mh\
values are slightly higher for BaSTI than for Padova00. Therefore when comparing
the BaSTI model predictions to observational data we expect the derived
ages and metallicities to be slightly higher than when using the Padova00
model predictions. There are also differences for intermediate-aged stellar
populations where the AGB contribution is relevant, due to the differences in
the treatment of this late evolutionary stage (see
Section~\ref{sec:isochrones}).

\subsection{Line-strengths with varying abundance ratios}
\label{sec:indices_ass}

\begin{figure*}
\resizebox{1.0\textwidth}{!}{\includegraphics[angle=-90]{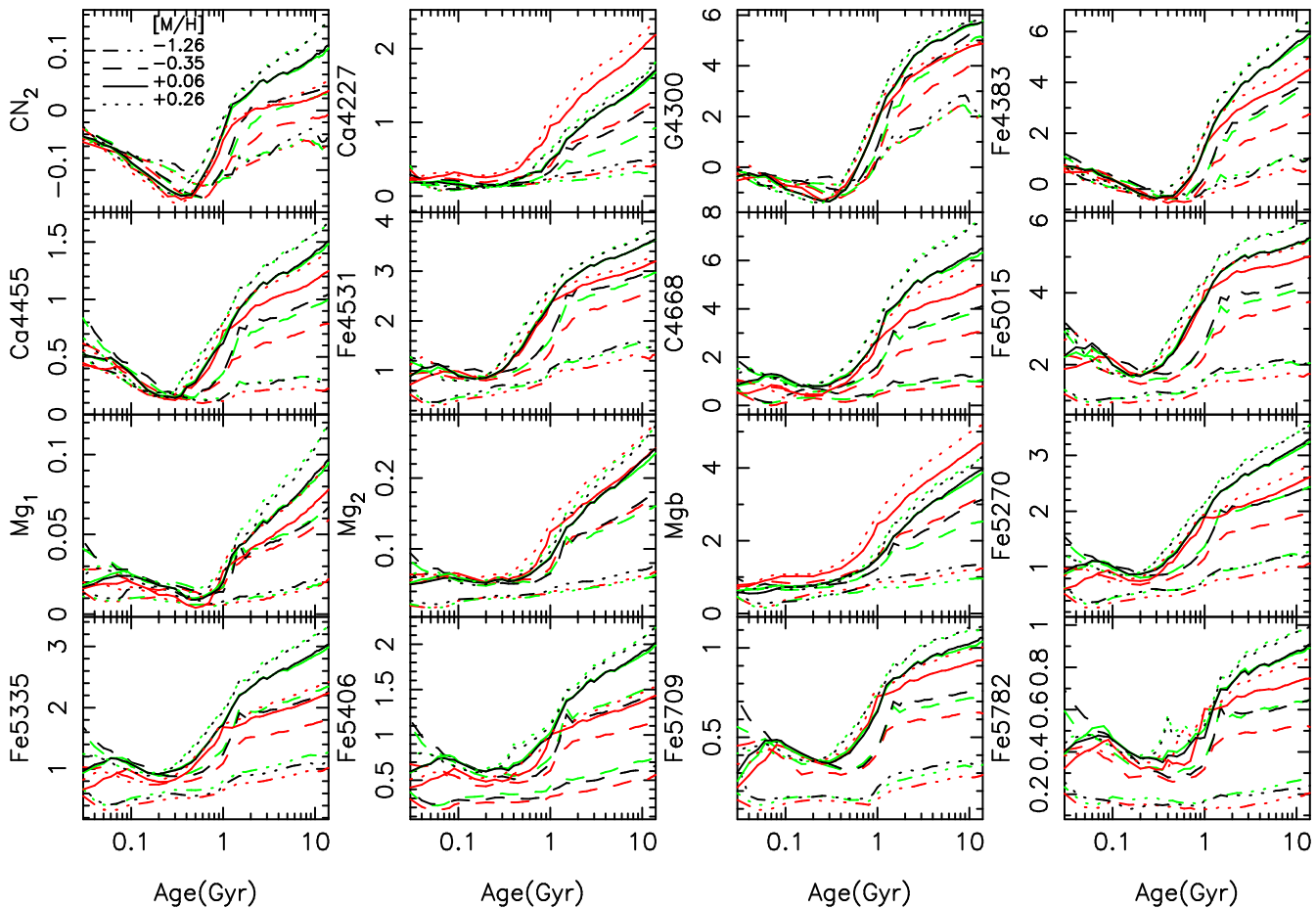}}
\caption{Variations of a representative set of Lick line-strength indices with
age for four different metallicities, \Mh=$-1.26$ (dot-dashed line);
\Mh$=-0.35$ (dashed line) \Mh$=0.0$ (solid line) and \Mh$=+0.26$ (dotted line).
Colours represent three different models built with the BaSTI isochrones; 
Red: \alfa-enhanced models, green: scaled-solar models and black: base models.
As in Fig.~\ref{isochrones:fig}, all the index measurements are performed in the LIS-8.4\,\AA\ system, i.e. at
$8.4$\,\AA\ (FWHM) and flux-calibrated response.}
\label{index_age}
\end{figure*}

\begin{figure}
\includegraphics[angle=270,width=\columnwidth]{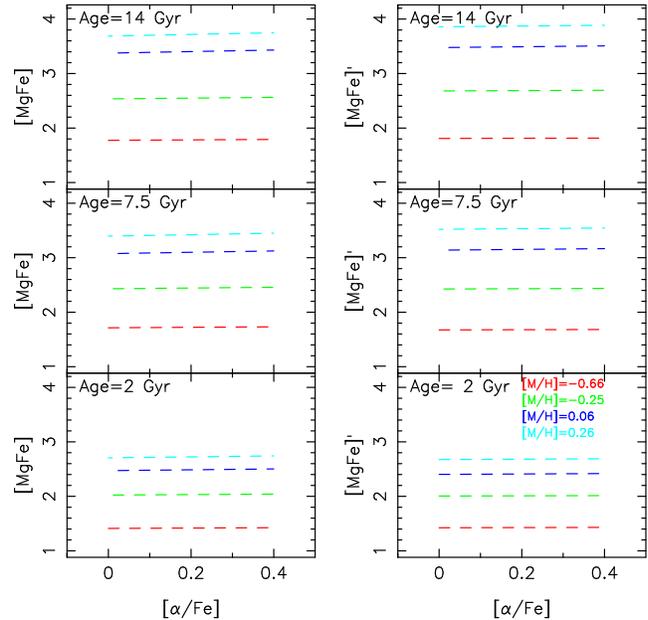}
\caption{\mgfe\ (left panels) and \mgfep\ (right panels) indices as a function
of \aFe, for SSP models with three different ages ($14$, $7.5$ and $2$\,Gyr) from top
to bottom panels, and four different metallicities (\Mh$=-0.66, -0.25, +0.06$ and
$+0.26$, in red, green, blue and cyan, respectively. All the models have a Kroupa
Universal IMF. Note that the two indices are almost insensitive to this abundance ratio.}
\label{mgfemgfe}
\end{figure}

Fig.~\ref{index_age} shows the predicted time evolution of our BaSTI-based SSP
models for a selection of line-strength indices measured from the scaled-solar, 
\alfa-enhanced and base models. 
All these models are computed with a Kroupa Universal IMF. We have
plotted the results for four different metallicities. Perhaps most
readily seen in this set of plots is that, for the majority of indices, the
effect of the different \aFe\ ratios is largest at old ages.  For some
indices (e.g., Mg$_1$, Fe\,$5015$, Fe\,$5335$, Fe\,$5406$, Fe\,$5709$,
Fe\,$5782$), differences also appear at very young ages ($<0.1$\,Gyr), with 
these differences increasing at high metallicities. These differences come from the 
influence of supergiant stars at these young ages. 

The base models have the abundance pattern of the Milky Way (MILES) stars,
although the isochrones are scaled-solar. Therefore, at low metallicities
($\Mh<-0.4$), the base models are more similar to the \alfa-enhanced models, while at high
metallicities they are almost identical to the scaled-solar models. 

As expected those indices sensitive to Fe (Fe\,$4383$, Fe\,$4531$, Fe\,$5015$,
Fe\,$5270$, Fe\,$5335$, Fe\,$5406$, Fe\,$5709$ and Fe\,$5782$) show lower values in the
\alfa-enhanced models than in the scaled-solar models.  
This occurs because the models are built at constant metallicity and, therefore, 
the \alfa-enhanced models have a lower Fe content 
(i.e, \Mh$-0.3$, according to Eq.~\ref{eq:MhFehrelation}). 
Note however that the effect on the Fe\,$5015$ index is relatively smaller
in comparison to the other Fe-sensitive indices, as a result of its sensitivity
to Ti abundance, which is enhanced in the \alfa-enhanced models
\citep{Lee09,Johansson12}. The same applies to Fe\,$4531$ for which we also see
that the \alfa-enhanced model lines are closer to the corresponding base model
lines (see also figure~9 in \citealt{Lee09})
This beahviour is also seen for fixed \Feh, as Mg, i.e an important electron
donor, tends to lower the true continuum \citep{Barbuy03,Coelho05}. 

It is worth noting that, while \mgb\ is higher in the \alfa-enhanced
models than the scaled-solar models, as expected due to its sensitivity to the Mg abundance, 
the same is not true for Mg$_1$ and Mg$_2$. In fact, the Mg$_1$ indices are stronger 
in the scaled-solar models at all metallicities. It is known that Mg$_1$ has a greater
sensitivity to C than Mg. In the \alfa-enhanced models, all the non-\alfa\ elements are depleted 
relative to the scaled-solar models, including C. This
behaviour of Mg$_1$ is similar to that of the CN$_2$ and C$_2$\,$4668$ indices,
which have higher sensitivities to variations of N and C abundances, respectively. 

Similar to the case for \mgb, Ca\,$4227$ is also much stronger in the
\alfa-enhanced models than in the scaled-solar models, as expected due to the
sensitivity of this index to variations of Ca. Notice that Ca\,$4227$ is also anti-correlated
with the CN feature that affects its the blue pseudocontinuum, hence  the Ca\,$4227$ 
index decreases in strength in the scaled-solar models. As with the \mgb\ index, the predictions
of the \alfa-enhanced models are similar to those of the base models at low metallicities.
This is as expected due to the high \aFe\ of the base models at low metallicities (reflecting
the intrinsic MILES abundance pattern). The Ca\,$4455$ index, however, does not depend on the 
Ca abundance, and behaves more like an Fe-index (e.g., \citealt{TripiccoBell95,Lee09}).

Finally, in Fig.~\ref{mgfemgfe} we make use of these new model predictions to
probe the sensitivity of the combined \mgfe\ and \mgfep\ indices to the \alfa-enhancement.
We confirm the results of \citet{Thomas03}. These authors employed
models that predict the strengths of a number of line indices with varying
abundances by implementing response functions derived from theoretical
spectra. We conclude that these two indices can be safely used to estimate
the total metallicity, \Mh, of stellar populations with non-solar \MgFe\
abundance ratios.

\subsection{Age-sensitive line-strength indices}
\label{sec:agelines}

\begin{figure*}
\resizebox{1.0\textwidth}{!}{\includegraphics[angle=-90]{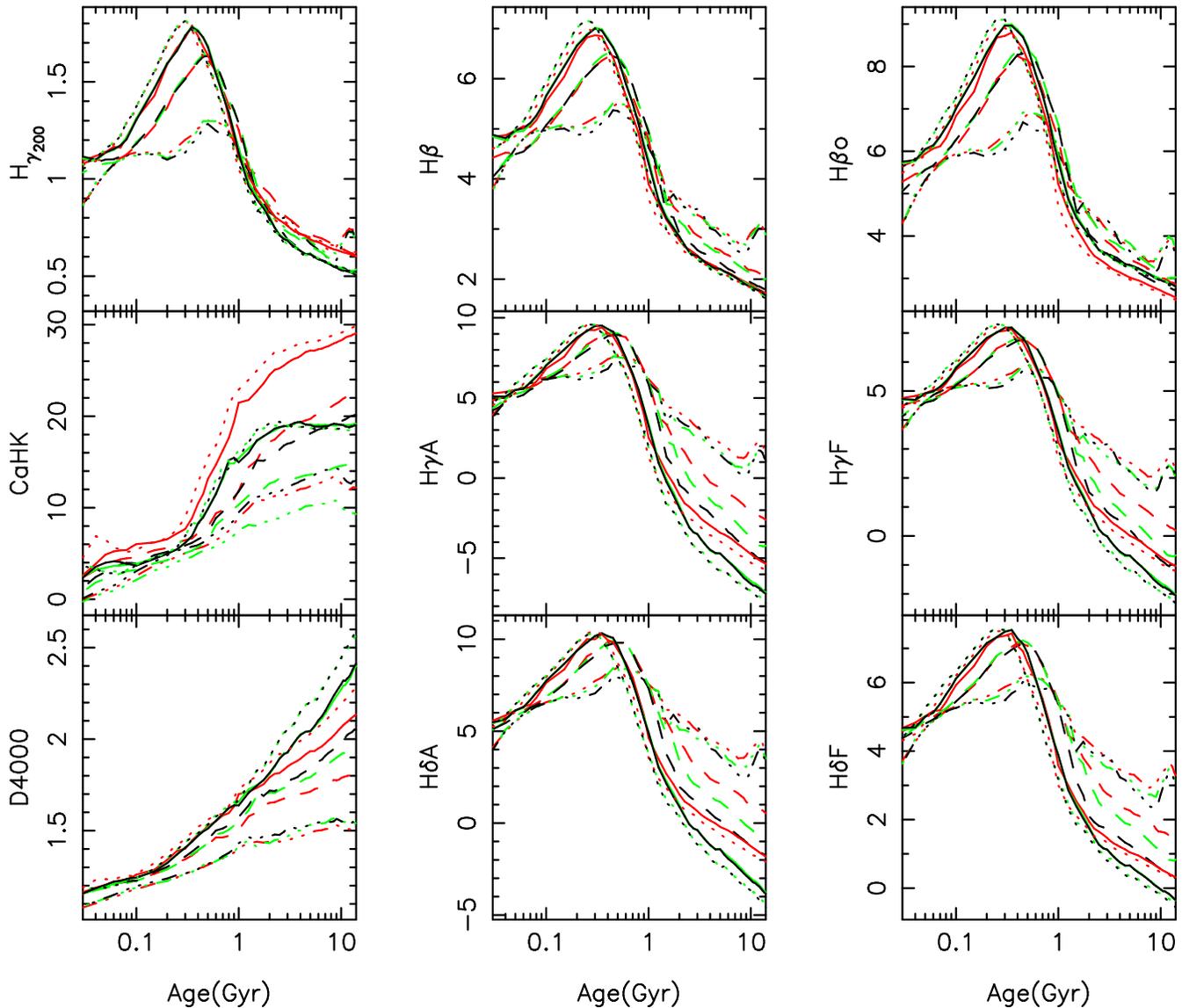}}
\caption{Predicted variation of various age-sensitive line-strength indices in
the LIS-8.4\,\AA\ system as a function of age for four different metallicities,
\Mh$=-1.26, -0.35, +0.06$, and $+0.26$ (dot-dashed, dashed, solid and dotted
line, respectively). Red line: \alfa-enhanced models, green: scaled-solar
models and black: base models. All the models have a Kroupa Universal IMF.}
\label{index_age:fig}
\end{figure*}

\begin{figure}
\resizebox{1.0\columnwidth}{!}{\includegraphics[angle=0]{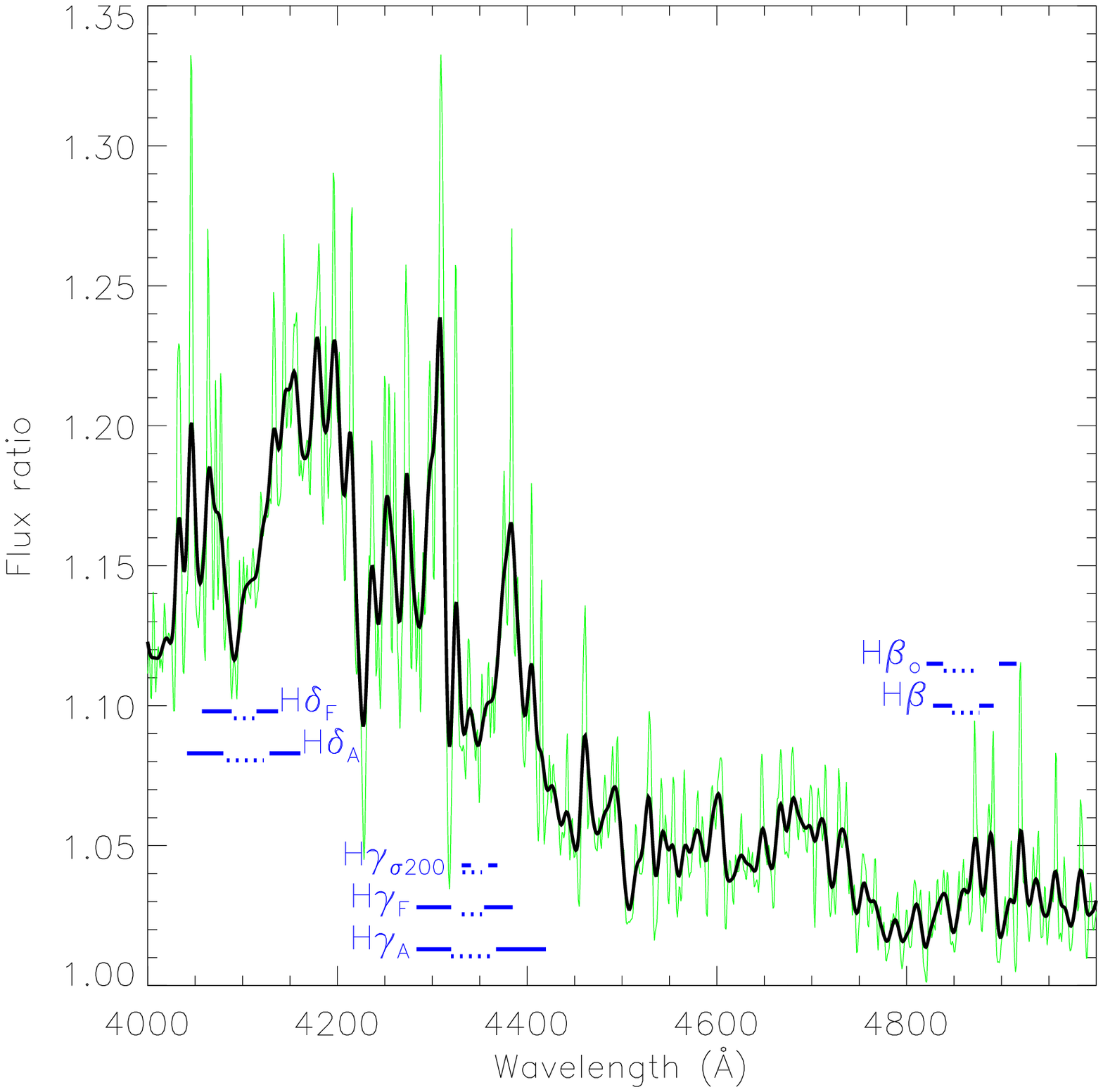}}
\caption{Flux-ratio obtained by dividing an \alfa-enhanced SSP model by a scaled-solar
SSP model, with age $10$\,Gyr, \Mh$=+0.06$ and Kroupa Universal IMF, in the spectral
region covered by H$\delta$, H$\gamma$ and H$\beta$. The flux-ratio is shown at
two resolutions FWHM$=2.51$\,\AA\ (i.e. the nominal resolution of the models) and
$8.4$\,\AA\ (i.e. the LIS-8.4\,\AA\ system), in thin green and thick black
lines, respectively. The pseudocontinua and feature bandpasses of various index
definitions for these features are shown in blue solid and dotted lines,
respectively.}
\label{fig:Balmeratio}
\end{figure}

Fig.~\ref{index_age:fig} shows the variation of different age-sensitive indices
with age predicted by our models. The sensitivity of the H$\beta$ index to
\alfa-enhancement is much less than that of the higher order Balmer lines (H$\delta$
and H$\gamma$). This is particularly true at high metallicities, where the 
base, scaled-solar and \alfa-enhanced models have similar values. 
The dependence of H$\beta_o$ on \aFe\ ($\sim0.2$\,\AA\ at solar metallicity and old ages) 
is higher than that of H$\beta$ (almost negligible), confirming the result of
\citet{CervantesVazdekis09}. Also the dependence of Ca{\sc II}\,H-K on \aFe\ is
very strong, especially for old ages. This dependence also increases with
metallicity. Note that Ca is another \alfa-element and, therefore, in the
adopted theoretical stellar spectra is also enhanced in the \alfa-enhanced
models. Similar behaviour is seen for the D\,$4000$ break. These variations
are expected given the residuals obtained in Fig.~\ref{fig:mhfehconstant}
(see also Section~\ref{sec:MILES_SSPs}). Note also the increase in the strength
of all the Balmer indices for the models with the oldest ages and the lowest
metallicity, i.e. $\Mh=-1.26$. This effect is due to the increased contribution
of the Horizontal Branch stars at such old ages and low metallicities (e.g.,
\citealt*{Lee00})  

Fig.~\ref{fig:Balmeratio} illustrates the effects of \alfa-enhancement on
various Balmer line index definitions. The figure shows the pseudocontinua and
feature bandpass definitions for the four indices of
\citet{WortheyOttaviani97}, (i.e. H$\delta_A$, H$\delta_F$, H$\gamma_A$,
H$\gamma_F$), one of the H$\gamma$ indices of \citet{VazdekisArimoto99}
(H$\gamma\sigma_{200}$, i.e. the one that is suitable for the resolution of the
LIS-8.4\,\AA\ system) and the standard and optimised index definitions for the
H$\beta$ feature of \citet{Trager98} (H$\beta$) and \citet{CervantesVazdekis09}
(H$\beta_o$), respectively.  

Fig.~\ref{fig:Balmeratio} confirms the claim of \citet{Thomas03} that the
higher order Balmer lines are more sensitive to the \alfa-enhancement than
H$\beta_o$. However we see that such an effect is mostly caused by the
sensitivity of the surrounding pseudocontinua of the H$\delta$ and H$\gamma$
lines to the abundance ratio, which is significantly lower for the H$\beta$
index (see also figures~17 and 18 in \citealt{Lee09}). Essentially, the narrower the index definition the smaller the sensitivity of the
index to the \alfa-enhancement. For example, H$\gamma_A$ has a larger
dependence on the \alfa-enhancement than H$\gamma_F$, while H$\gamma\sigma_{200}$ 
has the lowest sensitivity to this parameter as shown in Fig.~\ref{index_age:fig}.

Fig.~\ref{fig:Balmeratio} also shows two small, but significant, residuals
(between $\sim4850$\,\AA\ and $\sim4900$\,\AA) seen in emission in this
flux-ratio, affecting the H$\beta$ and H$\beta_o$ index definitions (see
 also the spectra of figures~12, 13 and 14 in \citealt{Lee09}). One of
these residuals (i.e., the bluest) affects the feature bandpass of the two
indices, whereas the second (the reddest) is placed within the red
pseudo-continuum corresponding to the H$\beta$ index. We see that the very small sensitivity
of the H$\beta$ index to the \aFe\ abundance ratio is due to the fact that the
reddest residual compensates in part the contribution to the feature bandpass from
the bluest residual. However, the red pseudo-continuum of the H$\beta_o$ index is
located in a valley within this flux-ratio, which has a similar flux level to
the blue pseudo-continuum. Therefore the H$\beta_o$ index includes the full
effect of the residual affecting the band-pass of this index. This residual
corresponds to Fe-peak elements, including Chromium, that affect the red wing of
the H$\beta$ absorption. 

The higher sensitivity of H$\beta_o$ to \aFe\ was
first shown in \citet{CervantesVazdekis09}, also employing the \citet{Coelho05}
model stellar spectra. However these authors also found the opposite behaviour, i.e.
a larger sensitivity of the H$\beta$ index to \aFe\ than H$\beta_o$, when using
the~\citet{Munari05} stellar spectra to obtain the differential correction.
This disagreement possibly originates in differences in the line lists
adopted for computing these two sets of theoretical stellar spectra.

\subsection{IMF-sensitive line-strength indices}
\label{sec:gravitylines}

\begin{figure}
\resizebox{1.0\columnwidth}{!}{\includegraphics[angle=-90]{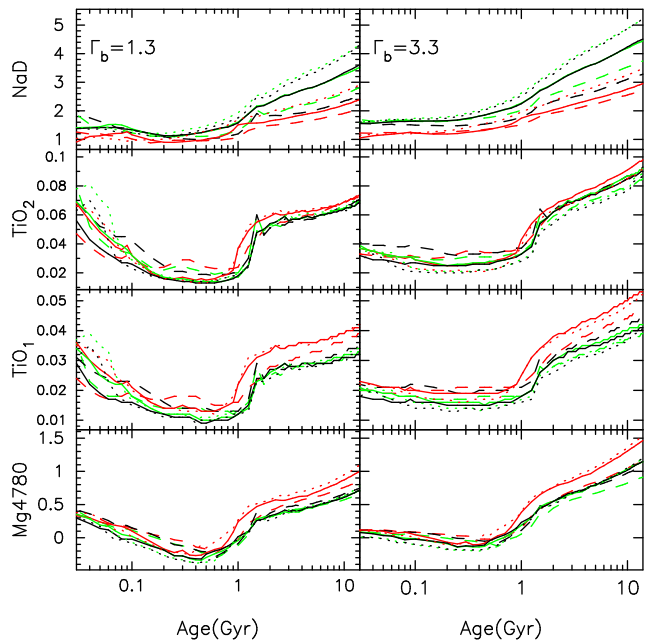}}
\caption{Variation of some IMF-sensitive indices with age of SSPs with a bimodal 
IMF with slope $\Gamma_b=1.3$ (left panels) and $\Gamma_b=3.3$ (right panels)
for the three different models presented here; \alfa-enhanced (red),
scaled-solar (green) and base (black). The index measurements are shown for
three different metallicities with different line-styles: \Mh$=-0.35$ (dashed),
\Mh$=+0.06$ (solid) and \Mh$=+0.26$ (dotted). All the index measurements are in the LIS-8.4\,\AA\ system} 
\label{fig:imf_sensitive}
\end{figure}

Fig.~\ref{fig:imf_sensitive} shows the variation of four gravity-sensitive
indices \citep{LaBarbera13} with age for the three models presented here (i.e.
base, scaled-solar and \alfa-enhanced models). The figure shows the index measurements
for SSPs with a bimodal IMF with slope $\Gamma_b=1.3$, similar to Kroupa
Universal, and $\Gamma_b=3.3$, i.e. a bottom-heavy IMF. These values cover the range
of bimodal IMF slopes required to match SDSS ETGs with velocity dispersions in
the range $100$ -- $300$\,\kms\ \citep{LaBarbera13}. The index measurements are
shown for three different metallicities. The two TiO indices show the highest
sensitivities to the IMF, whereas the NaD index shows the lowest sensitivity
to this parameter. However, the NaD index does show the largest
sensitivity to \aFe, whereas the TiO$_2$ index is almost insensitive to this abundance
ratio. These results are in agreement with recent works that employ either a
different set of model stellar spectra than the ones used here, e.g.,
\citet{Spiniello14}, or fully empirical evidence based on stacked galaxy
spectra from the SDSS \citep{LaBarbera13}.

The highest sensitivity to the IMF is seen for the models with the oldest ages.
For models with $\Gamma_b=1.3$, we find a plateau, with only a modest increase
with age, for ages between $2$ and $10$\,Gyr for the TiO$_1$ and TiO$_2$
indices and, to a lesser degree, for Mg\,4780. For this age regime these indices
mostly depend on the IMF, whereas for older populations there is also a
dependence on age (see \citealt{LaBarbera13}). Interestingly, the
\alfa-enhanced models show a slightly larger dependence on metallicity in
comparison to the scaled-solar and base models. We also find that the TiO$_2$
index, and the other IMF-sensitive indices, which show little sensitivity to
the age for the standard IMF slope (i.e $1.3$), become more sensitive to this
parameter for larger IMF slopes, such as those shown in the right panels of
Fig.~\ref{fig:imf_sensitive}.

Another interesting aspect is the high values of the TiO$_1$ and TiO$_2$
indices for ages below $0.1$\,Gyr for the models with $\Gamma_b=1.3$, index values 
which are similar to those obtained for ages around $\sim$8\,Gyr. This behaviour is
attributed to the influence of supergiant stars at these young ages. Such an
increase is likely to continue to even younger ages where the contribution of
this evolutionary phase peaks at around $10$\,Myr. Similar behaviour is seen
for other IMF-sensitive features such as the Ca{\sc II} triplet around
$\sim8500$\,\AA\ and the CO band-head at $2.3$\,$\mu$ (e.g. \citealt{Mayya97}).
It is not surprising to see that there is no such increase at these young
ages for the bottom-heavy IMF models. The steeper the IMF, the larger the
fraction of mass locked in low-mass stars. In principle, it could be appealing
to consider that the high TiO values observed in massive ETGs, which suggest a
bottom-heavy IMF (e.g. \citealt{LaBarbera13}), may result in part by
a contribution from such young stellar populations.
However such a contribution would have a significant impact on the
measurements of other line indices in the optical range which is not seen.   

\section{Colours}
\label{sec:SSPcolours}

\begin{figure*}
\includegraphics[width=0.48\textwidth]{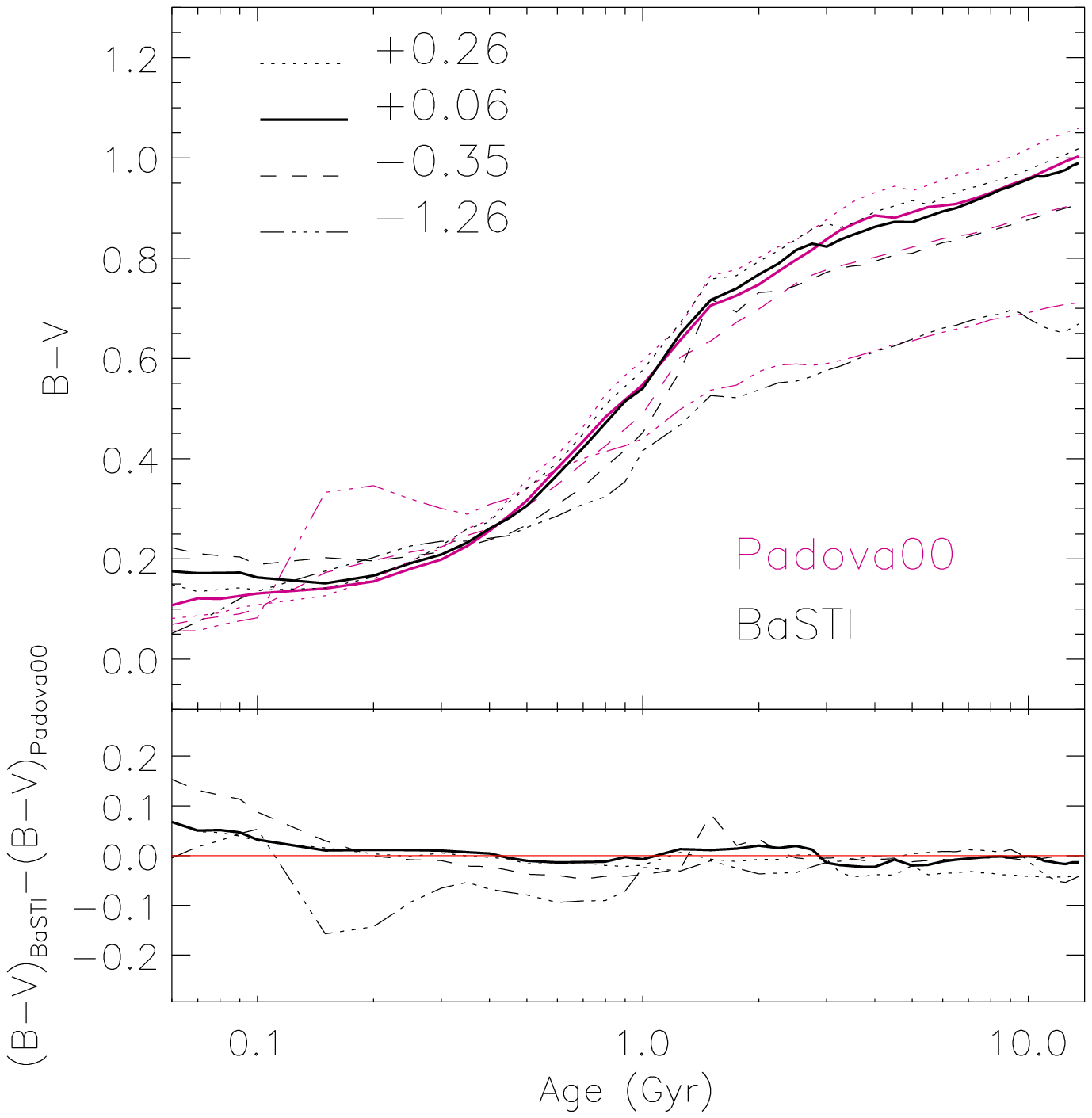}
\includegraphics[width=0.48\textwidth]{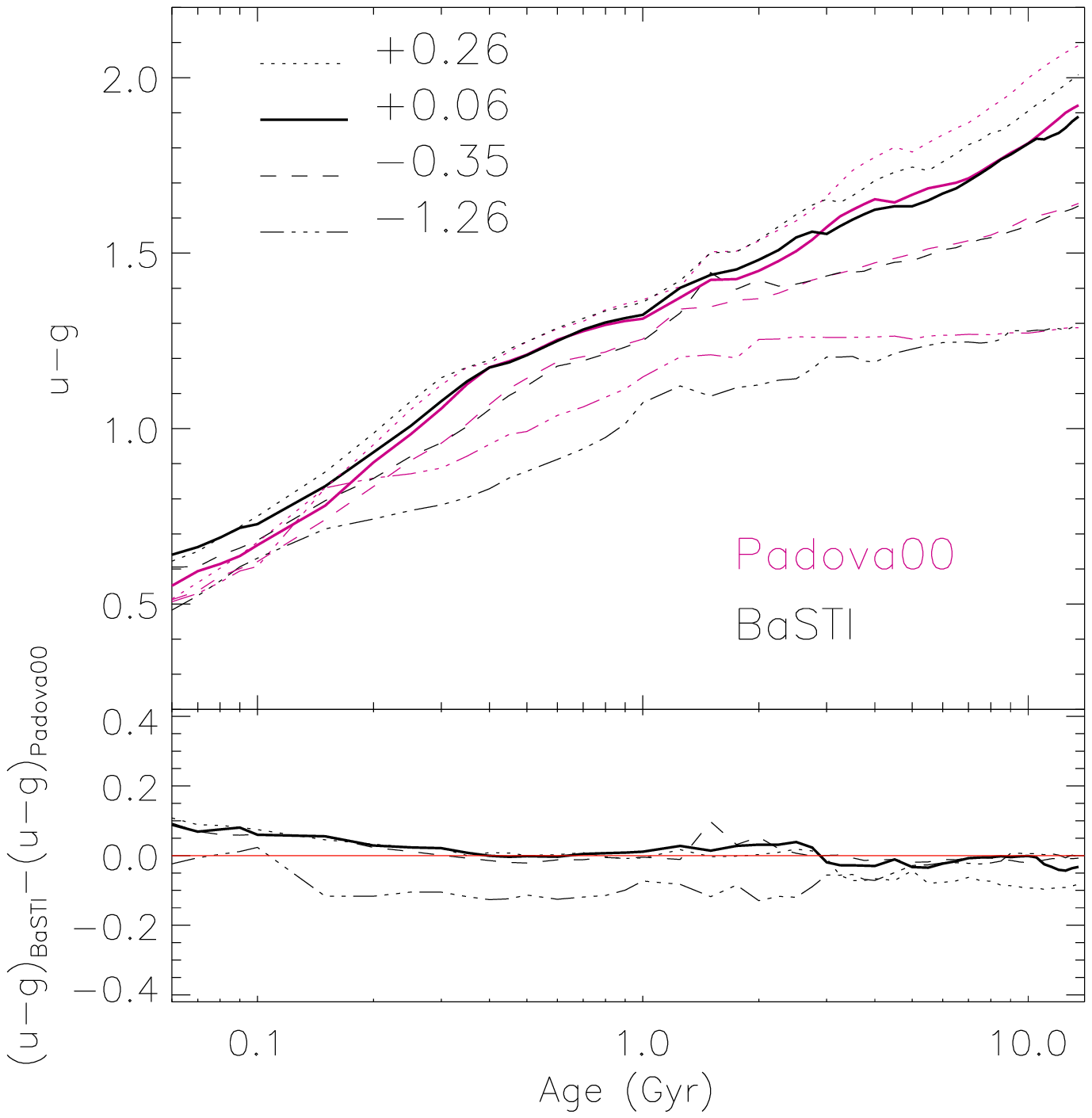}\\
\includegraphics[width=0.48\textwidth]{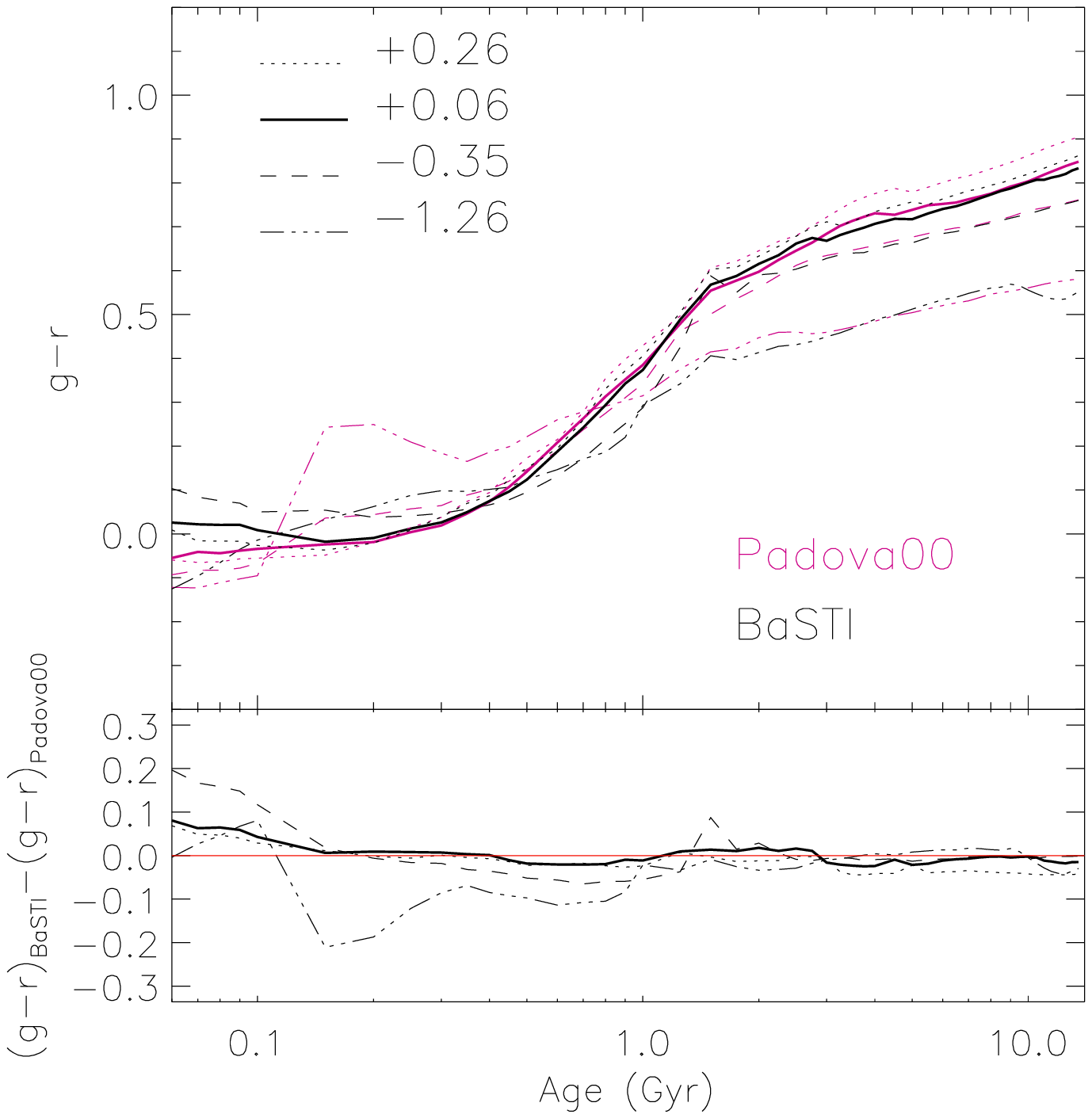}
\includegraphics[width=0.48\textwidth]{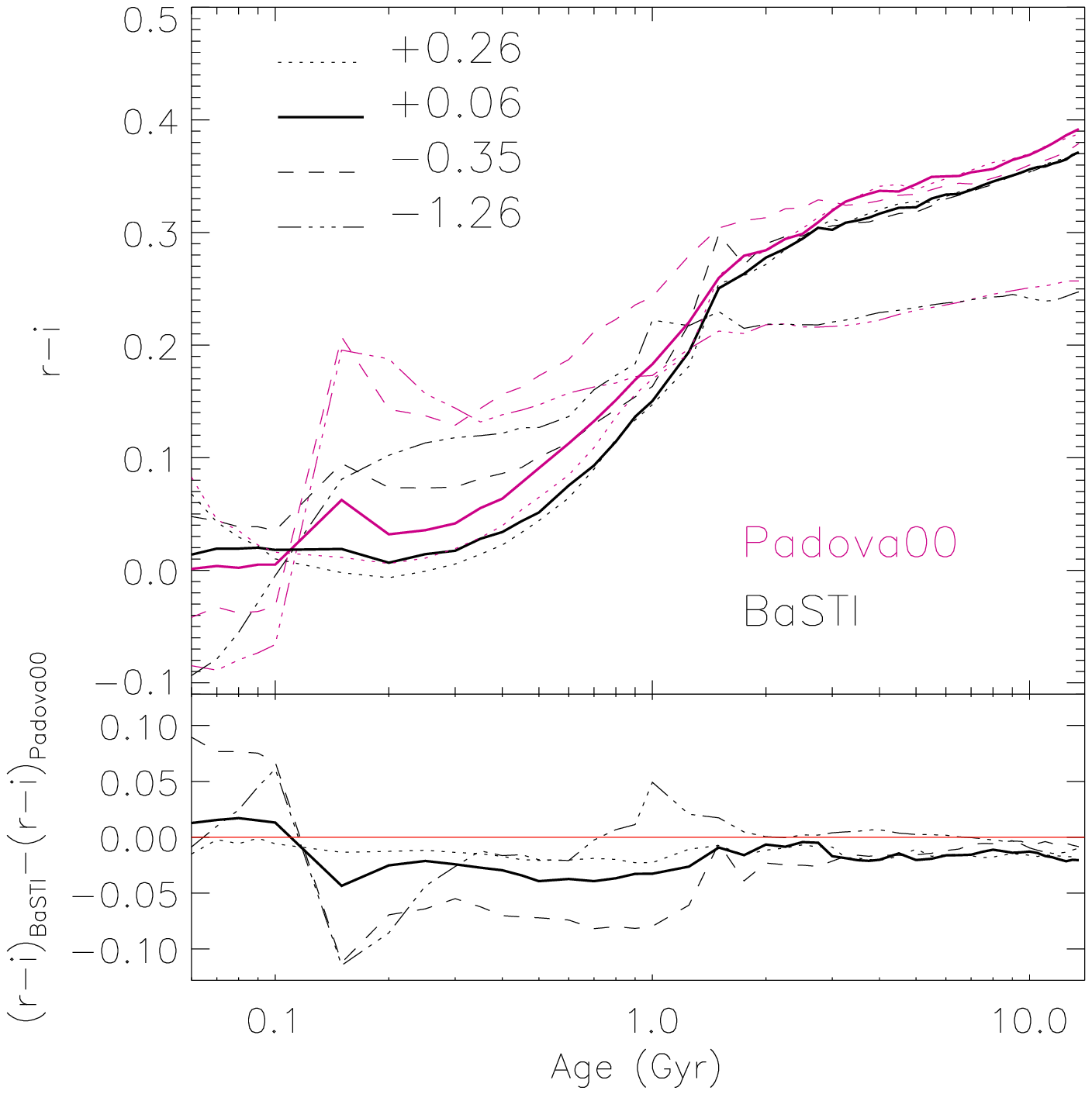}
\caption{Dependence of broad-band colours on the isochrones adopted. Colours
 are displayed as a function of age for four different metallicities:
 \Mh$=-1.26$ (dot-dashed line), $-0.35$ (dashed line), $+0.06$ (solid line) and
 $+0.26$ (dotted line). The magenta lines indicate the models computed with the
 Padova00 isochrones, whereas the black lines show the models based on BaSTI. 
 The B$-$V colour (upper-left panel) is computed in the Vega magnitude system,
 whereas the SDSS colours u$-$g (upper-right), g$-$r (lower-left) and r$-$i
 (lower-right)  are computed in the AB system.  The residual colours are also
 shown in each panel. All the models have a Kroupa Universal IMF }
\label{col_iso} 
\end{figure*}

\begin{figure*}
\includegraphics[width=0.49\textwidth,height=0.31\textheight]{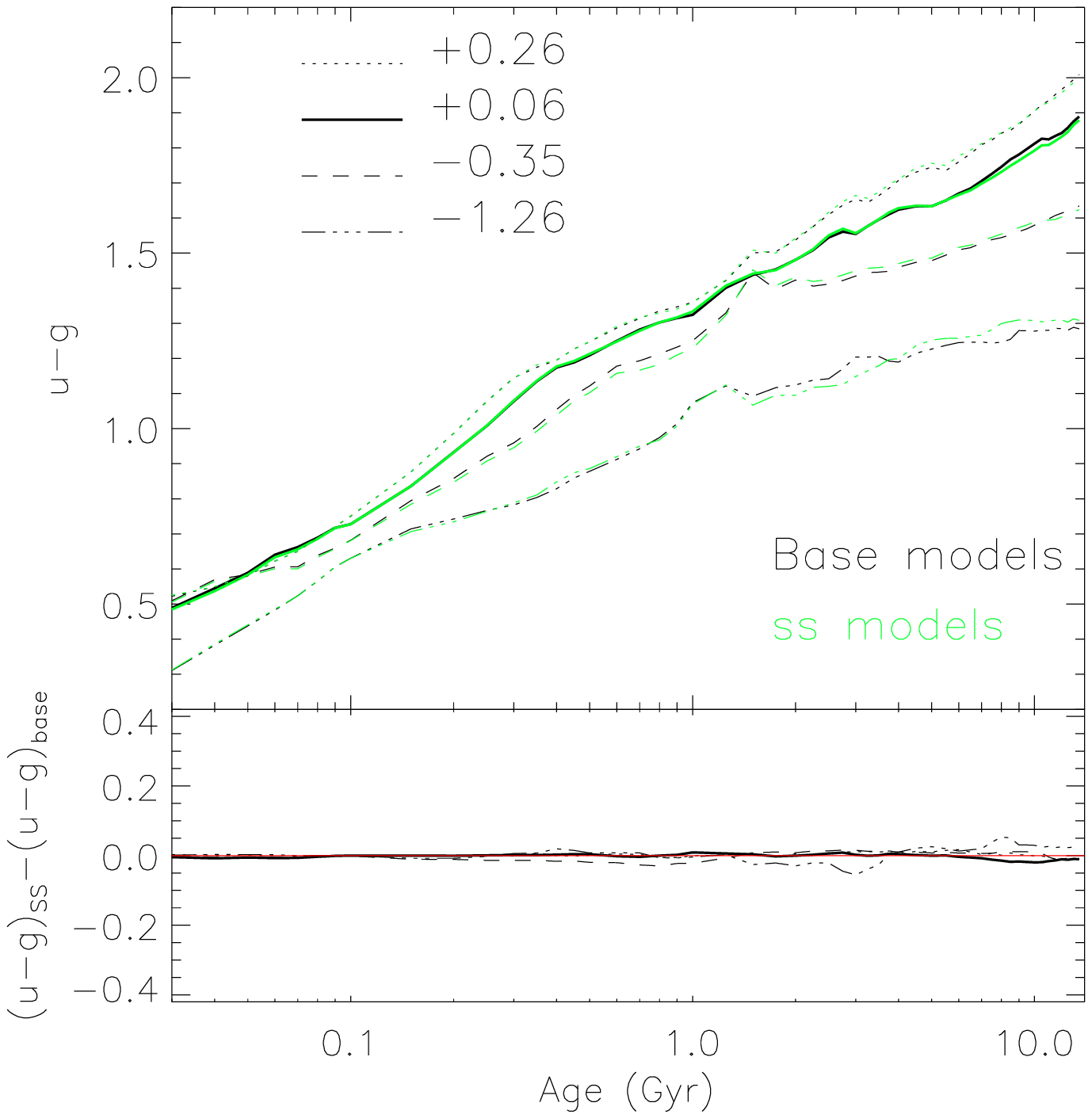}
\includegraphics[width=0.49\textwidth,height=0.31\textheight]{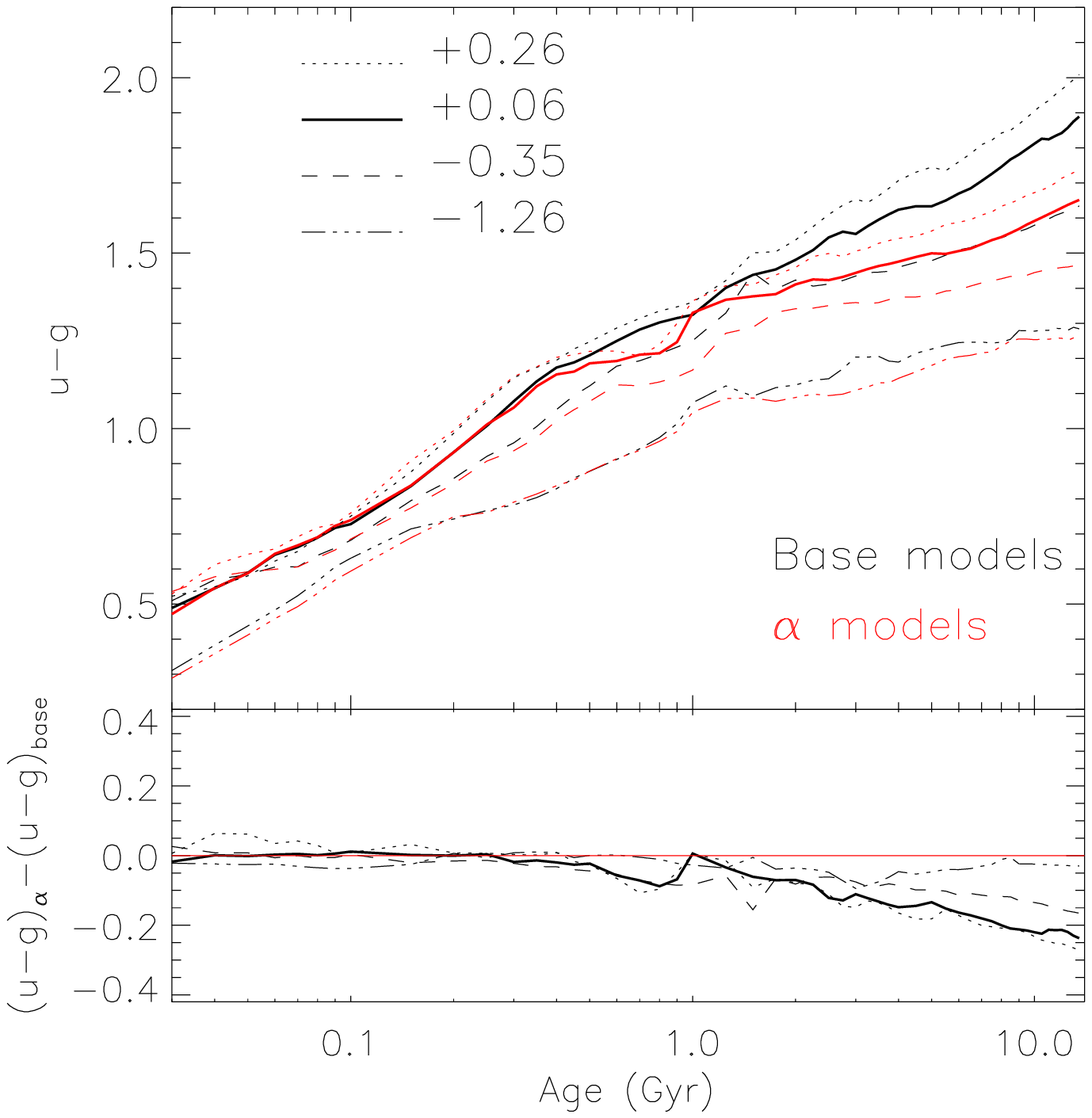}\\
\includegraphics[width=0.49\textwidth,height=0.31\textheight]{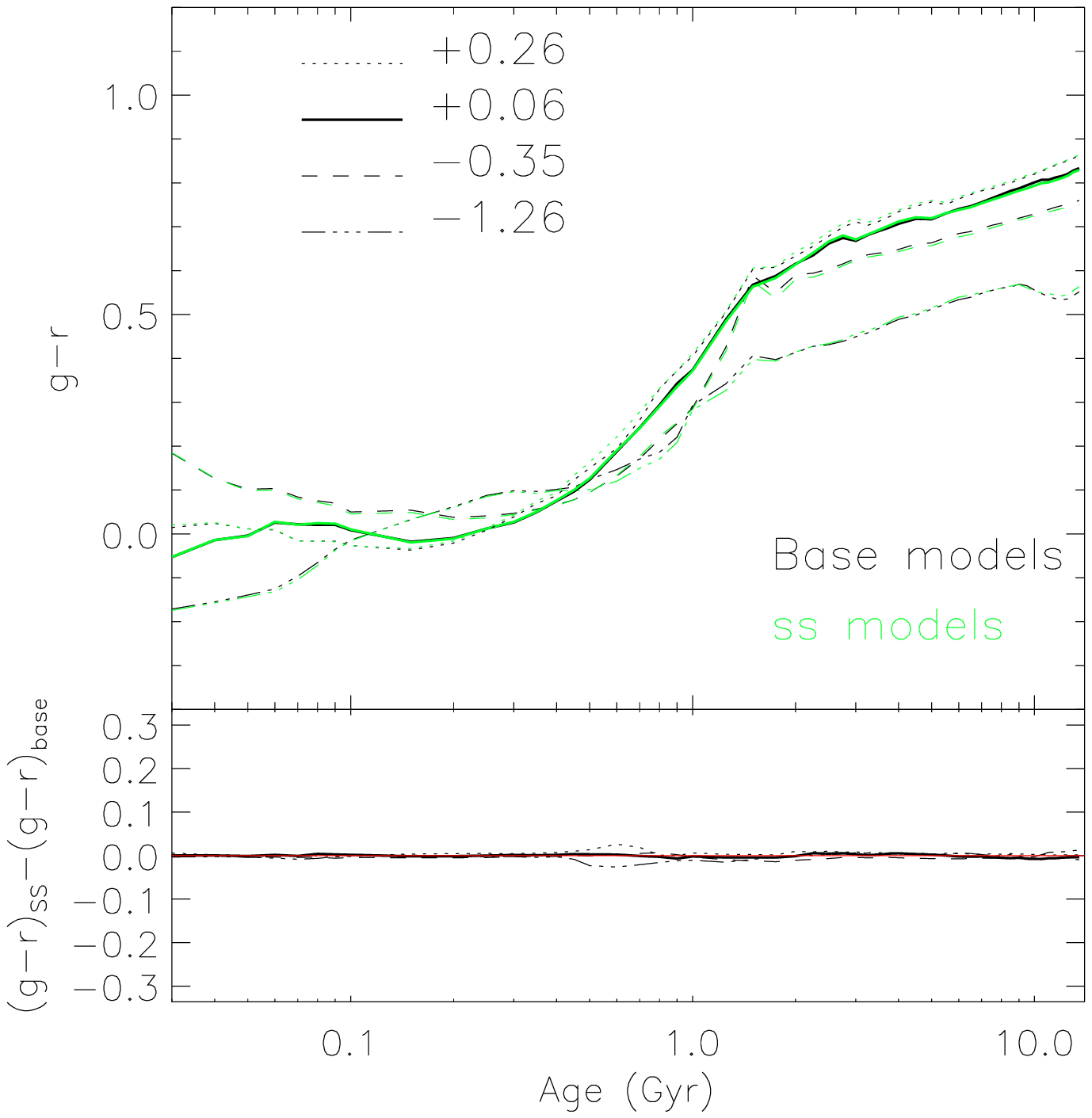}
\includegraphics[width=0.49\textwidth,height=0.31\textheight]{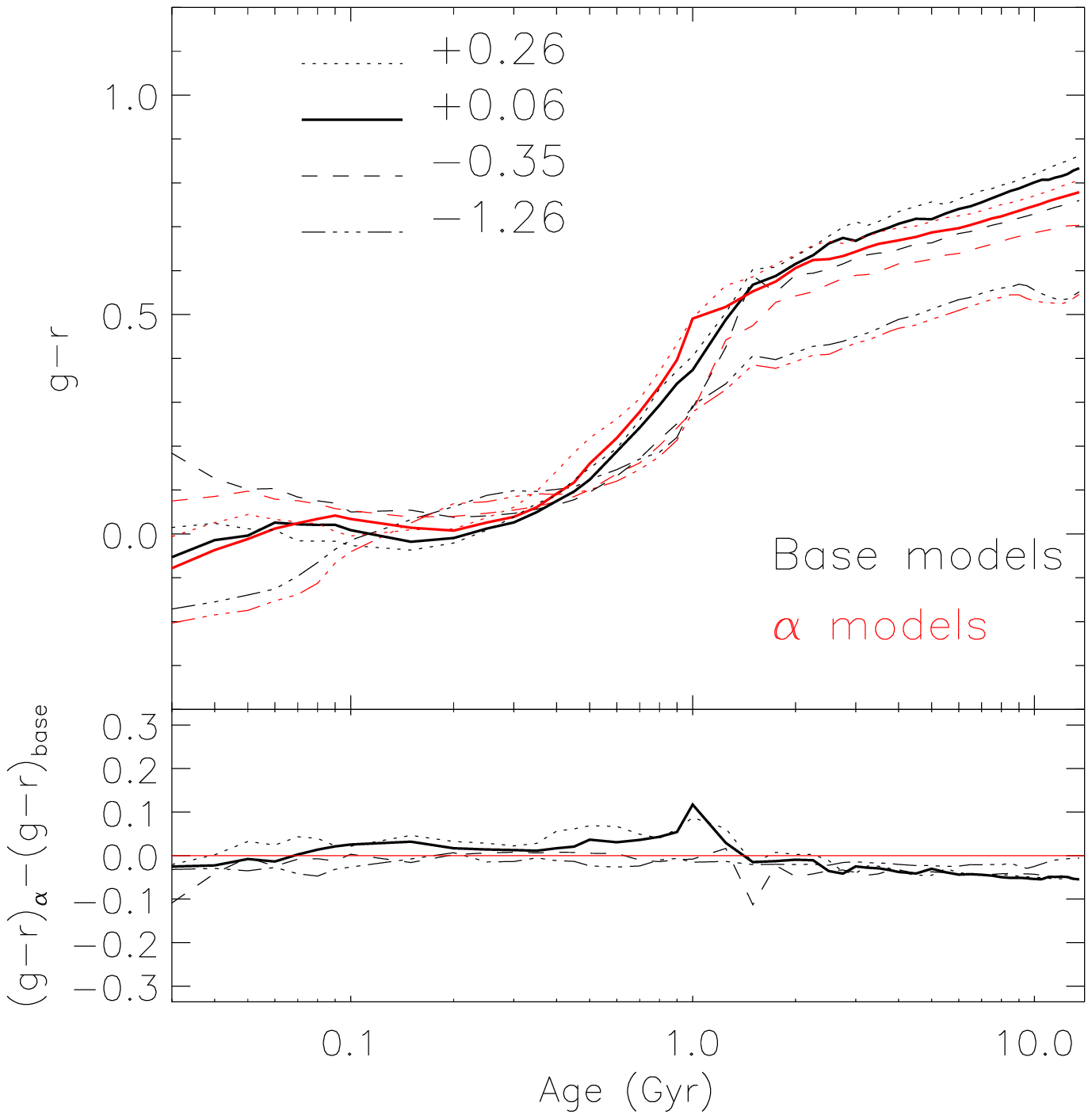}\\
\includegraphics[width=0.49\textwidth,height=0.31\textheight]{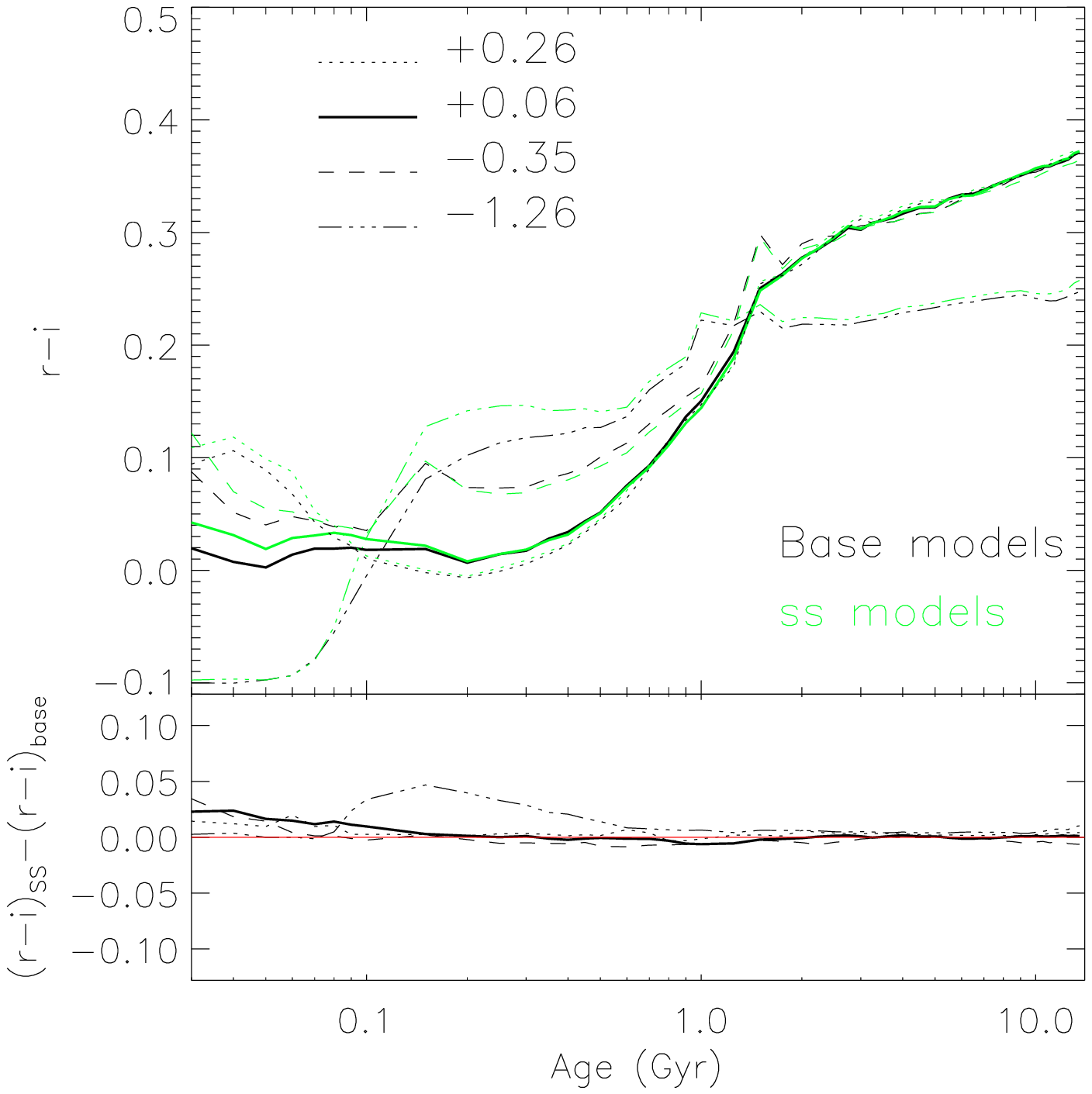}
\includegraphics[width=0.49\textwidth,height=0.31\textheight]{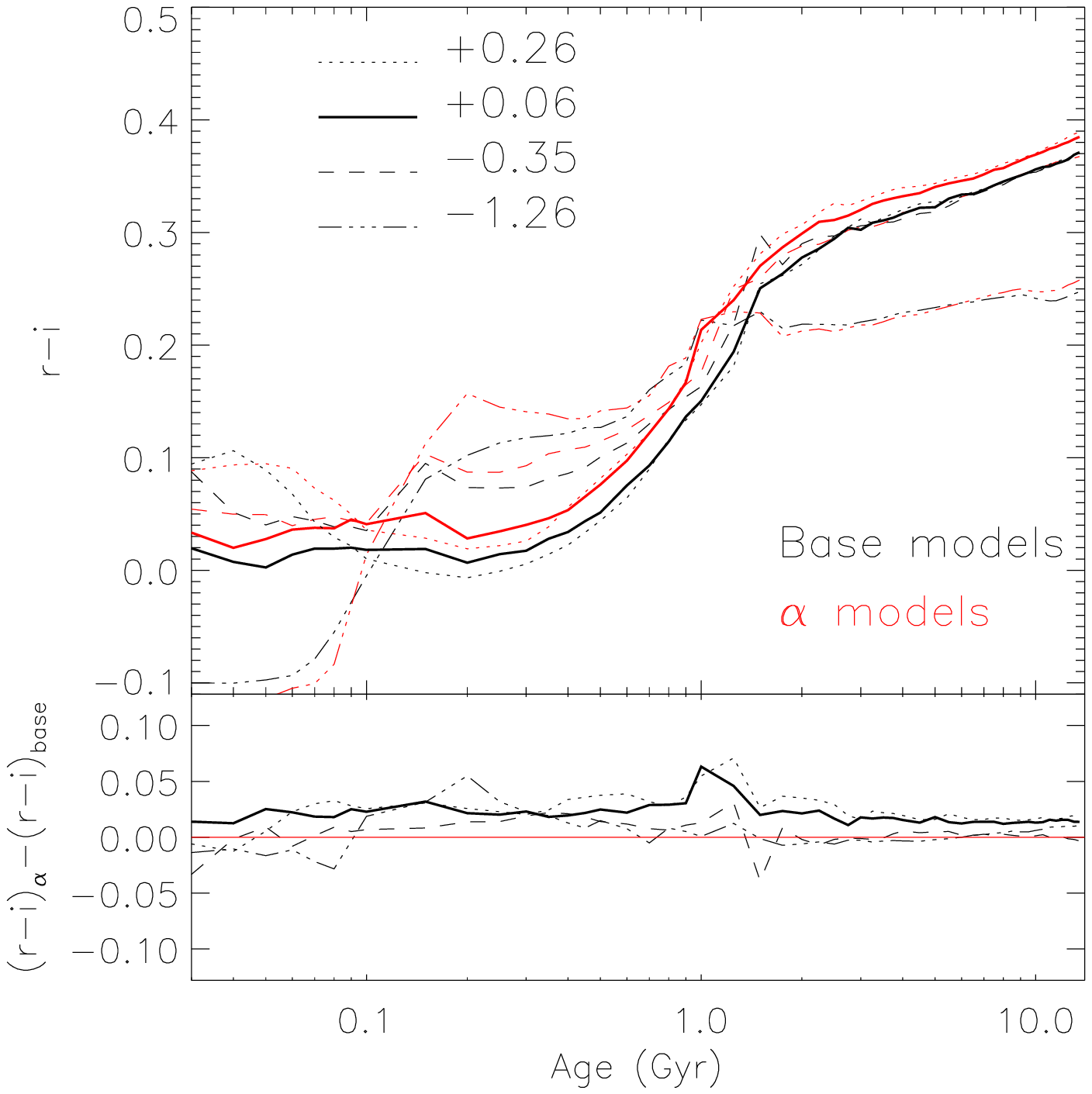}\\
\caption{Left-hand panels: comparison between colours in the SDSS bands of
 scaled-solar models (green lines) and the base models (black lines). Right-hand
 panels: comparison between  \alfa-enhanced models (red lines) and base models
 (black lines). The colour variation is shown as a function of age for four
 different metallicities: \Mh=$-1.26$ (dot-dashed line), \Mh=$-0.35$ (dashed
 line), \Mh$=+0.06$ (solid line) and \Mh$=+0.26$ (dotted line). The residual
 colors are also shown in each panel.}
\label{col_alpha} 
\end{figure*} 

In this Section we focus on the photometric predictions from our new set of 
models. To derive the synthetic colours, we adopt Vega magnitudes for the  
Johnson bands, B and V (\citealt{BuserKurucz78}; see also
\citealt{FalconBarroso11} for further details on our definition of Vega
magnitudes) and the AB system for the SDSS filters: {\it u}, {\it g}, {\it r}
and {\it i}.

\subsection{Synthetic {\it u} and {\it i} magnitudes}
\label{sec:ui}

The MILES spectral range ($3540.5$ $-$ $7409.6$\,\AA) does not fully cover
the wavelength range spanned by the {\it u} and the {\it i} SDSS filters, as the
{\it u} pass-band starts at $\sim3005$\,\AA\ and the {\it i} band extends up to
$8605$\,\AA. Therefore, in order to have reliable synthetic magnitudes in these
bands, we need to correct for the missing flux blue-ward and red-ward of the MILES
spectral range. We do this by using a similar methodology to that adopted in
\citet{MIUSCATII}, which consists of computing the magnitude in the part of the 
filter covered by the model and then correcting it for the missing flux. 

To obtain the missing flux blue-ward of the starting point of MILES, i.e. between
$3005$\,\AA\ and $3540.5$\,\AA, we use new model SEDs that employ the NGSL
stellar library, as prepared in \citet{NGSLI} and \citet{NGSLII}. These models, 
which are computed with the same code used here, extend the spectral range down
to $\sim1680$\,\AA. Thus, for a given MILES SSP, the {\it u} magnitude is
obtained from the flux through the {\it u} filter red-ward $3540.5$\,\AA\, and
then correcting this value by the missing flux obtained from the corresponding NGSL-based SSP.
Note that the missing flux correction in the blue is $\lesssim25$\,\% for
most of the SSPs covered here. Strictly speaking, this flux correction is only
valid for the base models, which are of the same type as these newly synthesised
NGSL-based SSP spectra. However the difference in flux due to abundance variations is
far smaller than our correction for the missing flux. 

The same approach is used for calculating the missing flux in the {\it i} 
filter, but this time using our published models extension, MIUSCAT 
\citep{MIUSCATI}. For the MIUSCAT models we employ the Indo-US stellar library
\citet{Valdes04} to fill-in the gap left in between the MILES range (Paper~I)
and Ca{\sc II} triplet range \citep{CATIV}. Since the effects of the
\alfa-enhancement on the continuum red-ward the V band is small (as shown by the
flux-ratios of Fig.~\ref{fig:ageratio} and Fig.~\ref{fig:metratio}), the
extension based on MIUSCAT is appropriate for this specific purpose.

\subsection{Base models with varying isochrones}
\label{sec:Basecolours}

Fig.~\ref{col_iso} shows the comparison between the broad-band colours based
on Padova00 (magenta lines) and BaSTI (black lines) isochrones for the base
models. Metal-poor models based on BaSTI tend to have bluer colours in all
bands with respect to the Padova00 models, by up to $0.2$\,mag for the lowest
metallicity considered here (\Mh$=-1.26$). For solar and supersolar metallicities,
a significant offset between the two sets of models is seen only in the r$-$i
colour, where the BaSTI models are bluer  by $\sim0.02$\,mag  for old
populations and by $\sim0.04$\,mag for young populations. Overall the agreement
is good for the old stellar populations, whereas for younger ages, these colour
differences reflect the differing prescriptions for AGB stars.

\subsection{scaled-solar versus \alfa-enhanced SSP models}
\label{sec:ssaacolours}

The comparison between base, scaled-solar and \alfa-enhanced models is shown in
Fig.~\ref{col_alpha} for the SDSS colours. No significant differences are
observed between the base models and the scaled-solar ones, with the exception
of the u$-$g colour for the most metal-poor SSPs. Such difference must be
attributed to the fact that at such low metallicity the MILES stars are
\alfa-enhanced. Conversely, \alfa-enhancement has a remarkable impact on the
resulting broad-band colours, having a differential effect on the different
spectral ranges (see also \citealt{Coelho07}). The \alfa-enhanced models are
bluer in the blue (by $\sim0.2$\,mag in u$-$g and $\sim0.05$\,mag in g$-$r for
old populations) and redder in the red (by $\sim0.02$\,mag in r$-$i) with
respect to the base models, as one can also infer from the flux ratio of the two
SSP spectra (Fig.~\ref{fig:metratio}). Note that the best agreement between
the base and \alfa-enhanced models is reached at low metallicities.

\section{Comparison with other models}
\label{sec:othermodels} 

\begin{figure}
\includegraphics[angle=0,width=1.0\columnwidth]{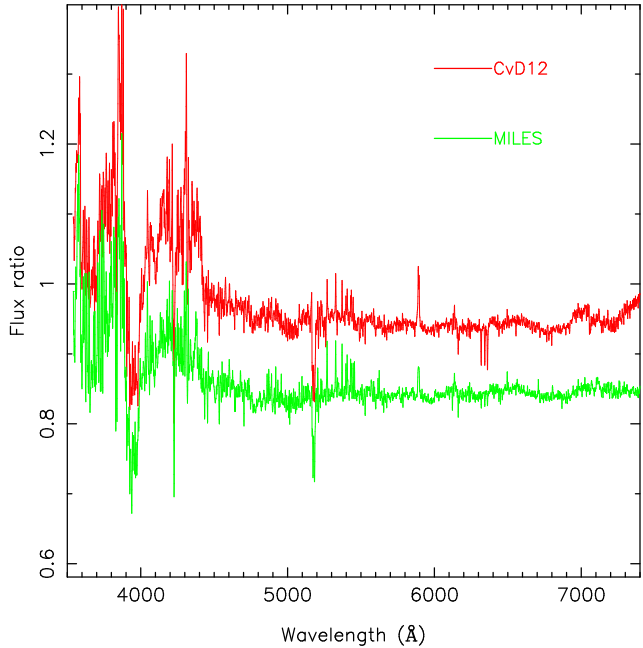}
\caption{
Ratio of \alfa-enhanced to scaled-solar models, at fixed  $\Feh=0.0$, for 
our models (green) and the \citet{CvD12models} (red) SSP models with an age of $13.5$\,Gyr and 
a Salpeter IMF. The green curve has been arbitrarily shifted downwards by $0.1$
in order to allow a better visual comparison with the red curve.
}
\label{comparemodelspectra}
\end{figure}

\begin{figure}
\includegraphics[angle=0,width=1.0\columnwidth]{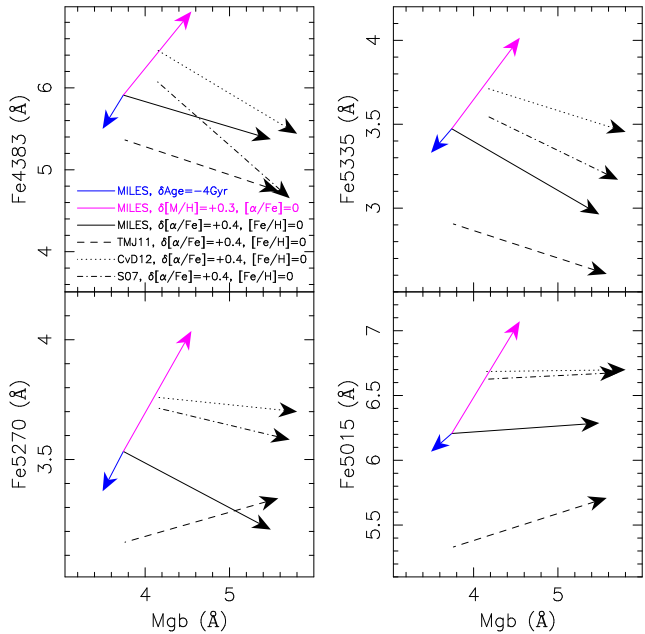}
\caption{
Line-strengths for different Fe lines ($\rm Fe\,4383$, $\rm Fe\,5270$, $\rm
Fe\,5335$, and $\rm Fe\,5015$; from top to bottom, and left to right) are
plotted as a function of \mgb\ ($\lambda \sim 5177$\,\AA). Black arrows, plotted
with different line types, show the effect of changing  \aFe, at fixed $\Feh=0.0$,
for SSP models with the same age of $13.5$\,Gyr and the same (Salpeter) IMF,  from
different authors, i.e. this work (solid), \citet[dashed]{Thomas11},
\citet[dotted]{CvD12models}, and~\citet[dot-dashed]{Schiavon07}, as labelled in
the top--left panel. The blue and magenta arrows show, as reference, the effect
of decreasing age by $4$\,Gyr and increasing total metallicity by $+0.3$~dex
(for $\aFe=0.0$), respectively, in our SSP models.
}
\label{comparemodellinesmgfe}
\end{figure}

We compare the effect of varying \aFe\ in our models with predictions from 
other synthetic stellar population models in the literature. We focus on some
widely employed models that are based on empirical stellar spectral libraries,
i.e. \citet{CvD12models} [CvD12], \citet*{Thomas11} [TMJ11] and
\citet{Schiavon07} [S07]. Note that while CvD12 provide full SEDs at varying
abundance ratios, TMJ11 and S07 provide line-strengths of Lick spectral
features. We note however that whereas the predictions of TMJ11 are given at
the nominal resolution of MILES, the models of S07 are provided in the Lick/IDS
system \citep{WortheyOttaviani97}.

Fig.~\ref{comparemodelspectra} compares the ratio of \alfa-enhanced
($\aFe=+0.4$) to scaled-solar SSPs for our models and those of CvD12, at a fixed age of
$13.5$\,Gyr, and for the same (Salpeter) IMF. The CvD12 models are computed at
fixed  \Feh. Therefore, in order to perform a meaningful comparison, we have computed 
our \alfa-enhanced SSP for a total metallicity of $\Mh=+0.3$ (i.e.
$\Feh\sim0.0$, see Eq.~\ref{eq:MhFehrelation}), and taken its ratio with 
the scaled-solar model. Note also that both our models and the CvD12 models 
have the same spectral resolution (i.e. that  of MILES) in the spectral range 
considered here (from $\sim3500$ to $7400$\,\AA). 
The overall effect  of increasing \alfa-enhancement is similar
in the CvD12 models and our models, reflected by the fact that the flux ratios in the
Figure have a similar large-scale behavior, and similar ``absorption'' and 
``emission'' residual features (e.g., for Mg and Fe lines, respectively). While
the average flux ratio is about one for CvD12 models, it is less than one for
our models. This occurs because we adopt \alfa-enhanced isochrones for our \alfa-enhanced 
model predictions, whereas CvD12 always use scaled-solar isochrones. 
Note however that the impact of the isochrones on the
line indices is significantly less than the impact of the stellar spectra on the indices; see
Fig.~\ref{fig:isocspectra}. 

It is evident that the ratios of different features in the two curves
in Fig.~\ref{comparemodelspectra} are not precisely the same, indicating that the 
two sets of models differ to some extent in the relative effect of \aFe\ on different 
spectral features.  This is not surprising
since the theoretical stellar spectra employed by CvD12  differ from 
those used in our models (see Section~\ref{sec:coelho}). In particular, 
CvD12 obtained their differential corrections using theoretical stars computed
with the ATLAS12 code for model atmospheres and the SYNTHE spectrum synthesis
package (\citealt{Kurucz70,Kurucz93}; ported to  Linux by \citealt{Sbordone04})
and line list provided by Kurucz.  Another relevant difference is that
our differential corrections are obtained from SSP spectra computed with 
theoretical stellar spectra (see Section~\ref{sec:aassSSPs}), whereas the differential
corrections of CvD12 were obtained from individual stars.
Finally, as mentioned above, CvD12 employ scaled-solar isochrones for both their scaled-solar and 
\alfa-enhanced predictions, whereas we employ \alfa-enhanced isochrones for our \alfa-enhanced 
model predictions. However, we note that the choice of isochrones has a small 
impact on the index predictions.

\begin{figure*}
\includegraphics[angle=0,width=1.0\textwidth]{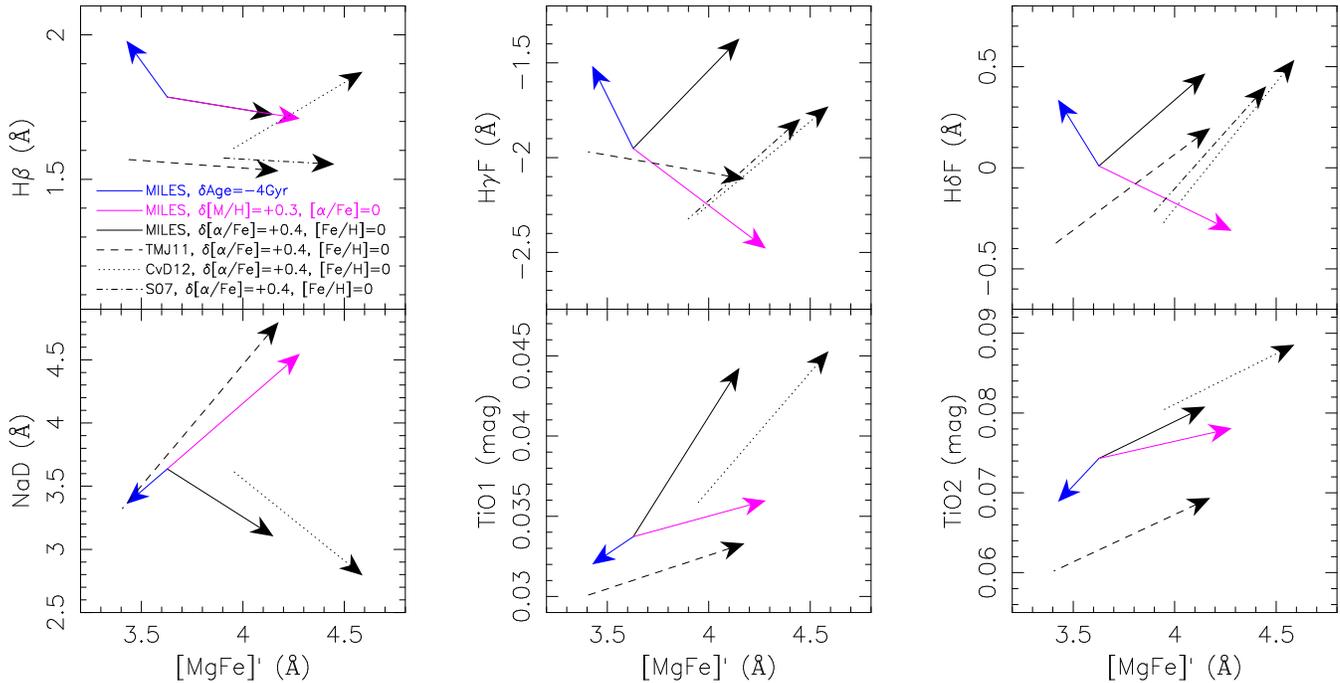}
\caption{
Same as Fig.~\ref{comparemodellinesmgfe} but plotting the line-strengths of
representative Balmer lines (top) and IMF-sensitive features (bottom) as a
function of the total metallicity indicator \mgfe'. Note that the IMF-sensitive
line-strengths are not provided by the \citet{Schiavon07} models.
}
\label{comparemodellines}
\end{figure*}

Some of the model differences seen in Fig.~\ref{comparemodelspectra} are quantified
in Fig.~\ref{comparemodellinesmgfe}, where we compare the effect of
\aFe\ on the Mg and Fe lines for our models and also for those of 
S07, TMJ11 and CvD12. Similarly, Fig.~\ref{comparemodellines}
shows the behaviour of the line-strengths of some representative Balmer lines (upper panels)
and IMF-sensitive (lower panels) indices as a function of the total metallicity
indicator \mgfe', for the different models. 
Each arrow in a given panel represents an individual model, with the black arrows
connecting  scaled-solar to $\aFe=+0.4$ SSP model line-strengths, at fixed \Feh.
Since our models and those of TMJ11 are computed at fixed total metallicity, the
comparison at fixed $\rm [Fe/H]$ is performed by increasing both \Mh\ and \aFe\
in these models, using Eq.~\ref{eq:MhFehrelation}, with $\rm A=0.75$ and $\rm
A=0.94$ (see~\citealt{Thomas03}) for models and the TMJ11 models, respectively. 
Our models, the TMJ11 models, and the CvD12 models all have the same native 
(MILES) spectral resolution, while the S07 line-strengths are given at the 
Lick/IDS resolution (see Table~7 of S07)\footnote{But note that the stellar library
used in the S07 models is of higher resolution than the Lick/IDS \citep{Jones99}}. 
For a given age, metallicity, and \aFe, we have determined the behaviour of the 
S07 line-strengths at the MILES resolution, by smoothing our models to match the
Lick/IDS resolution for each spectral feature, and then estimating the corresponding
variation of its line-strength at this resolution. 

It is evident in Fig.~\ref{comparemodellinesmgfe} that the arrows
corresponding to the various models do not have the same starting point in 
the index-index diagrams. These significant offsets among the
predictions of the various models at scaled-solar abundance are, however,
expected. These offsets reflect different model ingredients (e.g. the
isochrones) and different values for the solar metallicity (see
Section~\ref{sec:ingredients}) among other aspects of the model construction.
For instance, our (Mg, Fe, and Balmer lines) model predictions for an SSP with
$\Mh=0.0$, age of $13.5$\,Gyr, and  \aFe$=0.0$,  would be matched by the CvD12
models with a $\sim 3.5$\,Gyr younger age, and slightly lower ($\sim0.05$~dex)
total metallicity. Such differences strongly suggest that relative ages
are more robust than absolute ages in the models (e.g., \citealt{Gibson99,Vazdekis0147Tuc,Schiavon02},
\citealt*{Mendel07}). Note also that such differences decreases with decreasing age. 
However, a detailed discussion on the origin of these offsets between the models 
is beyond the scope of this paper. In the following, we 
focus on the relative responses of the various models to variations of
the \aFe\ abundance ratios (i.e., on the arrow directions, not their absolute 
values).

Fig.~\ref{comparemodellinesmgfe} shows that the \mgb\ index increases with 
\aFe, reflecting its strong sensitivity to \MgFe. The degree of increase is
larger, by $\sim0.2$\,\AA, in our models and in those of TMJ11, relative to the
S07 and CvD12 models. All the iron lines show a mild dependence on \aFe\ at fixed \Feh, as
these lines are believed to be mostly sensitive to Fe abundance (see,
e.g., S07, but also the discussion in \citealt{Barbuy03,Coelho05} about the effect of 
electron donors such as Mg on the continuum). While this does seem to be the case, Fig.~\ref{comparemodellines}
shows that differences exist among model predictions. All models agree that $\rm
Fe\,5335$ decreases with \aFe\ (the amount of decrease being slightly stronger
for our, relative to other, models). The behavior of $\rm Fe\,5270$ is more
model dependent,  as this index remains constant  with \aFe\ for the CvD12 models, increases for TMJ11, and
decreases in our models and in those of S07.  The $\rm Fe\,4383$ decreases with \aFe\ in
all models, with a milder variation for our models and TMJ11, relative to the CvD12 and
S07 models. Notice that, because of a possible dependence on Mg
abundance~\citep{Korn05}, the $\rm Fe\,4383$ index might not be as clean an Fe indicator as
as the $\rm Fe\,5270$ and $\rm Fe\,5335$ indices (see S07). Indeed,
Fig.~\ref{comparemodellinesmgfe} indicates that this index does not behave
differently than the other Fe lines. Finally, $\rm Fe\,5015$ is expected to have
some sensitivity to the abundance of \alfa-elements, in particular to
\MgFe\ and \TiFe\ \citep{Korn05}. However, no significant dependence of this
index on $\rm \aFe$ is seen in our models, CvD12, or the S07 models, while $\rm Fe\,5015$ actually 
increases with \aFe\ in the TMJ11 models.

Fig.~\ref{comparemodellines}, which includes the results for a number of age-
and IMF-sensitive indices, shows several interesting features:

\begin{description}
 \item[\hb\ line - ] TMJ11, S07 and our models point to little dependence
 of \hb\ on \aFe\ at fixed \Feh. In contrast, the CvD12 models predict
 that \hb\ increases significantly with \alfa-enhancement. Although the
 origin of this discrepancy is unclear, one should notice that \hb\ decreases
 with velocity dispersion in ETGs (see Section~\ref{sec:ETGs}), and at 
 fixed velocity dispersion, as seen in the stacked spectra 
 of~\citet{LaBarbera13}. \hb\ also mildly decreases with \MgFe, by less
 than $\sim0.1$~dex~\footnote{We have estimated this amount of variation using
 the \MgFe\ stacked spectra of ETGs, for the two highest velocity
 dispersion bins, defined in \citealt[see their Figure~7]{LaBarbera13}.}, from
 \MgFe$\sim0.1$ to \MgFe$\sim0.3$.  This seems to point
 to a genuine mild dependence of this index on enhancement, as predicted by S07,
 TMJ11, and our models. We return to this issue in
 Section~\ref{sec:ETGlines}. 
 \item[Higher order Balmer lines - ] All the models agree in that \hdf\ should
 increase with \aFe, although the amount of increase is larger for CvD12 and
 TMJ11 than  for our models and S07. For  \hgf, TMJ11 predict
 that this index should have little dependence on \aFe, while our, S07, and
 the CvD12 models predict an increase of \hgf\ with \aFe. 
 \item[TiO IMF-sensitive features - ] We compare in
 Fig.~\ref{comparemodellines} some representative IMF-sensitive features, i.e.
 NaD, TiO1, and TiO2, for which we can compare three different models
 (TMJ11, CvD12, and ours). A comparison between the CvD12 models and the 
base-models from~\citet{MIUSCATI} for other IMF-sensitive  
features  has recently been performed by~\citet{Spiniello14}. 
The bottom panels in Fig.~\ref{comparemodellines} show
 that all models agree on the prediction that TiO features should increase with
 \aFe. This holds at either fixed \Mh\ or fixed \Feh, as TiO bands are quite
 insensitive to total metallicity for old ages (see magenta arrows in the
 Figure, showing the effect of varying \Mh\ by $+0.3$ at fixed age). Also,
 both the CvD12 and our models agree that TiO2 should be less sensitive to \aFe,
 (when comparing the response to \aFe\ with that to age, i.e. black  and blue
 arrows in the Figure) than TiO1, making it a better IMF indicator than TiO1. 
The reader should be warned, however, that neither TiO1 nor TiO2 show any significant
 variation with \MgFe\ in ETGs at fixed velocity dispersion
 (see~\citealt{LaBarbera13}).
 Interestingly, although not plotted here, we find that the sensitivity of TiO2 to \aFe\
 is significantly smaller when a bottom-heavy IMF is adopted for computing the
 models. This is the case both for our models and also the CvD12 models.
 \item[NaD IMF-sensitive feature - ] This feature is found to decrease with
 \aFe\ at fixed \Feh\ in agreement with the CvD12 models. In our models, NaD also
 decreases with \aFe\ at fixed total metallicity, while at high \aFe\ the
 increase of NaD with \Mh\ is milder than at scaled-solar (see top--left panel
 of Fig.~\ref{fig:imf_sensitive}). The net effect is that NaD decreases with
 \aFe, at either fixed \Feh\ or \Mh. It has been recently pointed out that an
 increase of total metallicity and \NaFe\ (by $+0.3$~dex) might be able to
 explain the high NaD line-strengths observed in massive ETGs~\citep{Jeong13},
 with no need for a bottom-heavier IMF in these galaxies. As noted by
 \citet{Jeong13}, this result is based on the predictions from the TMJ11 models. 
 Our new models support the prediction that NaD decreases with \alfa-enhancement 
 as in the CvD12 models. This suggests that it is necessary to invoke another source 
(e.g. IMF) in order to increase the model line-strengths sufficiently to match the 
high NaD of massive ETGs. 
 However some caveats should be noted. For example, the NaD feature can be contaminated
 by interstellar absorption, although we note here that for low velocity
 dispersion ETGs with a Kroupa-like IMF (and selected to have low internal
 dust) our base model predictions for the NaD match exactly the observed
 strengths for this index (see \citep{LaBarbera13}). In addition, we find that NaD increases,
 rather than decreases, with \MgFe\ at fixed velocity dispersion in massive ETGs.
 This is possibly due to a correlated variation of \NaFe\ and \MgFe\ abundances in
 these systems~\citep{LaBarbera13}.
\end{description}

In summary, Fig.~\ref{comparemodellines} shows that although differences remain
among predictions from different models on the dependence of the Balmer lines on
\aFe, our new models agree well with the expectations from most models compared
here, i.e. that \hb\ remains constant (as in S07 and TMJ11), \hgf\ increases (as
in CvD12 and S07),  and \hdf\ increases (for all models) with \aFe. Also, our
models agree remarkably well with CvD12 models on the expected sensitivity of 
commonly used IMF-sensitive features (TiO1, TiO2, and NaD) to \aFe.

\section{Comparison with Globular Clusters and Early-Type Galaxies}
\label{sec:data}

\subsection{Galactic Globular Clusters}
\label{sec:GCs}

\begin{figure*}
\includegraphics[angle=0,width=1.0\textwidth]{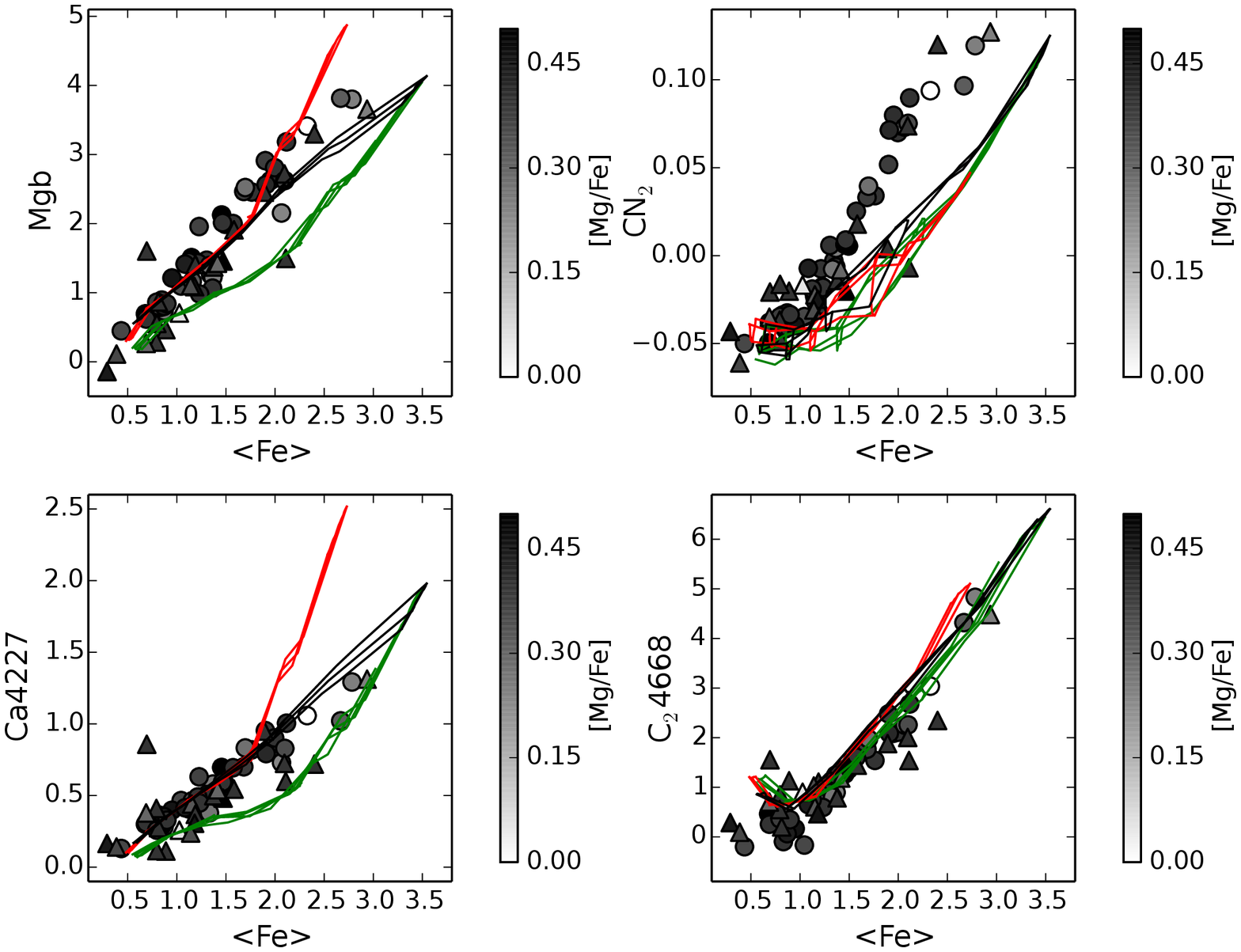}
\caption{Comparison of selected line-strength indices measured from our
scaled-solar (green), \alfa-enhanced (red) and base (black) model SEDs with
those measured from the Galactic GC libraries of \citet{SchiavonGC} (circles) and
\citet{KimGC} (triangles). Both the model SEDs and GC data have been smoothed to match a
common LIS resolution of $5$\,\AA\ (FWHM). Models with metallicities
$-2.27\leq \Mh \leq+0.06$, ages $8\leq t \leq14$\,Gyr and Kroupa Universal IMF
are shown. Clusters are colour-coded by [Mg/Fe] taken from the compilation of
\citet{Roediger14} supplemented with data from \citet{Johnson05} (NGC\,6205),
\citet{Behr03} (NGC\,6341), \citet*{Kacharov13} (NGC\,6864),
\citet{RodgersHarding90} (\CaFe: NGC\,6981), \citet{Kraft98} (NGC\,7006).}
\label{GCs}
\end{figure*}

In Fig.~\ref{GCs} we compare selected line-strength indices measured from our
scaled-solar and \alfa-enhanced model SEDs with those measured from the Galactic
GC (GGC) libraries of \citet{SchiavonGC} and \citet{KimGC}. Both the model SEDs
and GGC data have been smoothed in order to achieve a common LIS resolution of
$5$\,\AA\ (FWHM).

The use of GGCs to test SSPs is particular useful; the majority of the GGCs in
these two libraries have well-defined CMD-derived ages, and metallicities (\Feh)
and light element abundances such as \MgFe, \CFe, \NFe\ and \CaFe\ obtained from
high-resolution spectroscopy (see e.g. \citealt{Roediger14} for a useful compilation of
these quantities).

For all the indices shown in  Fig.~\ref{GCs}, at low metallicities
(\Mh$<-1.0$) the \alfa-enhanced and base models are very similar. This is to
be expected since the uncorrected MILES library stars for the base models in
this regime are \alfa-enhanced to a level similar to that of the \alfa-enhanced
models. Additionally, at the lowest metallicities the index predictions for the
scaled-solar models also converge towards the base- and \alfa-enhanced models.
Such behaviour is not surprising as the sensitivity to the \alfa-enhancement
decreases with decreasing metallicity (see Fig.~\ref{fig:metratio}).
This shows that the use of base models is also appropriate for studying
metal-poor GGCs.

The typical \MgFe\ for GGC stars is $\sim+0.4$ \citep{Roediger14} and this is
in good agreement with location of the majority of the GGCs in the \mgb - \fe\
panel of Fig.~\ref{GCs}. Most of the GGCs lie on the model \MgFe$= +0.4$ locus. A
couple of the most metal-rich GGCs (e.g. NGC\,6528 and NGC\,6553) lie between
the scaled-solar and \alfa-enhanced grids suggesting \MgFe$\sim+0.2$ -- $+0.3$
which is consistent with the high-resolution \MgFe\ measurements of
$+0.25\pm0.11$ \citep{Carretta01,Coelho01} and $+0.32\pm0.08$ \citep{Barbuy92}
respectively. Our result for these two clusters are also in good agreement with
\citet{Lee09b} (see their figure~13).

\CaFe\ is also enhanced in the majority of GGCs ($\sim+0.3$; \citet{Roediger14})
and this reflected in the location of GGCs in the Ca\,4227 - \fe\ panel of
Fig.~\ref{GCs}. [Ca/Fe] in GGCs is typically $\sim0.1$ dex lower than \MgFe\
for a given cluster and this is well reproduced in the figure. The fact that
\MgFe\ and \CaFe\ in GGCs behave similarly when compared to the \alfa-enhanced
grids is both reassuring and not surprising -- both Mg and Ca are
\alfa-elements. However, as pointed out by \citet{Roediger14}, the lack of any
correlation between \MgFe\ and \CaFe\ in GGCs seems to suggest that, while both
are indeed \alfa-elements, they are possibly produced in different
nucleosynthetic sites. Interestingly, \CaFe\ does not seem to directly track
\MgFe\ in massive galaxies (e.g., 
\citealt{Vazdekis97,Trager98,Cenarro03,Thomas03,Conroy13rev}).

 In terms of other indices, CN$_2$ (and CN$_1$ --not shown) is clearly stronger
in the GGCs than the models, and the inclusion of \alfa-enhanced SEDs does not
improve the situation. It is well established that GGCs (e.g.
\citealt{Burstein84,Cannon98,Roediger14}) and extragalactic GCs (e.g.
\citealt{Burstein84,BrodieHuchra91,Kuntschner02a,Strader03,Beasley04,Cenarro07GCs,Schiavon13})
show enhanced CN when compared to solar neighbourhood stars. Fig.~\ref{GCs}
provides support for the early results of \citet{LeeWorthey05} (see also
\citealt{Burstein84}), which found a significant discrepancy when comparing
Ca4227, CN$_1$, CN$_2$ and NaD index strength from their \alfa-enhanced SSP
models and data for GCs in the Milky Way and in M\,31. This shortcoming of
the current generation of \alfa-enhanced SSP models can be understood when taking
properly into account the phenomenon of Multiple Populations (MPs) in Galactic
GCs (see \citealt*{Gratton12};\citealt{Piotto12,CassisiSalaris13}, and references therein
for detailed reviews on this topic). 

Essentially, the MP phenomenon is related to the presence in each individual GC of 
two (almost coeval) stellar generations, the first one born with a canonical \alfa-enhanced heavy element
distribution, the second generation characterised by anti-correlated CN-ONa
abundance patterns, and by moderate (and - only - in a few cases very large) He
enhancement. Due to the impact on the SED of
molecular bands associated with both C and N (see the detailed discussion made by
\citealt{Sbordone11} and \citealt{Cassisi13}), one might anticipate that the 
light-element (anti-)~correlations characteristic of the 
second generation stars in Galactic GCs will affect the strength of specific
indices such as CN$_1$ and CN$_2$. This issue has been explored by
\citet*{Coelho11} given certain assumptions\footnote{They
adopted extreme -although still realistic - light-element chemical patterns in
order to maximise the possible expected impact of these chemical peculiarities
on SSP modes.} about the chemical abundance patterns of second generation stars.
These authors found that - regardless of the exact value of the He enhancement -
as a consequence of the stronger CN and NH molecular bands in the spectra of
second generation stars CN indices are significantly affected. Variations of the
order of $\sim0.08$\,mag were seen between the CN indices predicted for a pure 
first generation stellar population and those predicted for population
of pure second generation stars (with extreme values for the light-element
anti-correlations). 

Finally, unlike CN indices, the C$_2$4668 index of the GGCs is in
good agreement with the model grids. As evidenced by the superposition of the
scaled-solar, \alfa-enhanced and base models, it is clear that this index has
little sensitivity to the \alfa-abundance.

In summary, we conclude that the LIS \mgb\ and iron indices of the GGCs are
well-reproduced by the \alfa-enhanced MILES models. Good agreement is also found
for Ca\,4227 in GGCs.  The model CN indices are largely insensitive to the
\alfa-abundance, and fail to reproduce the high CN values seen for GGCs at a
given metallicity. The model C$_2$\,4668 index is also insensitive to the
\alfa-abundance, but unlike CN reproduces the behaviour of the GGC C$_2$\,4668
measurements.


\subsection{Early-Type Galaxies}
\label{sec:ETGs}

\subsubsection{Early-Type Galaxy Colours}
\label{sec:ETGcolours}

\begin{figure*}
\includegraphics[ width=1.0\textwidth]{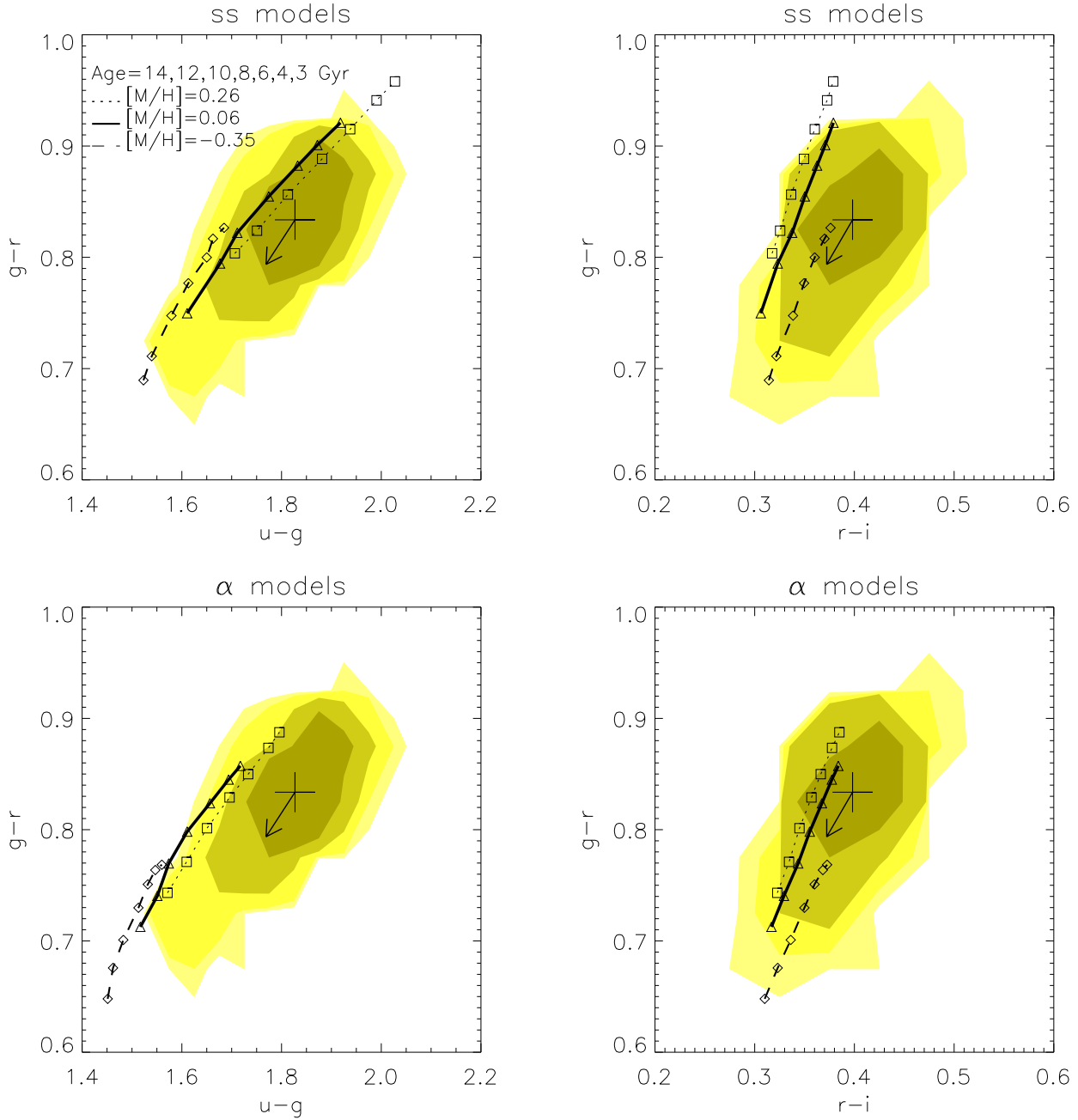}
\caption{Observed colour distribution in the SDSS bands of a sample of ETGs at
  z$\simeq0.04$. The contours enclose $20$, $40$, $60$ and $95$ per cent of the
  galaxies. The cross indicates the location of the median colours.  The
  synthetic colours derived from scaled-solar models (upper panels) and
  \alfa-enhanced models for a Kroupa Universal IMF are overplotted. Only SSP models with
  ages in the range $3$--$13.5$\,Gyr and metallicities: $-0.35, +0.06, +0.26$ have been
  considered. 
 }
\label{col_etg} 
\end{figure*} 

\begin{figure*}
\includegraphics[width=0.93\textwidth,height=0.93\textheight]{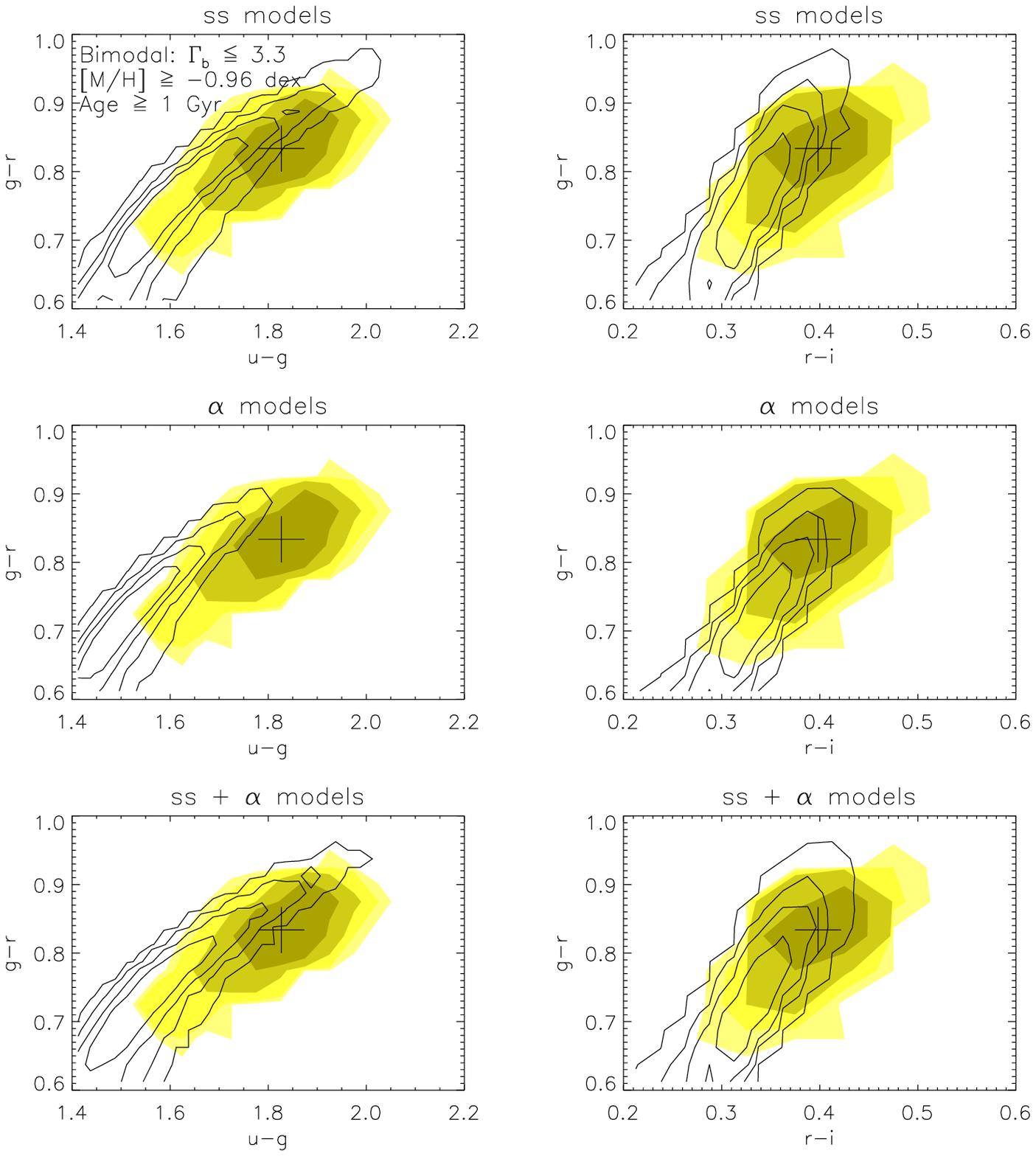}
\caption{The same ETG colour distributions of Fig.~\ref{col_etg} are compared
  with the predictions from composite stellar population models. The black
  contours enclose $20$, $40$, $60$ and $95$ per cent of the MC simulations of multiple
  bursts models,  where age, metallicity, IMF (bimodal) slope and strength of
  the bursts are allowed to vary. The upper panels show the predictions for the
  scaled-solar model, the middle panels display those of \alfa-enhanced models,
  whereas the bottom panels show the predictions from MC simulations where a
  random combination of the two set of models is considered. 
 }
\label{col_etg_mc} 
\end{figure*} 

A strong constraint on the reliability of our models comes from the optical
colours of nearby ETGs. Current stellar population models, both based on
theoretical stellar spectra and empirical libraries, still have difficulties in
reproducing the optical colour distribution of ETGs, with theoretical models showing 
the strongest discrepancies~\citep{Maraston09,ConroyGunn10,MIUSCATII}.

In Fig.~\ref{col_etg} we present the colour distribution in the SDSS bands of a
sample of z$\sim$0.04 ETGs, as selected in \citet{MIUSCATII}. 
These data are compared with our scaled-solar and \alfa-enhanced
SSP models for a range of metallicities and ages assuming a Kroupa Universal IMF.  The 
scaled-solar models (upper panels) have the same difficulties as highlighted in
\citet{MIUSCATII} for the base-models. This behaviour is expected since the scaled-solar
models are very similar to the base models in the high metallicity regime
required to fit luminous ETGs (see Fig.~\ref{col_alpha}), a region where the MILES stars have
approximately scaled-solar abundance ratios (see Fig.~\ref{milone}). 
The colours for the scaled-solar models are too red in g$-$r and too blue in 
u$-$g and r$-$i. This is a problem that cannot be solved by superimposing a young 
stellar population on top of the main population, since this would shift the whole set of colours 
to the blue and thereby worsen the mismatch with the u$-$g and r$-$i colours. 

As shown in Section~\ref{sec:ssaacolours}, the \alfa-enhanced models are bluer
in the blue bands, and redder in the red, as seems to be required by the
data. Indeed, the comparison with the \alfa-enhanced models (lower panels of
Fig.~\ref{col_etg}) show a much better agreement with data in the g$-$r/r$-$i
plane.  However, the shift to the blue in u$-$g appears too dramatic to reconcile the
model with these data, suggesting that an \alfa-enhancement of \aFe$=+0.4$ (as in the
models presented here) is too high when compared to the observed colours of
ETGs. However, the best solutions obtained with these models are more consistent
than those obtained with the scaled-solar models. It is also worth recalling here
that our models with varying abundance ratio do not extend to the blue end of
the u filter response, where we used our NGSL-based base model predictions. The
\alfa-enhanced models point to old ages in the colour-colour diagrams, whereas
the scaled-solar models require significantly younger ages for the u$-$g versus
g$-$r diagram.

To explore the effect of more complex star formation histories, we perform a set
of Monte Carlo (MC) simulations, where we allow all the stellar population
parameters to vary: age, metallicity and IMF slope (using the bimodal parameterization). 
We adopt three-burst star formation histories, where the mass fraction involved in each
burst is also a free parameter. The resulting predictions for $10000$
MC simulations are shown in Fig. \ref{col_etg_mc} for scaled-solar models (upper
panels), for \alfa-enhanced models (middle panels) and for a random combination of the
two (lower panels). The mixture of scaled-solar and \alfa-enhanced models gives better
results as the solutions obtained for the two colour-colour diagrams are more
consistent each other. The improvement in the u$-$g/g$-$r plane with respect to
the case where only SSPs are used, is mainly due to the contribution of young
stars and metal-poor populations. Nevertheless, the models have difficulties in
reproducing the part of the diagram lying at high u$-$g. The match in the
r$-$i/g$-$r plane also improves significantly. Indeed, we see a clear reddening
of r$-$i, that is due to both the use of \alfa-enhanced models and to varying
the IMF slope (see Section 6.1 in \citealt{MIUSCATII}). The obtained solutions
are consistent with recent studies based on absorption line-strengths, which
find steeper IMF slopes \aFe\ values around $\sim0.2$ and old ages for massive
ETGs (e.g., \citealt{LaBarbera13,Spiniello14,CvD12}).

\subsubsection{Full Spectrum Fitting}
\label{sec:ETGspectra}

\begin{figure}
\includegraphics[width=\columnwidth]{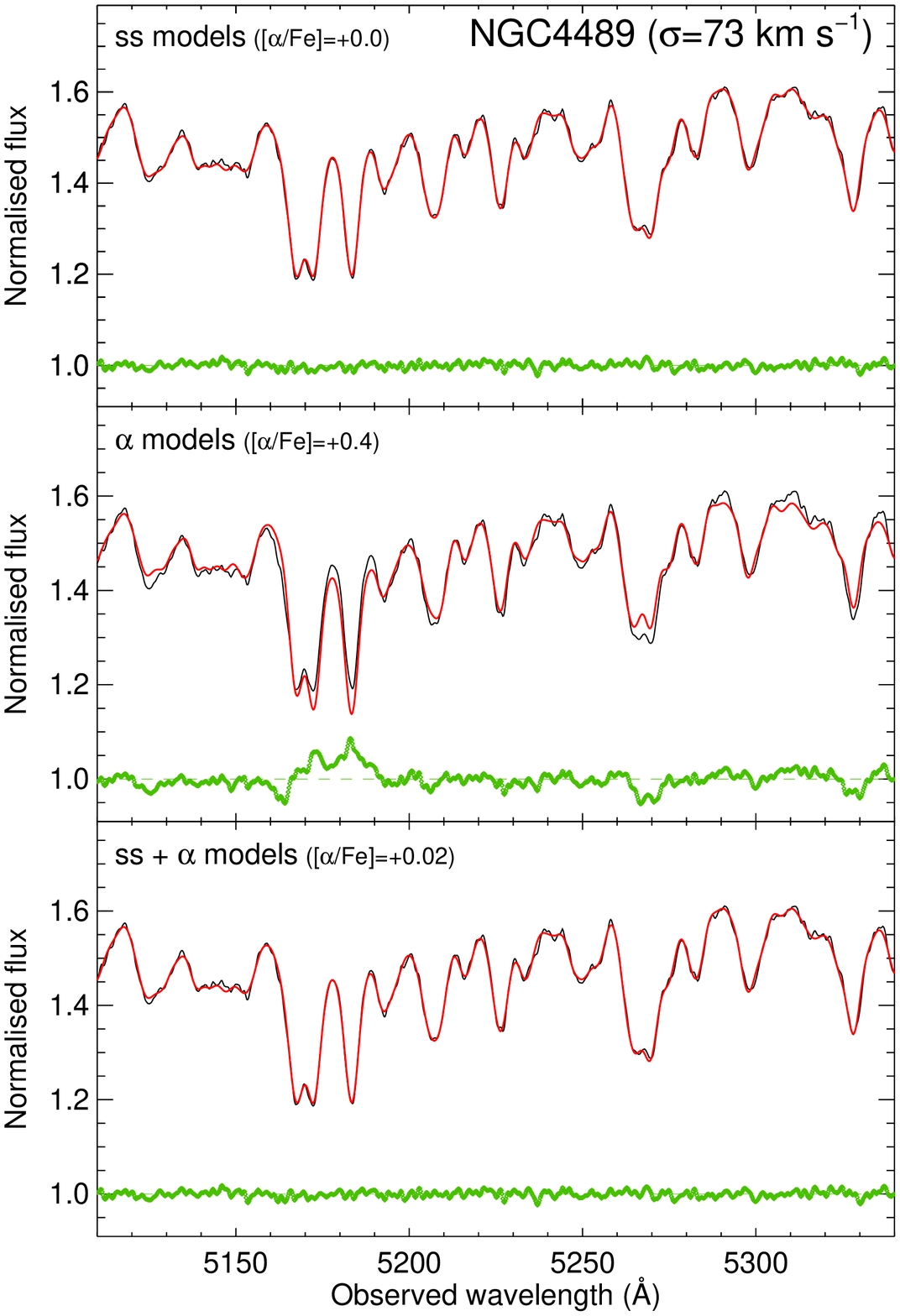}
\caption{Spectral fitting of the low velocity-dispersion galaxy NGC\,4489. The upper
panel shows the fit obtained with scaled-solar models. The middle panel shows the
fit obtained with the \alfa-enhanced models with \MgFe$=+0.4$. The bottom panel
shows the fit with both set of models included in the database. The residuals
are shown (in green), keeping the same scale.}
\label{mgfemgfe1}
\end{figure}

\begin{figure}
\includegraphics[width=\columnwidth]{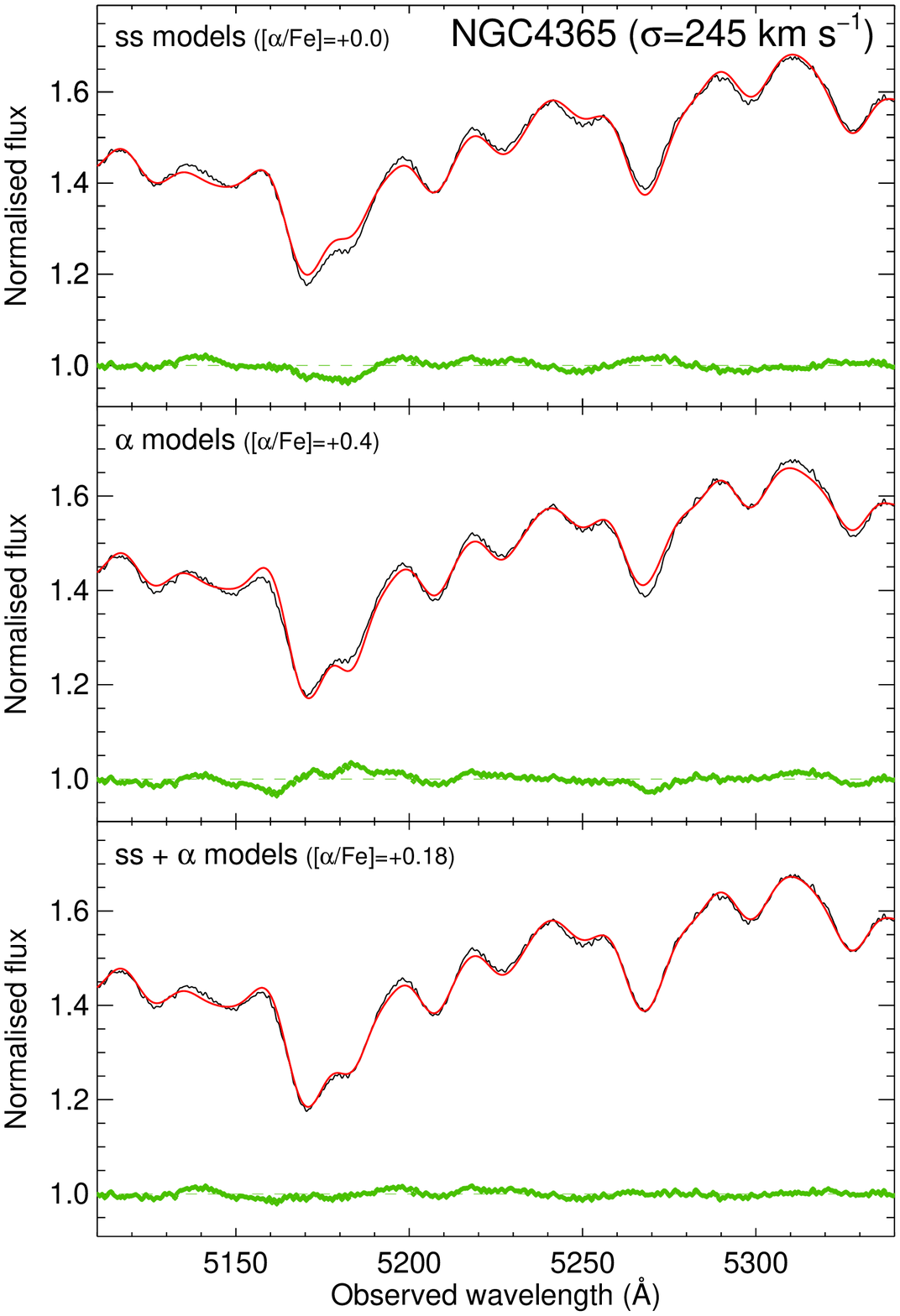}
\caption{Spectral fitting of the giant elliptical galaxy NGC\,4365. See
Fig.~\ref{mgfemgfe1} for the details on the panels.}
\label{mgfemgfe2}
\end{figure}

The solar-scaled and \alfa-enhanced model SEDs can be used in full spectral 
fitting techniques that can, in principle, improve the overall fits to data and
be used to derive  (in particular) the \MgFe\ overabundance observed 
in massive galaxies. We have attempted to follow this approach by using the
penalised pixel-fitting routine developed by \citet{CappellariEmsellem04}. We
carried out this exercise on data for Virgo galaxies published in
\citet{Yamada06}. We perform our spectral fits on two ETGs, one
relatively low-mass (NGC\,4489), and the other a giant elliptical (NGC\,4365) 
which have scaled-solar and \MgFe-enhanced abundances, respectively.

At first we tried to perform spectral fits over large wavelength ranges  (e.g.
more than $1000$\,\AA), but we were unable to reproduce the \MgFe\ values and 
trends within galaxies reported by \citet{Yamada06}. This is not surprising given
the fact that the information contained in relevant Mg- and  Fe-sensitive
features gets diluted with increasing wavelength range (e.g.
\citealt{Walcher09}). We therefore turned to fit only those regions where we
expected the effect to be most noticeable (i.e. the \mgb\ -- Fe\,$5335$
range; see for example Fig.~\ref{fig:ageratio} and Fig.~\ref{fig:metratio}). We
therefore performed the fit within the range $\lambda\lambda$ $4800 - 5500$\,\AA.
The results of this exercise are shown in Figs.~\ref{mgfemgfe1} 
and~\ref{mgfemgfe2}. Each figure shows our attempts to fit the observed spectrum
using (i) scaled-solar models only, (ii) \alfa-enhanced models only, and (iii) a
combination of the two using the same age and metallicity for the two populations.
In Fig.~\ref{mgfemgfe1} we see that, for NGC\,4489 - a low velocity dispersion galaxy - pure
scaled-solar and the mixed models reproduce the spectrum fairly well, although
pPXF has some preference for the true scaled-solar models. Fits with
\alfa-enhanced  models only cannot match all the observed absorption features.
The case of the massive galaxy NGC\,4365 ($\sigma=245$\,\kms) is different (see
Fig.~\ref{mgfemgfe2}). Here \alfa-enhanced models are preferred over the scaled-solar
models. It is important to note that the higher velocity dispersion of this galaxy 
makes it harder to identify any mismatches with the templates, as can been seen in the residual 
spectra.

The overall conclusion from this exercise is that our models together with
full-spectral fitting can be used to estimate the \MgFe\ enhancement. However
the spectral fits are sensitive to the wavelength range employed to
estimate the \alfa-enhancement. Short wavelength regions around the main
spectral features (e.g. \mgb, Fe\,$5270$) are preferred over long baselines
(see also the discussion in \citealt{Walcher09} and \citealt{Cezario13} on the
dependence of spectral fitting results with the wavelength range). 

\subsubsection{Line indices}
\label{sec:ETGlines}

\begin{figure}
\includegraphics[width=0.32\textwidth,angle=-90]{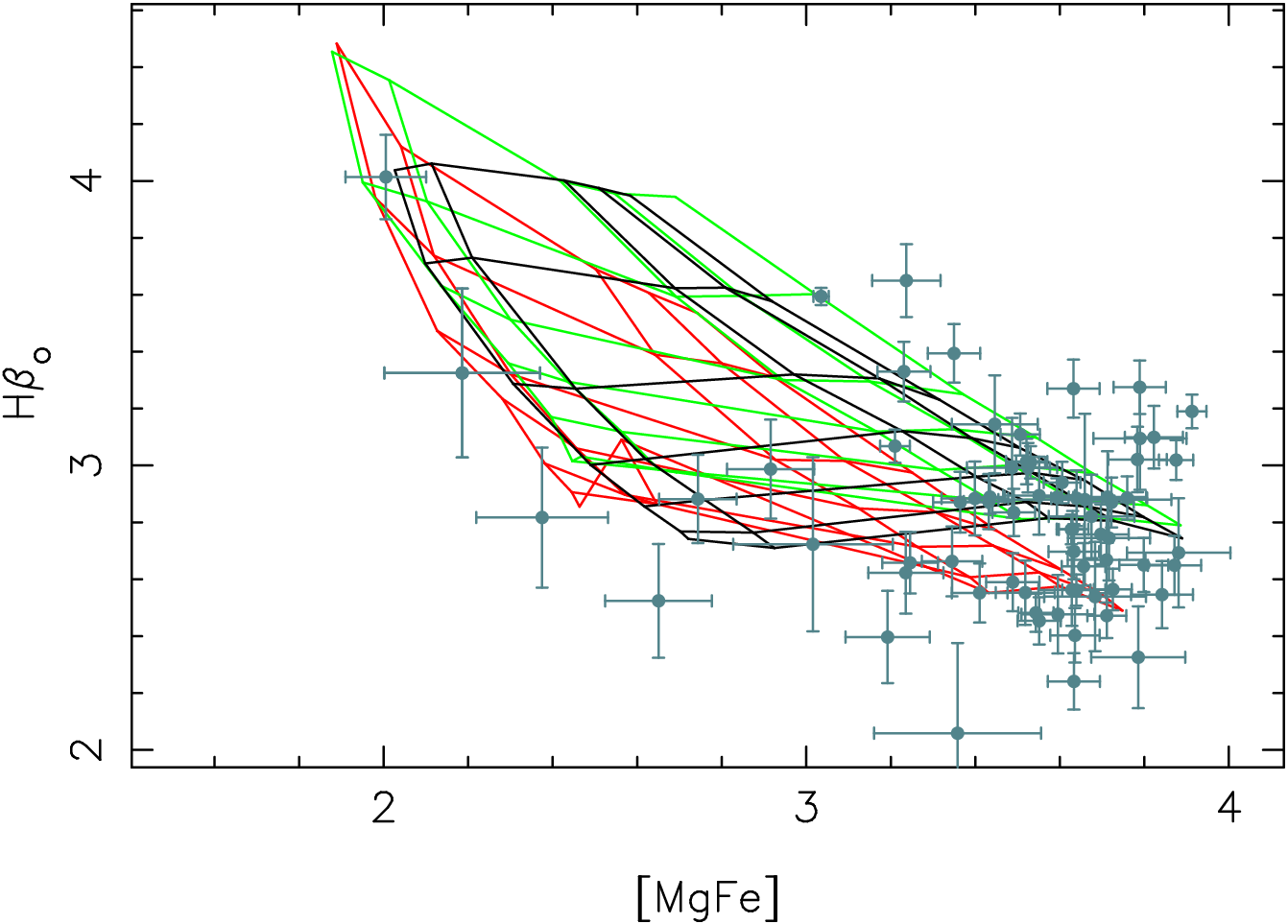}\\
\includegraphics[width=0.32\textwidth,angle=-90]{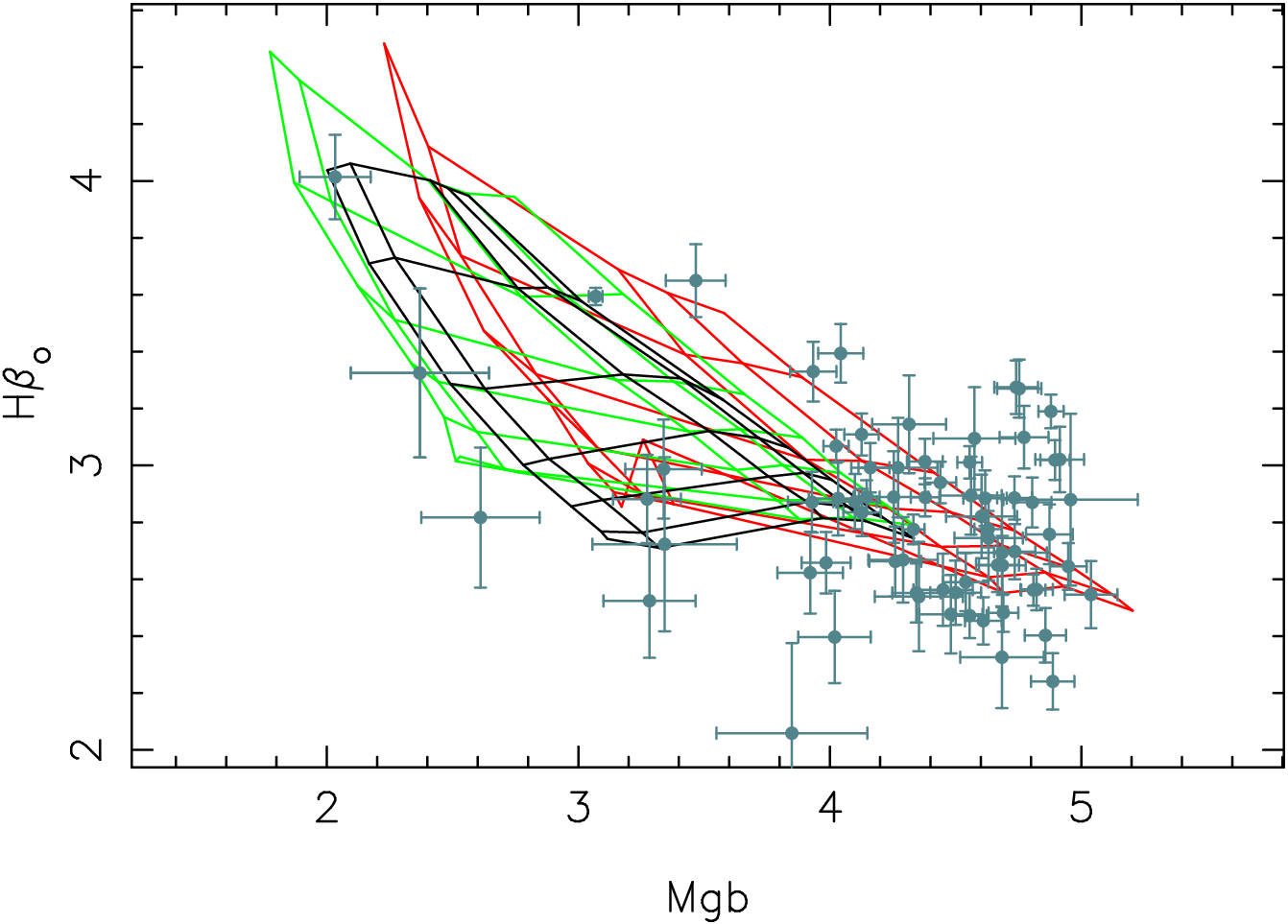}\\
\includegraphics[width=0.32\textwidth,angle=-90]{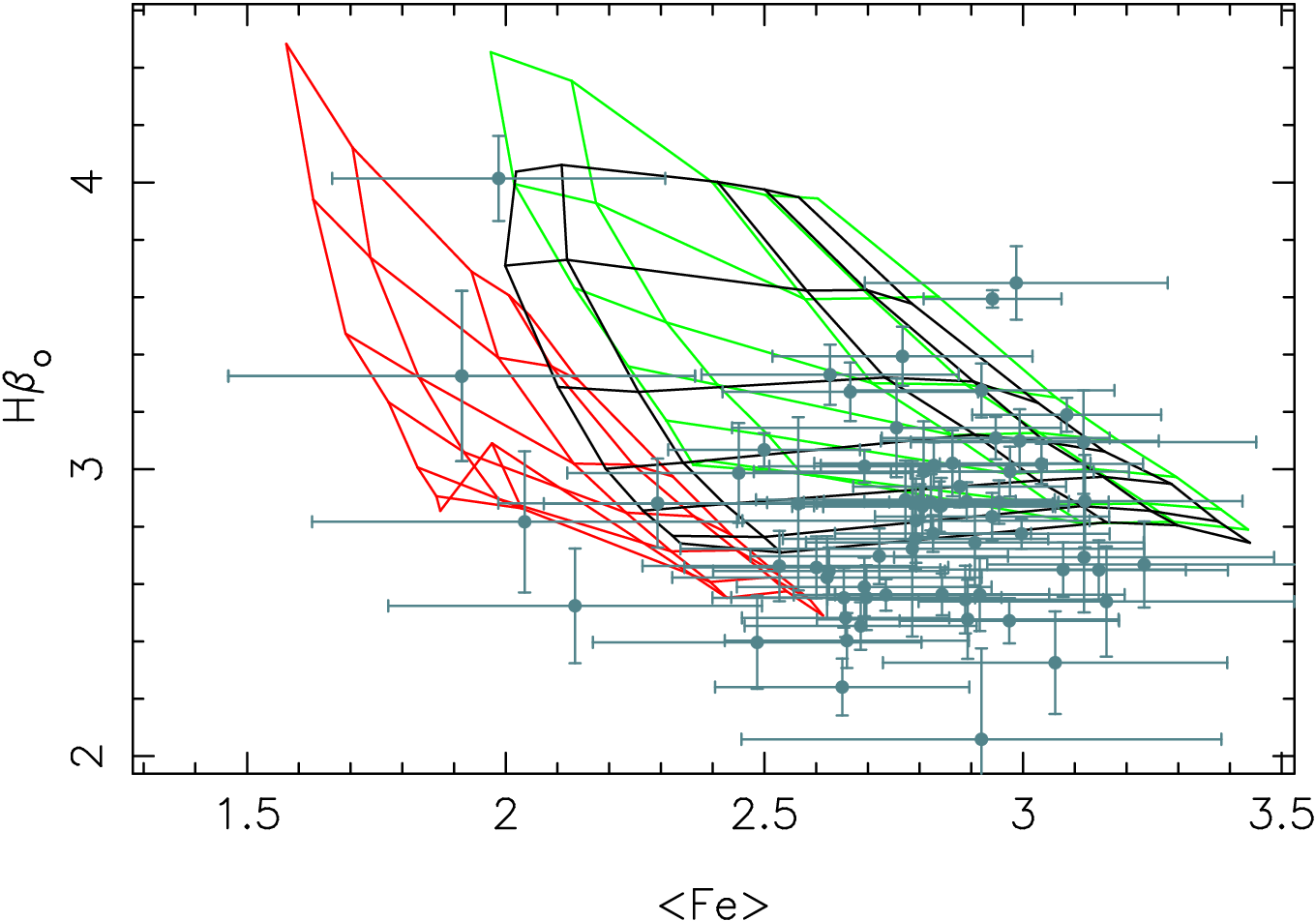}\\
\caption{
Various Mg and Fe index-index diagrams for three different model sets; the base
models (black), the scaled-solar models (green) and the \alfa-enhanced models (red), with
$\MgFe=+0.4$. Overplotted are the line-strength indices measured in the sample
of \citet{SB06a}. The metallicity increases from left to right $\Mh=-0.35,
-0.25, +0.06, +0.15, +0.26$, whereas the age increases from top to bottom $2.0,
2.75, 5.0, 7.5, 10.0, 12.5, 14.0$\,Gyr. All the models have a Kroupa Universal
IMF and have been smoothed to match the LIS-8.4\,\AA.
}
\label{index:fig}
\end{figure}

Fig.~\ref{index:fig} shows three classical index-index diagrams with the
predictions of the three models, base, scaled-solar and \alfa-enhanced.
Overplotted are the index measurements in the LIS-8.4\,\AA\ system for the
early-type galaxy sample of \citet{SB06a}. The model grids obtained
when plotting \hbo\ against \mgb, \fe\ and \mgfe\, indices, which are
sensitive to Mg, Fe and total metallicity, respectively, provide similar
age/metallicity solutions for the most massive galaxies, i.e. those galaxies with the
largest abundance ratios and largest \mgfe\ index values. 

\begin{figure}
\includegraphics[angle=0,width=\columnwidth]{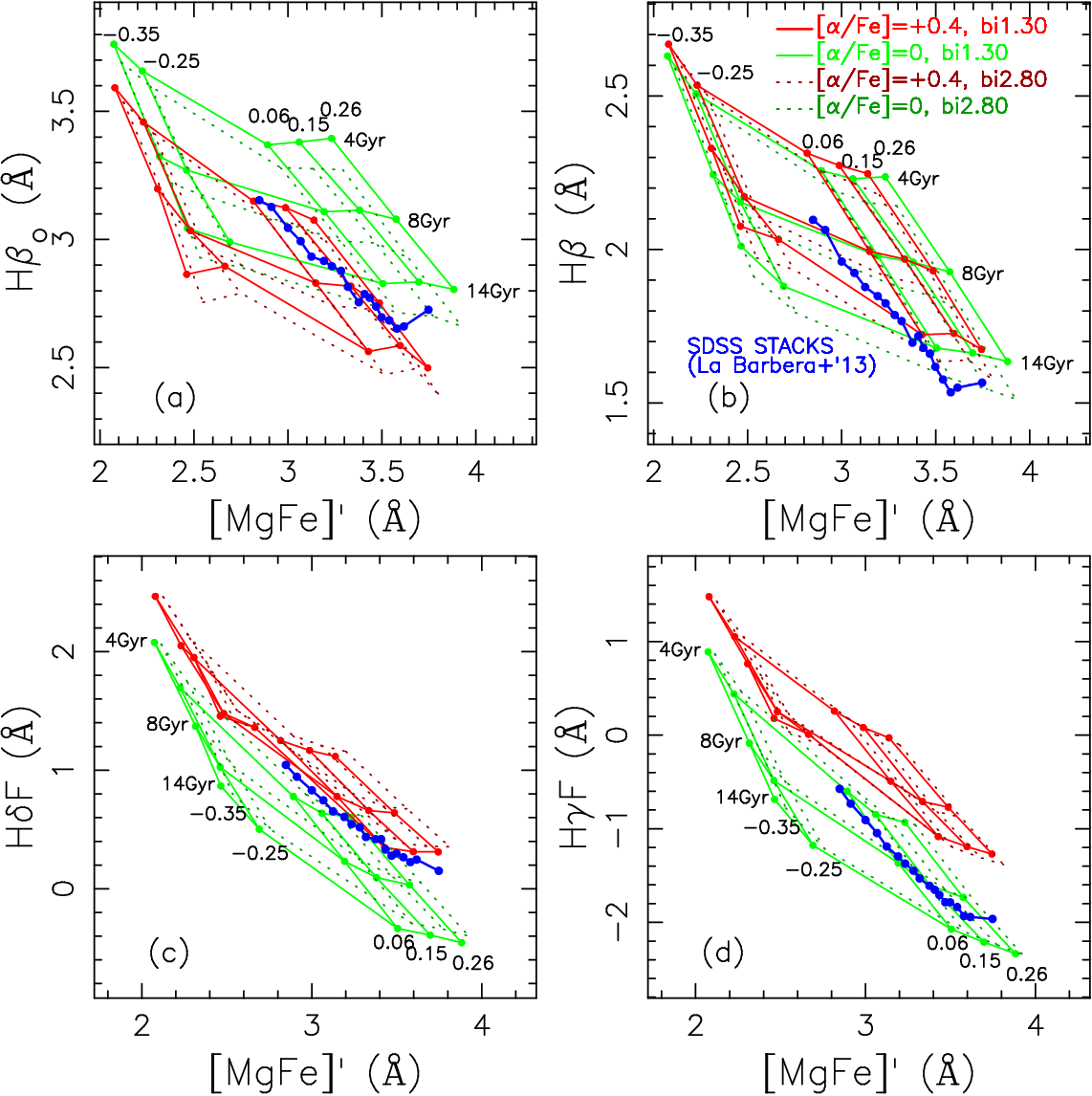}
\caption{Line-strengths of Balmer lines, \hb , \hbo , \hdf , and \hgf\ are
plotted as a function of the total metallicity indicator \mgfep , for
scaled-solar (green) and \alfa-enhanced SSP models (red). Light-solid and
dark-dotted grids correspond to models with standard (bimodal; 
$\Gamma_{b}=1.3$) and bottom-heavy (bimodal; 
$\Gamma_{b}=2.8$) IMF, respectively. The grids show the effect of changing
total metallicity (moving from left to right in each panel) and age (increasing
from top to bottom in each panel), respectively, as labelled. Blue dots,
connected by a blue line, mark the line-strengths for $18$ SDSS stacked spectra of
ETGs, from~\citet{LaBarbera13}, in the range of galaxy velocity dispersion
from $\sim100$ to $\sim300$\,\kms. Line-strengths have been corrected for all
models and spectra to the same $\sigma$ of $200$\,\kms\ (i.e. the same
resolution employed in \citealt{LaBarbera13}), summed up in quadrature to
the SDSS spectral resolution.}
\label{fig:Balmerstacks}
\end{figure}

We see that most of the massive galaxies fall below the horizontal line
corresponding to the oldest stellar population (14\,Gyr) of the scaled-solar
model grids. The same applies to the base model grids, which at high
metallicities have an approximately scaled-solar \MgFe\ abundance ratio. 
However, the locus of these galaxies falls in between the lines corresponding to the oldest
SSPs of the scaled-solar and the \alfa-enhanced models, consistent with old
ages, total metallicity above solar and an abundance ratio around \MgFe$\sim+0.2$.
Note that we only employed models with a Kroupa Universal IMF for the grids of
Fig.~\ref{index:fig} as our present purpose is not fit these galaxies,  but rather to
illustrate the advantages of the new models presented here. We could
have obtained slightly younger ages for the most massive ETGs had we employed
grids with steeper IMF slopes, according to the relation between the slope
and the velocity dispersion according to
\citep{Ferreras13,LaBarbera13,Spiniello14} (see also the results we show and 
discuss shortly in Fig.~\ref{fig:Balmerstacks}).

It also is worth noting in Fig.~\ref{index:fig} (top panel) that the scaled-solar line that corresponds to solar
metallicity, i.e. the third vertical line starting from the right, is very similar to that
corresponding to the base model.  Similar behaviour is seen for the more metal-rich
lines with \Mh$=+0.15$ and $+0.26$, although in these cases the lines corresponding
to the base models are slightly shifted toward the solar metallicity value with
respect to the scaled-solar models. We attribute this small difference to
the fact that many of the MILES stars have \MgFe\ values slightly below
scaled-solar at these high metallicities \citep{Milone11}. For lower
metallicities, the vertical lines corresponding to the base models are shifted toward a higher
metallicity with respect to the two, the scaled-solar and the \alfa-enhanced
model lines (top panel, Fig.~\ref{index:fig}). 
Notice that for the base models the stars are selected according to
their \Feh\ values. However at these metallicities the MILES stars are
over-enhanced in the \MgFe\ abundance. Therefore the stars have a total
metallicity that is higher than that of the stars employed in the scaled-solar
and \alfa-enhanced models according to Eq.~\ref{eq:MhFehrelation}.

Fig.~\ref{fig:Balmerstacks} is similar to Fig.~\ref{index:fig} but shows
representative Balmer lines, \hbo , \hb , \hdf , and \hgf\ (panels a--d),  as a
function of the total metallicity indicator \mgfep . The grids 
with dotted linetype in the Figure show models with a bottom-heavy, rather than Kroupa-like,
IMF. Green and red grids correspond to scaled-solar and \alfa-enhanced  models,
respectively. In each panel, we overplot data-points for 18 stacked spectra of
ETGs from~\citet{LaBarbera13}, ranging from a galaxy velocity dispersion of
$\sim100$ to $\sim300$\,\kms. Both model and data line-strengths are plotted at
the same spectral resolution of $200$\,\kms,  added in quadrature to the
native resolution of the SDSS spectrograph ($\sim 2.8$~\AA\ at  $\lambda
\sim5000$\,\AA). The comparison of models and data can be summarised as follows.

\begin{description}

 \item[i. ] The model \hb\ and \hbo\ indices behave differently with 
 \alfa-enhancement, in that the former is nearly independent of \aFe , while \hbo\
 decreases with \alfa-enhancement. As a result, one can explain the observed \hbo\ of our
 stacked spectra of ETGs with $\alpha$-enhanced  models, while for \hb\ a
 bottom-heavy  IMF (see dotted grids in panel b of the Figure) seems to be
 required  to account for the observed line-strengths. However, as noted above
 (Sec.~\ref{sec:othermodels}), the behaviour of \hb\ with \aFe\ is also quite
 model dependent, as the CvD12 models predict that this index will increase, rather than
 remain constant, with \aFe . Also, as reported in~\citet{CervantesVazdekis09},
 different theoretical stellar libraries lead to different predictions for the
 response of \hb\ and \hbo\ to \aFe . Indeed, using ~\citet{Munari05}, rather
 than~\citet{Coelho05} stellar models, one would predict the opposite
 \aFe\ dependence of \hb\ and \hbo\ than that seen in 
 Fig.~\ref{fig:Balmerstacks} (see section~4 of~\citealt{CervantesVazdekis09}).
 These differences originate, at least in part, in the residual seen
 toward the red end of the feature bandpass of the \hb\ and \hbo\  index
 definitions, as illustrated in the flux-ratio of Fig.~\ref{fig:Balmeratio}. This
 residual corresponds to a metallic contamination within the red wing of the
 \hb\ feature, which includes Fe-peak elements such as chromium (see also
 Section~\ref{sec:agelines}). The discrepancies between the predictions for the
 \hb\ and \hbo\ lines based on the ~\citet{Munari05} and ~\citet{Coelho05}
 stellar spectra likely stem from differences in the adopted
 line lists employed for the computation of these two sets of model spectra,
 and are presently being investigated (Sansom \& Coelho, in prep.). 

 \item[ii. ] The position of the stacked spectra with respect to scaled-solar
 model grids is approximately the same in both the \hb\ vs. \mgfep and \hbo\
 vs. \mgfep\ diagrams. This, together with the fact that both \hb\ and \hbo\
 exhibit a similar variation with \aFe\ at fixed velocity dispersion in  the
 stacked spectra of ETGs (see Sec.~\ref{sec:othermodels}), suggests that both
 \hb\ and \hbo\ have more similar dependencies on \aFe\ than suggested by
 the models.  

 \item[iii. ] The \hdf\ index behaves consistently with \hbo\ in the sense 
that both indices, in the stacked spectra of ETGs, could be explained by a metal-rich, old and
 significantly \alfa-enhanced ($\aFe\sim+0.3$~dex) SSP.

 \item[iv. ] In contrast, the behaviour of \hgf\ with \aFe\ seems to be too
extreme in the models. The  observed \hgf\  of the stacked spectra
would require \aFe$\sim0.2$~dex for an old ($\sim14$\,Gyr) and metal-rich
($\Mh\sim+0.3$) SSP, in contrast to the higher \aFe\ implied by
\hbo\ and \hdf. The position of the stacked spectra with respect to the
scaled-solar \hgf\ grid suggests that the ``true''  dependence of \hgf\ on
\aFe\ should indeed be mild. Note however that the blue pseudo-continuum of the
\hgf\ line is located on a strong CH absorption, i.e. the G-band at
$\sim4300$\,\AA. Therefore, non-solar carbon (and nitrogen) abundance ratios
reported in massive galaxies (e.g. \citealt{Conroy14}; \citealt*{Worthey14}), which might
depend on environment \citep{SB03,Carretero04,Carretero07}, may have a
noticeable effect on measured \hgf\ index strength. 

\end{description}

\begin{figure*}
\includegraphics[width=\textwidth]{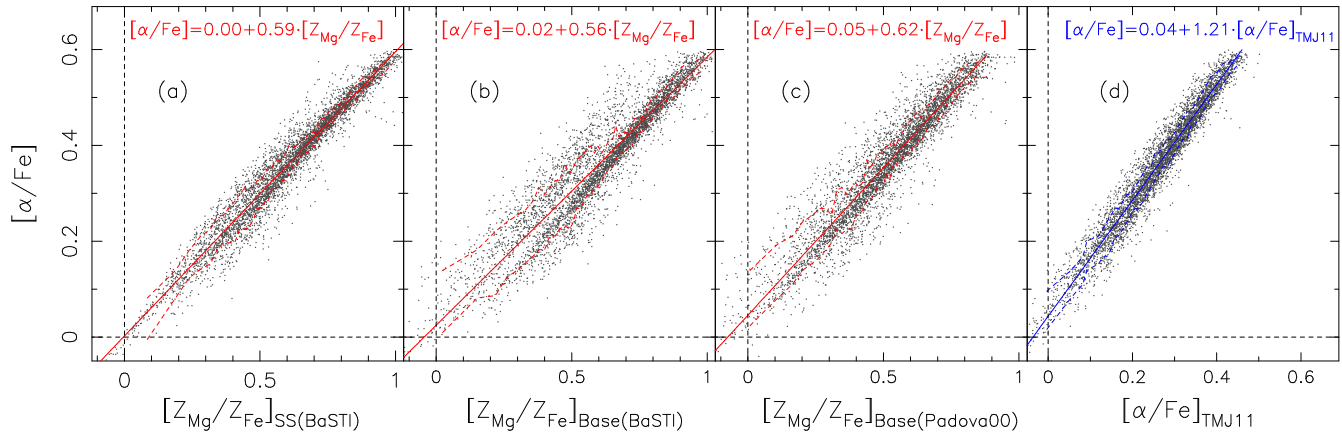}
\caption{
The ``true'' \aFe, as estimated by our new models (see the text), is compared
to different proxies for \aFe\ (panels a--c), and the \aFe\ estimated with the TMJ11
models (panel d). We compare \aFe\ to the proxy for scaled-solar (a), base-BaSTI
(b) and base-Padova (c) models. For each panel, grey dots correspond to a
sample of $4,671$ ETGs selected from the SPIDER sample (see the text), solid
curves are the best-fitting lines to the data (using \aFe\ as dependent variable
in the fitting), dashed curves are the $\pm 1$-$\sigma$ scatter around the
median trends, while horizontal and vertical dashed lines mark the values of
zero on the x- and y-axes. The scatter in \aFe\ around the best-fitting lines
amounts to $0.025$, $0.05$, $0.04$ and $0.025$~dex for panels a to d,
respectively.
}
\label{proxy}
\end{figure*}

 In summary, Fig.~\ref{fig:Balmerstacks} points to the well-known issue that
 SSP model predictions for different Balmer lines (and indeed also different definitions of the same line,
 e.g. \hb\ and \hbo ) provide  different estimates of the age of the old stellar
 populations of ETGs (see also \citet{Schiavon04} for an alternative view, who argues that the 
disagreement between high order Balmer lines might be related to the presence of young stellar 
populations.) The Figure shows that neither the 
 \alfa-enhanced models nor models with a varying IMF seem able to solve this
 issue. Some clues come from the comparison of the \aFe\ dependence of Balmer
 lines among different models (see Fig.~\ref{comparemodellines}) and the
 observed line-strengths of ETGs, as reported above. However, these clues should
 be only interpreted in a qualitative manner and a more quantitative 
 comparison of models and data is desirable. In fact,
 the discrepancies between data and models in Fig.~\ref{fig:Balmerstacks} might
 occur due to different effects, such as a variation of abundances ratios of
 specific chemical species in the stellar atmospheres,  and/or more  complex
 star-formation histories than pure SSPs. Also, the observed differences in the
 predictions from the various models suggest differences in the theoretical
 stellar spectra that are employed to obtain the differential corrections,
 which might come in part from the adopted line lists.

\subsubsection{The proxy for \MgFe}
\label{sec:proxies}

As discussed above (Section~\ref{sec:ETGspectra}), our new \alfa-enhanced
models  can be used to estimate the \MgFe\ of ETGs through spectral fitting, 
provided  one defines carefully the  spectral range to fit. In this section, we
discuss alternative methods to estimate \MgFe,  by either comparing the
line-strengths of selected (Mg and Fe) spectral features to predictions of
either \alfa-enhanced  (as in, e.g., TMJ11) or base models
(see~\citealt{LaBarbera13}).  To this end, we rely on a subsample of $4,671$
nearby ($z\sim0.07$) ETGs from the SDSS-based SPIDER sample, restricting the
analysis to the best-quality spectra, with  $S/N>25$ per \AA\ in the spectral
regions of the \mgb, Fe\,$5270$ and Fe\,$5335$ features (see~\citealt{LaBarbera13} 
and references therein). For each spectrum, we obtain different estimates of
\MgFe\ as follows:

\begin{description}

 \item[``True'' \aFe\ --] We fit \mgfep, \mgb\ and $\rm Fe3$\footnote{$\rm
 Fe3=\frac{1}{3}(Fe 4383+Fe 5270+Fe 5335)$; \citet{Kuntschner00}.}  line-strengths
 with either TMJ11 or our $\alpha$-enhanced models, fixing the age of the SSP 
 models in the fit to the luminosity-weighted age measured with the software
 STARLIGHT for each galaxy  (see~\citealt{LaBarbera13} for details). 

 \item[Proxy for \aFe\ --] At fixed (STARLIGHT) age, we estimate the
metallicity by fitting alternately the \mgb\ and Fe3 line-strengths using
either the base or the scaled-solar models, without the differential
correction for the alpha-enhancement effect (i.e. \Mh$_{\rm Mgb}$ and
\Mh$_{\rm{Fe3}}$). The difference of the two metallicity estimates,
\proxy$=$\Mh$_{\rm Mgb}$$-$\Mh$_{\rm Fe3}$, is the proxy for
\MgFe\footnote{Note that we keep  for the proxy the same notation we employed
in our previous works (e.g., \citealt{Yamada06,LaBarbera13}).}. The computation
is repeated for three sets of reference models, i.e. our scaled-solar, base
models and base models constructed with the Padova00 isochrones. We refer to
these three proxies as  \afepSS, \afepB\ and \afepBP, respectively.

\end{description}

\begin{figure}
\includegraphics[angle=0,width=\columnwidth]{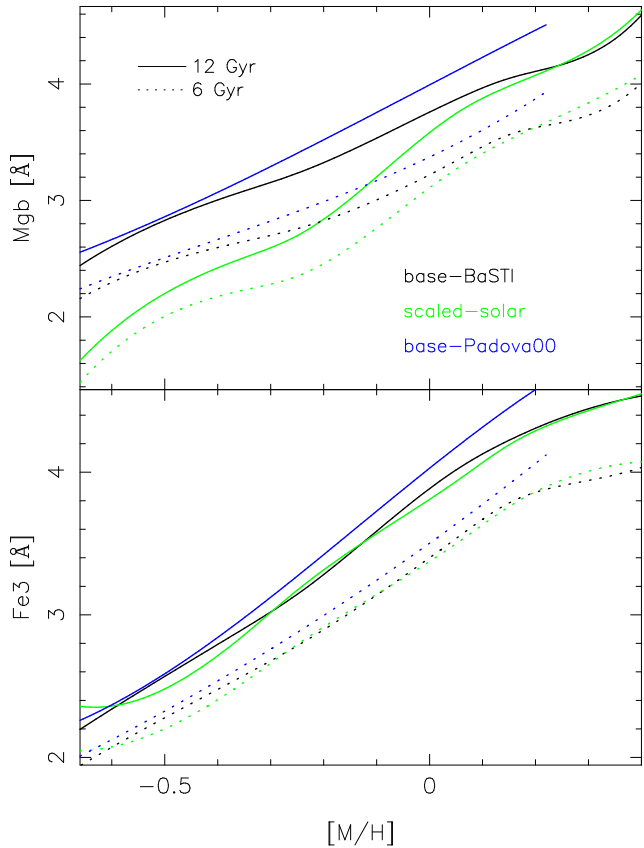}
\caption{Behaviour of \mgb\ (top) and Fe3 (bottom) as a function of total
metallicity for the base-BaSTI,  scaled-solar, and base-Padova models, as labelled in
the upper panel. Solid and dotted curves correspond to ages of $12$ and $6$\,Gyr,
respectively, typical for the spectra of ETGs used for Fig.~\ref{proxy}.
}
\label{fig:furtherproxy}
\end{figure}

As discussed in \citet{LaBarbera13}, since for an $\alpha$-enhanced population
one has  \Mh$_{Mgb}$ $>$ \Mh$_{Fe3}$, estimating the proxy may require some
extrapolation of the reference models  above the highest available \Mh\
in the models. This may hamper the estimate of the proxy  for extremely high
\aFe . On the other hand, as shown in figure~6 of~\citet{LaBarbera13},  the
\afepBP\ proxy correlates remarkably well with the \aFe\ estimated from models
(TMJ11) that take  \aFe\ explicitly into account. 
This suggests that for several
applications one does not need to use $\alpha$-enhanced models to estimate the
\MgFe, provided that a suitable calibration of the  proxy into the ``true''
\aFe\ is performed. To this effect, Fig.~\ref{proxy} expands on figure~6 
of~\citet{LaBarbera13}, comparing the \aFe\ estimate from our new models with
the three proxies  \afepSS, \afepB and \afepBP\ (panels a-c, respectively), and
comparing it to the \aFe\ from the TMJ11 models (\aFe$_{TMJ11}$; panel d). Solid
lines in the Figure show best-fitting lines to the data, while dashed lines mark
the $\pm$1-$sigma$ scatter around median trends. The Figure confirms the 
remarkable correlation of the true \aFe\ with the proxies. However, differences
do exist among different  proxy estimates, in that (i) the slope and zero-point of
the best-fitting line are slightly different from one proxy to another, the
former ranging from $\sim0.56$ to $\sim 0.62$\footnote{Note that in figure~6 of~\citet{LaBarbera13} we
reported a gradient of $\sim0.55$, which is different from the values
in Fig.~\ref{proxy}. In that paper we compared the proxy  to the TMJ11
\aFe\ estimates, and did not apply the same $\rm S/N$ cut to select ETGs,
as in the present  work.}, and the latter from $\sim
0$ to $\sim 0.05$. Notice also that the scatter around the proxy -- \aFe\
best-fitting lines is smallest for our scaled-solar models ($\sim0.025$~dex in
\aFe), and increases to $\sim0.04$~dex for \afepBP, and to $\sim0.05$~dex for \afepB. 
The scatter is largest for the base--BaSTI models because of a prominent
tail of objects having too low \afepB\ at given \aFe\ in panel b (relative to
panel a) of Fig.~\ref{proxy}. The origin of this tail can be understood 
through  Fig.~\ref{fig:furtherproxy}, comparing the dependence of \mgb\ and Fe3
line-strengths on total metallicity for different non \alfa-enhanced models. While
the slope of the Fe3--\Mh\ relation turns out to be approximately
the same for all models, the behaviour of \mgb\ vs. \Mh\ is more
model dependent,  with the shallowest slope for the base-BaSTI models. As seen by
comparing the green and black curves in the top panel of
Fig.~\ref{fig:furtherproxy}, at high \Mh\ ($\gtrsim 0$) one obtains similar
\Mh$_{Mgb}$ and \Mh$_{Fe3}$ estimates when using either base--BaSTI
or scaled-solar models. In contrast, at low \Mh, the \Mh$_{Mgb}$ estimate
is always lower (for given \mgb\ line-strength) for base--BaSTI  than scaled-solar
models, leading to lower \proxy's, explaining the tail of low
proxy values in panel b, relative to panel a, of Fig.~\ref{proxy}. 

Fig.~\ref{proxy}, panel d, also shows that a tight correlation exists between
the ``true'' \aFe\ estimates from our models and from the TMJ11 \alfa-enhanced models, with a
scatter of only $~0.025$~dex in \aFe$_{\rm TMJ11}$. Notice that the slope of the
best-fitting relation between the two estimates of \aFe\  is larger than one
($\sim1.2$), implying that using our models leads to values of \aFe\ larger
than those from the TMJ11 models.

In summary, we find that relative estimates of \MgFe, with accuracy
$<0.05$~dex,  may not require (for most applications) stellar population models
with enhanced \aFe, while absolute estimates of \aFe\ require ``truly''
\alfa--enhanced models. However, absolute estimates of \aFe\ are affected by
differences in the adopted stellar populations models (consistent with what
found, e.g., by~\citealt{Conroy14}).

\section{The MILES webpage}
\label{web}

In an effort to facilitate the use of our models, we have integrated 
these new predictions into our dedicated website (http://miles.iac.es).
The main new feature is the ability to compute spectra for more complex 
SFHs, with varying IMF and \alfa-enhancement. As 
for the current set of models, our website will deliver spectra at the desired 
instrumental setup decided by the user, as well as provide line-strength and 
colour predictions among other observables.  

\subsection{Complex models with multiple bursts}
\label{Complex}

The complex SFHs used for Fig.~\ref{col_etg_mc} have been obtained through the
SFH  facility implemented in the MILES website. This tool allows the user to
retrieve the model spectrum corresponding to a given SFH, having an arbitrary
number of bursts, with different ages, metallicities and burst-strengths (see
Section~3 in \citealt{MIUSCATII}). The SFH  required by the user can either
follow some parametric description or be user-defined. For the parametric SFHs
four different cases are allowed: multiple bursts (analogous to the user-defined
case but limited to 5 bursts only), a truncated SFH, in which the star formation rate
(SFR) is assumed to be constant between the time of formation and truncation,
an exponential SFH and an exponential plus bursts, where up to five bursts
can be added on top of the exponential SFR. The user-defined SFHs are designed
to handle complex SFHs, with an arbitrary number of bursts permitted. In the new web
implementation, the user-defined SFH has been upgraded to include also variable
IMF slope and \alfa-enhancement.

Given the mass fraction, the metallicity, the IMF slope and the \alfa-enhancement at different ages, 
the tool computes the corresponding SED, $S_{\lambda}$, according to:

\begin{equation}
S_{\lambda}=\sum_{i=1}^{N_b}f_{m_{i}}S_{\lambda}(t_i,\Mh_i,\aFe_i,\Phi(\Gamma_i),I_{\alpha_i})
\end{equation}

\noindent where $N_b$ is the number of input bursts, $f_{m_{i}}$ is the fraction
of mass formed in the i-th burst, at age $t_i$, with metallicity $\Mh_i$, IMF
slope $\Gamma_i$ (or $\Gamma_{b_{i}}$ in the case of a bimodal IMF) and
enhancement $\aFe_i$, which in the present study can be
$+0.0$ and $+0.4$. $S_{\lambda}(t_i,\Mh_i,\aFe_i,\Phi(\Gamma_i),I_{\alpha_i})$ is the i-th
SSP spectrum, which is computed with an isochrone, $I_{\alpha_i}$, with similar
\alfa-enhancement.

\section{Summary}
\label{summary}

This paper is the second, and last, of a series where we present evolutionary
stellar population synthesis models based on the empirical library MILES
\citep{MILESI,MILESII}, which represents a significant improvement over previous
empirical libraries. In the first paper (Paper~I) we developed "base models",
employing the Padova00 scaled-solar isochrones \citep{Girardi00} and MILES to
predict spectra of intermediate- and old-aged stellar populations in the range
$\lambda\lambda$ $3540.5$--$7409.6$\,\AA, at moderately high resolution
(FWHM=2.51\,\AA). Following the abundance pattern of the solar neighborhood the
computed model SEDs are enhanced in \alfa-elements at low metallicity, whereas
around solar metallicity these predictions are approximately scaled-solar. The base
models are therefore inconsistent at low metallicities. In the present work we
further developed our models based on the MILES library, as well as theoretical 
stellar spectral libraries used in a differential way, to provide
self-consistent predictions at all metallicities. 
We emphasise that our models can
be regarded as self-consistent in the sense that both ingredients, the
theoretical stellar spectra and the isochrones are computed with the same overall
\alfa-enhancement.  
We summarise below these model developments and results:

\begin{description}

 \item[{\bf i)}] We computed fully-theoretical models based on the stellar library
of \citet{Coelho05}, with its extension to cool stars of \citet{Coelho07}, to
obtain SSPs-based differential spectral corrections
\citep{Cervantes07,Walcher09}.  We employed the BaSTI isochrones
\citep{Pietrinferni04,Pietrinferni06}, calculated with similar \aFe\ abundance
ratios to the stellar atmospheres, converted to the observational plane using extensive empirical
photometric stellar libraries, mostly from \citet{Alonso96,Alonso99}. To
improve the metallicity coverage around the solar value we computed a new
set of BaSTI isochrones with metallicity $Z=0.024$ ($\Mh=+0.15$). The obtained
theoretical differential corrections are applied to the model SEDs computed with
MILES, taking into account the \MgFe\ determinations of the library stars of
\citet{Milone11}. Therefore, at solar metallicity the scaled-solar models are
truly scaled-solar and fully based on MILES, whereas for the \alfa-enhanced
models, we apply the differential corrections to the model SEDs computed with
MILES. At low metallicities, the opposite is true; the MILES library is 
intrinsically \alfa-enhanced and we correct the model SEDs using differential
corrections to obtain scaled-solar SEDs at low metallicity.
We therefore compute self-consistent models for two abundance ratio values:
\aFe$=0.0$ and \aFe$=+0.4$. 

 \item[{\bf ii)}] In addition to the scaled-solar and \alfa-enhanced models, and to
the models we published in Paper~I using the Padova00 isochrones, we also computed
base models employing the BaSTI scaled-solar isochrones. 

 \item[{\bf iii)}] We have assessed the quality of the new models, establishing
the parameter space where the models are safe. The new
models cover the metallicity range $-2.3\leq\Mh\leq+0.4$, although the model 
predictions, and in particular those of the scaled-solar and alpha-enhanced
models, are generally safe within the range $-1.49\leq\Mh\leq+0.26$ (within the
constraints listed in Table~\ref{tab:SEDcoverage} and shown in
Fig.~\ref{fig:Q_BI} and Fig.~\ref{fig:Q_UN}). Also, the spectral range blue-ward
of $3800$\,\AA\ is unsafe for the scaled-solar models with sub-solar metallicities
and for the \alfa-enhanced models with solar and higher metallicities. We
strongly suggest the interested user of these predictions to read the notes on
the unsafe model parameter ranges that we provide in 
Table~\ref{tab:SEDproperties} and Table~\ref{tab:SEDcoverage}.  It is
particularly important to do so when the obtained fits to the data fall within
these unsafe ranges.

 \item[{\bf iv)}] We provide SSP spectra for a suite of IMF shapes: single-power law 
(unimodal), low-mass ($\lesssim0.5$\,\Msol) tapered IMF (bimodal) as
defined in \citet{Vazdekis96,CATIV} and the Kroupa Universal and Revised IMFs of
\citet{Kroupa01}. For the unimodal and bimodal IMFs we have varied the
logarithmic slope from $0.3$ (top-heavy) to $3.3$ (very bottom-heavy). A
unimodal shape with slope $1.3$ matches approximataley the Salpeter IMF, whereas a bimodal
shape with the same slope is very close to the Kroupa Universal IMF. The
predictions for the models adopting a unimodal IMF are safe for slopes smaller
than $2.3$, taking into account the above constraints. 

 \item[{\bf v)}] We have characterised the behaviour of the newly synthesised SSP
model spectra. We find that, at constant total metallicity, \Mh, the
\alfa-enhanced models, which have lower iron content and therefore have lower
opacities, show an excess of flux blue-ward of $\sim4500$\,\AA\ with respect to the
scaled-solar models. Besides this excess in flux, the \alfa-enhanced and the
scaled-solar models differ significantly in prominent features that are mainly
related to the Ca{\sc II}\,H-K and several CN and CH molecular absorptions.
Both the blue flux excess and the differences seen in these features, which 
mostly originate in the stellar atmospheres, increase with increasing age and
increasing metallicity. In comparison to the effects of the stellar spectra the
effects of the isochrones are of second order, and mostly impact on the
continuum over the whole spectral range covered by the models, confirming the
results by \citet{Coelho07}.

 \item[{\bf vi)}] We confirm previous results that the combined \mgfe\ and \mgfep\ are virtually
insensitive to the \aFe\ abundance ratio, and therefore are suitable for
measuring the total metallicity.  

 \item[{\bf vii)}] We also confirm previous findings that the
H$\gamma$ and H$\delta$ Balmer line indices are more sensitive to \aFe\ than
H$\beta$. However, we show that such sensitivity is due to the pseudo-continua
employed for the higher order Balmer index definitions, which are placed on
relevant features within the blue flux excess of the \alfa-enhanced models. 
The narrower the Balmer index definition, the lesser the sensitivity to \aFe. 

 \item[{\bf viii)}] As expected, the Fe-dominated indices, such as Fe\,4383, Fe\,5015, Fe\,5270 and
Fe\,5335 are weaker, and \mgb\ stronger in the \alfa-enhanced models. This
behaviour becomes more pronounced with increasing age and increasing metallicity. Such
trends do not apply to Mg$_1$ and Mg$_2$, which are stronger in the scaled-solar
models, as these indices show a greater sensitivity to carbon than \aFe. These and
all the other non-\alfa\ elements are depleted with respect to the
scaled-solar models. A similar behaviour is seen for C$_2$\,4668 and CN$_2$ (and
CN$_1$), which depend on C and N abundances respectively. 
Also in our models the NaD index behaves as the Fe-dominated indices. 

\item[{\bf ix)}] We also see that at ages less than $\sim0.1$\,Gyr there is a
significant increase in the strengths of various lines, most noticeably for
the IMF-sensitive indices. This is as a consequence of the influence of
supergiant stars at these young ages. Among the IMF indicators we find that,
unlike TiO$_1$, the TiO$_2$ index is almost insensitive to \aFe, in good
agreement with other authors. We also find that the TiO$_2$ index, and other
indices that show little sensitivity to age when employing a standard IMF slope
(i.e $1.3$), become more sensitive to this parameter for bottom-heavier IMFs.

 \item[{\bf x)}] The base models, which approximately follow the abundance pattern of the
Milky Way (MILES) stars, show index values that are more similar to those of
the \alfa-enhanced models at low metallicities ($\Mh\lesssim-0.4$), despite the
fact that the employed isochrones are scaled-solar. On the other hand, at high
metallicities the base models provide index values that are very close to the
scaled-solar ones. In general, the indices of the base models computed with
BaSTI and Padova00 isochrones are very similar, with the line-strengths obtained
with BaSTI being slightly larger for the Balmer lines and slightly smaller for
the metallicity indices. Therefore, for a given dataset,  somewhat older ages 
and higher metallicities would be inferred from the BaSTI-based models than 
from Padova00-based  models.

 \item[{\bf xi)}] A similar behaviour is seen for the colours, with the models based
on BaSTI being bluer than those based on Padova00. Overall the agreement is good
for the old stellar populations. For ages where the AGB contribution
is more significant we find the larger discrepancies as a result of the
varying descriptions for this evolutionary phase. 

 \item[{\bf xii)}] We use the base models as a reference for assessing the behaviour
of the scaled-solar and \alfa-enhanced colours. We find that the \alfa-enhanced
models are bluer in the blue, by $\sim0.2$\,mag in u$-$g and $\sim0.05$\,mag in
g$-$r for old populations, and redder in the red, by $\sim0.02$\,mag in r$-$i,
with respect to the base models. The best agreement between the base and
\alfa-enhanced models is reached at low metallicities. On the other hand, for the
scaled-solar models the best agreement is reached at high metallicities.

 \item[{\bf xiii)}] We compared our newly-synthesised SSP spectra and line-strengths
with other models in the literature. In particular we find that there is an
overall agreement with the SSP spectra of the CvD12 models. We find that the
overall effect of increasing \aFe\ is similar to our models, although there are
significant differences in the relative effect on different spectral features.
We also compared our line-strength indices with the predictions of the TMJ11
and S07 models. As expected we find significant offsets among the predictions
of the various models, even at scaled-solar abundance. At fixed \Feh\ the
amount of increase of the \mgb\ index is larger, by $\sim0.2$\,\AA, for our
models and the TMJ11 models, relative to the S07 and CvD12 models. The iron lines show a mild
dependence on \aFe, but we find that differences exist among the various
models. Whereas all the models agree on $\rm Fe\,5335$ and $\rm Fe\,4383$,
which decrease with \aFe\ (although with varying sensitivities), the $\rm
Fe\,5270$ and $\rm Fe\,5015$ indices are more model dependent. Although differences
remain among predictions from different models on the dependence of Balmer
lines on \aFe, our new models agree well with expectations from most models
compared here, i.e. that \hb\ remains constant (as in S07 and TMJ11), \hgf\
increases (as in CvD12 and S07),  and \hdf\ increases (for all models) with
\aFe. Finally our models agree remarkably well with CvD12 models on the
expected sensitivity of commonly used IMF-sensitive features (TiO1, TiO2, and
NaD) to \aFe. 

 \item[{\bf xiv)}] We compared the line-strengths of the new models with Galactic
Globular Cluster data. We find good agreement with \MgFe\ determinations
obtained from individual stars in these clusters. The clusters with the highest
metallicities fall in between the \alfa-enhanced and the scaled-solar model
predictions in the \mgb\ -- \fe\ index-index diagram. At very low metallicities
the differences between the index values obtained from the various models, i.e.
base, \alfa-enhanced and scaled-solar, are far less prominent. This allows us to
conclude that the base models are also suitable for studying the line-strengths
of such metal-poor stellar systems. We find that all the models fit 
the C$_2$4668 index very well, but not CN$_2$ (CN$_1$). 
This is possibly related to the general presence 
of two stellar generations in GCs, the first with a canonical
\alfa-enhanced heavy element distribution and the second characterised by an
anti-correlated CN-ONa abundance pattern.  

 \item[{\bf xv)}] We match reasonably well the $u-g$ vs. $g-r$  and $r-i$ vs. $g-r$ optical
colour-colour diagrams of luminous SDSS ETGs simultaneously if we use appropriate
combinations of \alfa-enhanced models and a varying IMF. Note that this is
not possible using base models with a standard IMF. This result is in good
agreement with our previous results obtained with the MIUSCAT base models
\citep{MIUSCATII}.

 \item[{\bf xvi)}] We obtain good fits to Early-Type Galaxy spectra with varying
mass and \alfa-enhancement. The SSP spectra are able to match the Fe- and
Mg-dominated features with great accuracy. However, we also find that in order 
to recover the \alfa-enhancement with this fitting technique it is
necessary to perform the fit within a spectral range that encompasses these
\alfa-enhancement sensitive features. Within a larger spectral range this
information is in part washed out.

 \item[{\bf xvii)}] The model grids obtained when plotting H$\beta_o$ against \mgb,
\fe\ and \mgfe\, indices, which are sensitive to Mg, Fe and total metallicity
\Mh, respectively, provide consistent age/metallicity solutions for the most
massive galaxies. 

 \item[{\bf xviii)}] Using SDSS ETG spectra stacked by velocity dispersion
we find that different Balmer line indices (and also different definitions of
the same line) provide different estimates of the age for old stellar
populations. Neither the \alfa-enhanced models nor models with varying IMF are
able to solve these discrepancies, which might be due to different
effects such as a variation of abundance ratios of specific chemical species in
the stellar atmospheres, and/or more complex SFHs than pure SSPs. In addition 
these differences may originate in the theoretical stellar spectra used to obtain the
differential corrections employed to synthesise the \alfa-enhanced SSP spectra.

 \item[{\bf xix)}] Using $\sim5000$ high quality SDSS ETG galaxy spectra, with
$S/N>25$ per \AA, we find a linear relation between a proxy for the abundance
obtained using solely scaled-solar models, \proxy$_{SS}$, and the \MgFe\ derived when
employing our new models with varying abundance. We obtain the following
relation 

\begin{equation}
\MgFe=0.59 \proxy_{SS(BaSTI)}
\end{equation} 

\noindent Very similar relations are obtained when using
the base model grids, either based on BaSTI or Padova00 isochrones. 
The method consists of plotting a good age indicator (such as e.g., H$\beta$ or
H$\beta_o$, or any other approach that provides us a reliable age
estimate) against \mgb, and against \fe, using a scaled-solar model grid as a
reference. The difference in metallicity obtained from these two diagrams, i.e.
the proxy \proxy, which is non-zero when the object is \MgFe\ enhanced, is
appropriate for estimating the "true" abundance, without making any use of  
\alfa-enhanced models. 

\end{description}

All these models can be retrieved from the MILES webpage at http://miles.iac.es.
The link includes a user-friendly web-tool facility that, among other things,
allows for the construction of SED models for a variety of SFHs. This includes a
user-defined SFH, where it is possible to combine bursts with varying IMFs and
\aFe\ abundance ratios.

\section*{Acknowledgments} 

We are indebted to the Padova group for making their isochrone
calculations available to us. We are grateful to A. de C. Milone for providing us with his
\MgFe\ estimates for the stars of the MILES library before publication. We also
are grateful to H.-S. Kim, J. Cho, R.~M. Sharples and S.-J. Yoon, who provided
us with Galactic Globular Cluster line-strength measurements prior to publication.
We are grateful to B. Barbuy, J.~L. Cervantes, I.~G. de la Rosa, M. Koleva, R.
Peletier and A. Sansom for very useful discussions. We thank the
referee H.-C. Lee for a careful reading of the manuscript and for very valuable
clarifications that helped us to improve the manuscript. We also would like to thank
J.~A. Perez Prieto for helping us in the construction of the web page for the
models. The MILES library was observed at the INT on the island of La Palma,
operated by the Isaac Newton Group at the Observatorio del Roque de los
Muchachos of the Instituto de Astrof\'{\i}sica de Canarias. This research has
made an extensive use of the SIMBAD data base and VizieR catalogue access tool
(both operated at CDS, Strasbourg, France), the NASA's Astrophysics Data System
Article Service. We made use of data retrieved from the  Sloan  Digital Sky
Survey archives ({\tt http://www.sdss.org/collaboration/credits.html}). Funding
for the  SDSS and SDSS-II  has been  provided by the Alfred P. Sloan Foundation,
the Participating Institutions, the National Science Foundation, the U.S.
Department of Energy, the National Aeronautics and Space Administration, the 
Japanese Monbukagakusho, the Max Planck Society, and   the Higher Education
Funding Council for England. PC acknowledges the financial support by FAPESP via
project 2008/58406-4  and fellowship 2009/09465-0. JFB acknowledges support from
the Ram\'on y Cajal programme by the Spanish Ministry of Economy and
Competitiveness (MINECO). This  work has been supported by the Programa 
Nacional de Astronom{\'{\i}}a y Astrof{\'{\i}}sica of MINECO, under grants 
AYA2013-48226-C3-1-P, AYA2013-48226-C3-2-P, AYA2013-48226-C3-3-P and by
the Generalitat Valenciana under grant PROMETEOII2014-069.

\bibliographystyle{mn2e}
\bibliography{papers}

\end{document}